%% file: main.tex
\documentclass[12pt]{article}

\usepackage{natbib}
\usepackage{newtxtext,newtxmath}
\usepackage{graphicx}

\usepackage{url}
\usepackage{csquotes}
\usepackage{tikz}
\usetikzlibrary{shapes.geometric, arrows.meta, positioning, fit, calc, backgrounds, shadows}
\usepackage{amsmath}
\usepackage{caption}
\usepackage{subcaption}

\newcommand{\emphasis}[1]{\textbf{#1}}

\definecolor{myblue}{HTML}{C9DFFF}
\definecolor{myred}{HTML}{FFD5D2}
\definecolor{mygreen}{HTML}{D4EFDF}
\definecolor{mygray}{HTML}{E0E0E0}
\definecolor{bordercolor}{HTML}{4A4A4A}
\definecolor{accentblue}{HTML}{5B9BD5}
\definecolor{textgray}{HTML}{333333}

\usepackage[letterpaper,margin=1in]{geometry}

\renewenvironment{abstract}
	{\quotation}
	{\endquotation}

\makeatletter
\renewcommand{\fnum@figure}{\textbf{Figure \thefigure}}
\renewcommand{\fnum@table}{\textbf{Table \thetable}}
\makeatother

\newenvironment{itquote}
  {\begin{quote}\itshape}
  {\end{quote}\ignorespacesafterend}

\def\scititle{
Towards Generalizable AI-Assisted Misinformation Inoculation: \\
Protecting Confidence Against False Election Narratives 
}
\title{\bfseries \boldmath \vspace{-1in}\scititle}

\author{
	Mitchell Linegar$^{1\ast}$,
	Betsy Sinclair$^{2}$,
	Sander van der Linden$^{3}$\and
        R. Michael Alvarez$^{1}$\\
	\small$^{1}$Linde Center for Science, Society, and Policy, California Institute of Technology, Pasadena \& 91125, USA.\and
	\small$^{2}$Department of Political Science, Washington University in St. Louis, St. Louis \& 63130, USA.\and
        \small$^{3}$Department of Pyschology, Cambridge University, Cambridge \& CB2 3EB, UK.\\
	\small$^\ast$Corresponding author. Email: \url{mlinegar@caltech.edu} 
}

\date{}

\usepackage{xcolor}

\begin{document}

\maketitle

\begin{abstract} 
\bfseries

We present a generalizable AI-assisted framework for rapidly generating effective ``prebunking" interventions against misinformation. Like mRNA vaccine platforms, our approach uses a stable template structure that can be quickly adapted to counter emerging false narratives. In a preregistered two-wave experiment with 4,293 U.S. registered voters, we test this framework against politically-charged election misinformation -- one of the most challenging domains for misinformation intervention. Our design directly tests scalability by comparing human-reviewed and purely AI-generated inoculation messages. We find that LLM-generated prebunking significantly reduced belief in election rumors (persisting for at least one week) and increased confidence in election integrity across partisan lines. Purely AI-generated messages proved as effective as human-reviewed versions, with some achieving larger protective effects, demonstrating that effective misinformation inoculation can be achieved at machine speed without proportional human effort, offering a scalable defense against the accelerating threat of false narratives across all domains.

\end{abstract}

\newpage

\noindent

\noindent {\bf Significance Statement}\\

\noindent Misinformation spreads faster than our ability to counter it. We present an AI-assisted framework that rapidly generates effective prebunking interventions. In a preregistered two-wave experiment (N=4,293 U.S. registered voters), we demonstrate that our modular approach, requiring only a stable prompt template and authoritative information, produces inoculation messages against diverse false narratives. By comparing human-reviewed versus purely AI-generated interventions, we directly test scalability. Fully automated messages proved as effective as human-reviewed versions at reducing belief in election misinformation and increasing electoral confidence, with effects persisting one week. This finding changes the resource requirements for combating misinformation: effective inoculation no longer demands proportional human effort. Our framework generalizes across domains, providing institutions with validated tools to counter false narratives at unprecedented speed.

\section{Introduction}

Misinformation has become a challenging problem in contemporary societies \citep{LEWANDOWSKY2017353}. Its reach extends across virtually every domain of public concern. Climate action stalls when false narratives about sustainability gain traction \citep{lewandowsky2021b}. Public health suffers when, as during the COVID-19 pandemic, vaccine misinformation undermines immunization efforts \citep{loomba2021measuring}. Perhaps most alarmingly, democratic processes themselves face disruption: from the Brexit referendum \citep{henkel2021destructive} to the 2016 US presidential election \citep{guess2020} to electoral contests in India \citep{BADRINATHAN_2021}, misinformation has become a recurring threat to informed civic participation.

Election misinformation is a particularly difficult problem and is considered as a persistent threat in many democracies, including the United States \citep{ecker_etal_2024}.  It can spread quickly, and in the days before an election can be difficult to counter.  It can erode the confidence of voters and stakeholders in the conduct of an election, undermining the legitimacy of an otherwise free and fair election. 
Election misinformation and concerns about election integrity can threaten the orderly transfer of power from one party to another \citep{levy2021winning, alvarez2021voting, Berlinski_etal_2023}. For example, the attack on the U.S. Capitol on January 6th, 2021 was motivated by false claims of election fraud and election manipulation in the 2020 presidential election \citep{selectcommittee2022}. From encouraging violent protests to justifying hate crimes, from the degradation of civic community to the establishment of terrorism, election rumors have a powerful capacity to destabilize democratic government\citep{Albertson2020Conspiracy,Piazza2024Allegations,Jungkunz2024Populist}.

Addressing this challenge is complicated by the asymmetric nature of the information environment: malicious actors can generate and disseminate numerous, rapidly evolving false narratives relatively cheaply, while defenses like fact-checking struggle to keep pace 
\citep{wardle2017information}. Psychological inoculation \citep{McGuire1964}, or ``prebunking," offers a promising proactive strategy by building cognitive resistance to misinformation prior to exposure \citep{roozenbeek2022, vanderLinden_2022}. However, traditional prebunking methods face a critical bottleneck: the need for experts to synthesize persuasive false rumors into attenuated exposures that confer resistance without causing harm. Here, we investigate whether generative artificial intelligence (AI) can overcome this limitation, enabling the rapid development and scaling of prebunking interventions necessary to counter the speed and diversity of modern misinformation. 

To overcome this scalability challenge, we propose an AI-assisted framework that functions analogously to mRNA vaccine platforms, maintaining a stable core structure while allowing rapid adaptation to new threats. Our system combines a rigorously tested prompt template with authoritative contextual information (such as official election security guidelines) to generate targeted inoculations against specific false narratives. This modular design enables rapid response to emerging misinformation while maintaining consistent quality and effectiveness. We test this using five common myths about the 2024 U.S. election, demonstrating that brief, AI-written arguments can preemptively counter misinformed rhetoric about election integrity for rumors \textit{previously unseen by the AI}, with effects lasting at least a week and without evidence of backlash.

The unique temporal dynamics of election cycles compound the challenges of combating misinformation. Election rumors gain potency precisely when democratic institutions are most vulnerable: in the charged atmosphere leading up to voting, when emotions run high and time for careful deliberation runs short. Traditional debunking strategies, which require extensive fact-checking and expert analysis, cannot match the speed at which false election narratives proliferate across social networks. Although our prebunking interventions demonstrate the additional benefit of bolstering participants' confidence in electoral processes, this represents a secondary outcome; our core focus is proactive inoculation against the specific myths that threaten informed democratic participation. Our scalable approach has thus been tested in a potentially challenging setting.

Mitigating election-related misinformation and disinformation has been the subject of an important and growing body of research. Recent work has explored various approaches to do so, including fact-checking and inoculation strategies \citep{Voelkel2024Megastudy}.
However, the inoculation literature has only recently begun to study election misinformation specifically \citep{lockhartetal2024}.
Traditional methods often lag behind the emergence of new narratives. The rapid spread of many different pieces of misinformation through social media and other channels thus necessitates innovative and scalable solutions. We argue that our scalable approach points to the development of more generalized AI-assisted misinformation inoculation.  The method that we develop below can be quickly and easily re-trained for testing and use on non-election rumors and false narratives.  This methodology can be generalized for use across issue domains, and thus could easily be deployed in other contexts where rapidly-evolving and spreading rumors and misinformation are a pressing problem.  

\section{Previous Research}
Prebunking has shown to be a successful approach to combating misinformation \citep{Lu2023Psychological, vanderLinden_2022}.
Inspired by inoculation theory in psychology, prebunking aims to build cognitive resistance to misinformation by exposing individuals to weakened forms of false claims along with factual counterarguments before they encounter more persuasive misinformation. Prebunking leverages the cognitive processing mechanisms by which individuals develop mental schemas and defenses against false but otherwise potentially persuasive claims, thereby establishing attitudinal resistance prior to exposure to manipulative content. It can do so both by preempting specific factual misconceptions and by building more general anti-misinformation skills like exposing disinformation tactics and developing media literacy skills. This method has shown potential in various domains, including climate change denial, conspiracy theories, and vaccine misinformation \citep{Traberg2022Psychological,Compton2021Inoculation,Lewandowsky2021Countering}. However, its practical application at the scale and speed required by modern information ecosystems remains a challenge.

While traditional inoculation theory suggests that prebunking is most effective when administered before exposure to misinformation \citep{compton2020}, recent research indicates that post-exposure ``therapeutic" inoculation can also be beneficial \citep{ivanov2017,vanderlinden2022}. The success of post-exposure inoculation is particularly important for Republicans and those that endorse other (non-election) conspiracy theories, as these groups are already more likely to have been exposed to (and believe in) election misinformation. The interaction between partisan identity and prebunking effectiveness remains underexplored, with theoretical ambiguity about whether Republican identifiers—who have shown heightened susceptibility to certain forms of misinformation—may exhibit differential treatment responses to prebunking interventions due to divergent information ecosystems, elite cuing effects, or motivated reasoning processes.

A critical bottleneck limiting the societal impact of traditional prebunking is the reliance on time-consuming human expertise to identify emerging narratives, craft specific weakened counterarguments, and produce tailored inoculation messages. This manual process inherently struggles to match the pace and sheer volume of misinformation generated online, particularly during dynamic events like election campaigns. Rumors proliferate faster than bespoke interventions can typically be developed and deployed. In an effort to establish a scaleable intervention, we introduce a novel element to the prebunking approach: the use of Large Language Models (LLMs) to assist in generating inoculation content. This application of artificial intelligence offers the potential for rapid, scalable production of prebunking materials tailored to emerging misinformation narratives, addressing the core limitation of manual methods.

The scalability of anti-misinformation interventions is therefore not just an improvement, but potentially a necessary innovation. Recent work on election misinformation suggests that human-written fact checks do not durably increase confidence in election administration \citep{carey_2024_targeted}. Those hoping to inform the public about the realities of the integrity of elections must currently respond to each false election rumor that arises. The sheer multiplicity of rumors makes a purely reactive, one-by-one debunking strategy insufficient.  
With their ability to take direction and to mimic examples, LLMs are a natural option for quickly and semi-automatically producing effective inoculation doses of misinformation, generated at a speed unachievable by human authors alone.

\section{Research Design}

To test the efficacy of LLM-generated prebunking, we conducted a two-wave experimental study prior to the 2024 U.S. general election. Participants were randomly assigned to read a persuasive, human-written article endorsing one of five commonly believed election myths. Before doing so, participants were randomly assigned to either a treatment group, which received an LLM-generated prebunking article addressing the myth, or a control group, who received an LLM-generated article about return-to-office policies. All participants were then exposed to a ``full exposure" of misinformation in the form of a persuasive article promoting the assigned election myth. We measured participants' beliefs in election myths, confidence in true election facts, and overall trust in election integrity immediately after the intervention and one week later.

Prompt creation is a crucial part of our experiment: if LLM-generated prebunking messages are to be scalable, it is important to develop prompts that are able to push against new and emerging disinformation. Our article generation process is described in Figure \ref{fig:prompt_process}. We break the process into two panels.  In the left panel, we describe how we use an iterative process combination of AI and human review to generate a high-quality prompt (which in the figure is $\text{Prompt}^*$). The LLM is given a ``full exposure" article and a prompt asking it to produce an inoculation article. After each generation, the input prompt was edited by the researchers. This process was repeated until the researchers determined no additional edits to the generated inoculation article were required. The right panel shows how we take $\text{Prompt}^*$ for each rumor and use it to produce a final high-quality inoculation article, which we use as treatment.

Next, our experimental design is described in Figure \ref{fig:survey_workflow}. Participants are randomly assigned to one of five election rumors. Our design makes it possible to compare the efficacy of prebunking articles written by a collaboration between human experts and LLMs, and those written solely by LLMs, by comparing treatment effects for participants given the human review procedure and a wholly automated one.

In this paper we test two main preregistered hypotheses.
First, \textbf{H1}: Participants exposed to prebunking of a specific election-related rumor will report lower confidence in that rumor compared to the control group.
Second, \textbf{H2}: Participants exposed to prebunking of a specific election-related rumor will report higher confidence that their votes will be accurately counted in the next election compared to the control group.
In addition, we report preregistered analyses of the temporal durability and heterogeneity of these effects.\footnote{We do not specifically focus on these hypotheses due to space constraints. Temporal durability is listed as H4 in our preregistration, and hypothesizes that treatment effects will persist for one week, but decline in magnitude. Treatment effect heterogeniety is listed as H6 in our pregistration. Refer to Section \ref{sec:surveymethods} for more details about our preregistration, and deviations from our preregistration.}

Three measures of election confidence are common in the literature, corresponding with belief that voters' own ballots were correctly counted, that ballots in their locality were correctly counted, and that ballots across the nation were correctly counted. In this paper we focus on confidence in the national election. Prior work has identified large gaps in perceptions of vote counting at the national level, where confidence is lower, and local levels, where confidence is higher \citep{sances2015partisanship}. Fraud at the level of the national election is thus more likely to be salient to voters, while also being more directly important to policy makers. 

Our findings demonstrate that LLM-generated prebunking can effectively reduce belief in specific election-related rumors and increase confidence in accurate election facts, with these effects persisting for at least one week. The intervention appears to be similarly effective across party lines and ideologies, suggesting its potential as a broadly applicable tool for combating election misinformation.

\clearpage
\newpage

\begin{figure}[ht!]
    \centering
    \tikzset{
        basenode/.style={
            rectangle, 
            rounded corners=4pt, 
            draw=bordercolor!60, 
            line width=0.7pt,
            text width=2.8cm, 
            minimum height=1.5cm, 
            align=center, 
            font=\small\sffamily,
            text=textgray,
            drop shadow={shadow xshift=1pt, shadow yshift=-1pt, opacity=0.15, fill=gray!50}
        },
        misinfo/.style={basenode, fill=myred, draw=myred!70},
        common/.style={basenode, fill=myblue, draw=myblue!70},
        review/.style={basenode, fill=mygreen, draw=mygreen!70, text width=3.2cm},
        rumorlist/.style={
            basenode, 
            fill=mygray, 
            draw=mygray!70,
            text width=2.2cm, 
            minimum height=1.0cm,
            inner sep=5pt,
            font=\small\sffamily
        },
        flowarrow/.style={
            -Stealth, 
            line width=0.8pt,
            color=bordercolor!70,
            shorten >=1.5pt,
            shorten <=1.5pt
        },
        bendarrow/.style={
            -Stealth,
            line width=0.8pt,
            rounded corners=8pt 
        },
        paneltitle/.style={
            font=\large\sffamily\bfseries,
            text=textgray
        }
    }

\vspace{-2cm}
    
    \begin{subfigure}[t]{0.58\textwidth} 
        \centering
        \begin{tikzpicture}[node distance=1.8cm and 2cm]
            \pgfmathsetmacro{\titleY}{6.5} 
            \node[paneltitle] at (3, \titleY) {Prompt Creation Process};
            
            \node (misinfo) [misinfo] at (0,3.8) {Full Exposure\\Article\\{\footnotesize (Human)}\\[3pt]{\small\emphasis{Voter fraud}}};
            \node (facts) [common] at (0,0.5) {Election\\Security Facts\\{\footnotesize (CISA)}};
            \node (prompt) [common] at (0,-2) {Prompt};
            
            \node (aiart) [misinfo] at (5.5,1.7) {Inoculation\\Article\\{\footnotesize (AI-generated)}\\[3pt]{\small\emphasis{Voter fraud}}};
            \node (review) [review] at (5.5,-1) {Human Review\\{\footnotesize \& Prompt Edits}};

            \draw[flowarrow] (aiart.south) -- (review.north);
            
            \draw[flowarrow] (misinfo.east) -- ++(0.8cm,0) |- ($(aiart.west)+(0,0.5cm)$);
            \draw[flowarrow] (facts.east) -- ++(0.4cm,0) |-  (aiart.west);
            \draw[flowarrow] (prompt.east) -- ++(0.8cm,0) |- ($(aiart.west)+(0,-0.5cm)$); 
            
            \draw[flowarrow] (review.south) |- ($(prompt.east)+(0,-0.4cm)$);

            \node[
                font=\footnotesize\sffamily, 
                text width=5cm, 
                align=center, 
                below=1.2cm of prompt,
                draw=accentblue!50,
                rounded corners=3pt,
                inner sep=8pt,
                fill=accentblue!5,
                text=textgray
            ] {Iterate until human reviewers approve\\the generated article without edits\\[2pt]{\small Final prompt: \emphasis{Prompt*}}};
            
        \end{tikzpicture}
        \caption{Iterative prompt refinement process}
        \label{fig:prompt_creation}
    \end{subfigure}%
    \hfill
    \begin{subfigure}[t]{0.41\textwidth} 
        \centering
        \begin{tikzpicture}[node distance=1.8cm and 2cm]
            \pgfmathsetmacro{\titleY}{6.5} 
            \node[paneltitle] at (2.5, \titleY) {Article Generation};
            
            \node (rumors) [rumorlist] at (0.5,4.3) {
                \textbf{Election Myths}\\{\footnotesize (5 total)}
            };
            
            \node (misinfohack) [misinfo, text width=2.8cm] at (0.5,1.2) {Full Exposure\\Article\\{\footnotesize (Human)}\\[3pt]{\small\emphasis{Hacking}}};
            
            \node (factsright) [common, text width=2.8cm] at (0.5,-1.8) {Election\\Security Facts\\{\footnotesize (CISA)}};
            
            \node (promptstar) [common, text width=2.8cm] at (0.5,-4.1) {\emphasis{Prompt*}};
            
            \node (aiartstar) [misinfo, text width=2.8cm] at (4.5,-1.8) {Inoculation\\Article*\\{\footnotesize (AI-generated)}\\[3pt]{\small\emphasis{Hacking}}};
            
            \draw[flowarrow, line width=1pt] (rumors.south) -- (misinfohack.north);
            \draw[flowarrow] (misinfohack.east) -- ++(0.3cm,0) |- ($(aiartstar.west)+(0,0.5cm)$);
            \draw[flowarrow] (factsright.east) -- (aiartstar.west);
            \draw[flowarrow] (promptstar.east) -- ++(0.3cm,0) |- ($(aiartstar.west)+(0,-0.5cm)$);
            
            \node[
                font=\footnotesize\sffamily, 
                text width=3.8cm, 
                align=center, 
                below=0.4cm of promptstar,
                text=textgray
            ] {Apply \emphasis{Prompt*} to generate inoculation articles for all myths};
        \end{tikzpicture}
        \caption{Scaled article generation}
        \label{fig:article_writing}
    \end{subfigure}
    
    \caption{\textbf{Article generation procedure.} 
    The left panel (a) describes the construction of the prompt. Given a human-written article endorsing one of the rumors (e.g., ``Voter fraud'') and an initial prompt, an LLM generates a potential inoculation article. This article is reviewed by humans, and the prompt is iteratively refined until the generated article requires no edits. Red indicates rumor-specific text or information (the human-written rumor-endorsing article and the corresponding AI-generated inoculation article). Blue indicates information available to the LLM for all rumors (the refined prompt and election security FAQs from CISA). Green indicates the human review and prompt editing stage. 
    In the right panel (b), once the refined prompt (\emphasis{Prompt*}) is finalized, it is used to generate inoculation articles for all five election myths considered: Voter fraud, Voter rolls, Hacking, Blue shift, and Voting machines. The panel illustrates this scaled generation process using ``Hacking'' as an example myth.}
    \label{fig:prompt_process}
\end{figure}

\clearpage
\newpage

\begin{figure}[ht!]
    \centering
    \tikzset{
        phasenode/.style={
            rectangle, 
            rounded corners=5pt, 
            draw=bordercolor!60, 
            line width=0.7pt,
            fill=myblue,
            text width=5cm, 
            minimum height=1.4cm, 
            align=center, 
            font=\small\sffamily,
            text=textgray,
            drop shadow={shadow xshift=1pt, shadow yshift=-1pt, opacity=0.15, fill=gray!50},
            inner sep=5pt 
        },
        articlenode/.style={
            rectangle, 
            rounded corners=5pt, 
            draw=myred!70, 
            line width=0.7pt,
            fill=myred,
            text width=4cm, 
            minimum height=1.3cm, 
            align=center, 
            font=\small\sffamily,
            text=textgray,
            drop shadow={shadow xshift=1pt, shadow yshift=-1pt, opacity=0.15, fill=gray!50},
            inner sep=4pt 
        },
        placebonode/.style={
            articlenode, 
            fill=mygreen!90,
            draw=mygreen!70
        },
        randomnode/.style={
            diamond, 
            draw=bordercolor!60, 
            line width=0.7pt,
            fill=mygray,
            aspect=2.2, 
            minimum width=3.8cm,
            align=center, 
            font=\small\sffamily,
            text=textgray,
            inner sep=3pt, 
            drop shadow={shadow xshift=1pt, shadow yshift=-1pt, opacity=0.15, fill=gray!50}
        },
        fig2flowarrow/.style={
            -Stealth, 
            line width=1pt,
            color=bordercolor!70,
            shorten >=2pt,
            shorten <=2pt
        },
        branchlabel/.style={
            font=\footnotesize\sffamily\bfseries,
            text=accentblue,
            fill=white,
            inner sep=3pt,
            rounded corners=2pt,
            draw=accentblue!50,
            line width=0.5pt
        }
    }
    \vspace{-2cm}

    \begin{tikzpicture}[node distance=0.8cm and 0.5cm] 
        \node (preq) [phasenode] {\emphasis{Pre-Treatment Questions}\\[2pt]{\footnotesize Demographics • Political Affiliation}\\{\footnotesize Election Beliefs • Confidence Measures}};
        
        \node (rand1) [randomnode, below=0.8cm of preq] {\emphasis{Randomize}\\to Myth\\[2pt]{\footnotesize 1 of 5 myths}};
        
        \node (rand2) [randomnode, below=0.8cm of rand1] {\emphasis{Randomize}\\Treatment vs\\Control};
        
        \node (treatinoc) [articlenode, below left=1.0cm and 2.8cm of rand2] 
            {\emphasis{Inoculation Article}\\[2pt]{\footnotesize AI-generated prebunk}\\{\footnotesize Target: Voter fraud}};
        \node[branchlabel, above=17pt of treatinoc] {Treatment Group};
        
        \node (controlplac) [placebonode, below right=1.0cm and 2.8cm of rand2] 
            {\emphasis{Placebo Article}\\[2pt]{\footnotesize AI-generated neutral}\\{\footnotesize Topic: Remote work}};
        \node[branchlabel, above=17pt of controlplac] {Control Group};
        
        \coordinate (midpoint) at ($(treatinoc.south)!0.5!(controlplac.south)$);
        \node (misinfoexp) [articlenode, below=0.1cm of midpoint, text width=5.5cm] 
            {\emphasis{Full Exposure Article}\\[2pt]{\footnotesize Human-written misinformation}\\{\footnotesize Example: Voter fraud claims from Breitbart}};
        
        \node (postq) [phasenode, below=0.8cm of misinfoexp] 
            {\emphasis{Post-Treatment Questions}\\[2pt]{\footnotesize Immediate assessment}\\{\footnotesize Election Beliefs • Confidence Measures}};
        
        \node (followup) [phasenode, below=0.8cm of postq] 
            {\emphasis{Follow-up Questions}\\[2pt]{\footnotesize Delayed assessment (1 week)}\\{\footnotesize Election Beliefs • Confidence Measures}};
        
        \draw[fig2flowarrow] (preq) -- (rand1);
        \draw[fig2flowarrow] (rand1) -- (rand2);
        
        \draw[fig2flowarrow] (rand2.west) .. controls +(-1.2,-0.6) .. (treatinoc.north); 
        \draw[fig2flowarrow] (rand2.east) .. controls +(1.2,-0.6) .. (controlplac.north);
        
        \coordinate (treat_out_fig2) at ($(treatinoc.south)+(0,-0.1cm)$);
        \coordinate (control_out_fig2) at ($(controlplac.south)+(0,-0.1cm)$);
        \draw[fig2flowarrow, rounded corners=6pt] (treat_out_fig2) |- (misinfoexp.west);
        \draw[fig2flowarrow, rounded corners=6pt] (control_out_fig2) |- (misinfoexp.east);
        
        \draw[fig2flowarrow] (misinfoexp) -- (postq);
        \draw[fig2flowarrow] (postq) -- (followup);
        
        \begin{scope}[on background layer]
            \node[
                fit=(preq),
                fill=myblue!8,
                rounded corners=8pt,
                inner sep=10pt
            ] {};
            \node[
                fit=(postq)(followup),
                fill=myblue!8,
                rounded corners=8pt,
                inner sep=10pt
            ] {};
        \end{scope}
    \end{tikzpicture}
    
    \caption{\textbf{Experimental design.} 
    Blue plates are common to all participants, regardless of assigned rumor. Gray diamonds represent where randomization occurs. Red plates indicate which articles participants are assigned to. We measure confidence in election myths and confidence in the administration of the election before and after treatment, and then again one week later. In the example, the participant is assigned to the treatment condition of the ``Voter fraud'' rumor, reading an LLM-written inoculation article before reading a human-written article claiming that voter fraud changed election results. Had the participant been assigned to the control condition they would have instead read an LLM-written article about remote work, and then read the misinformation-containing article.}
    \label{fig:survey_workflow}
\end{figure}

\clearpage
\newpage

\section{Results}\label{sec:results}

\begin{figure}[!hp]
    \centering
    \includegraphics[width=0.95\linewidth]{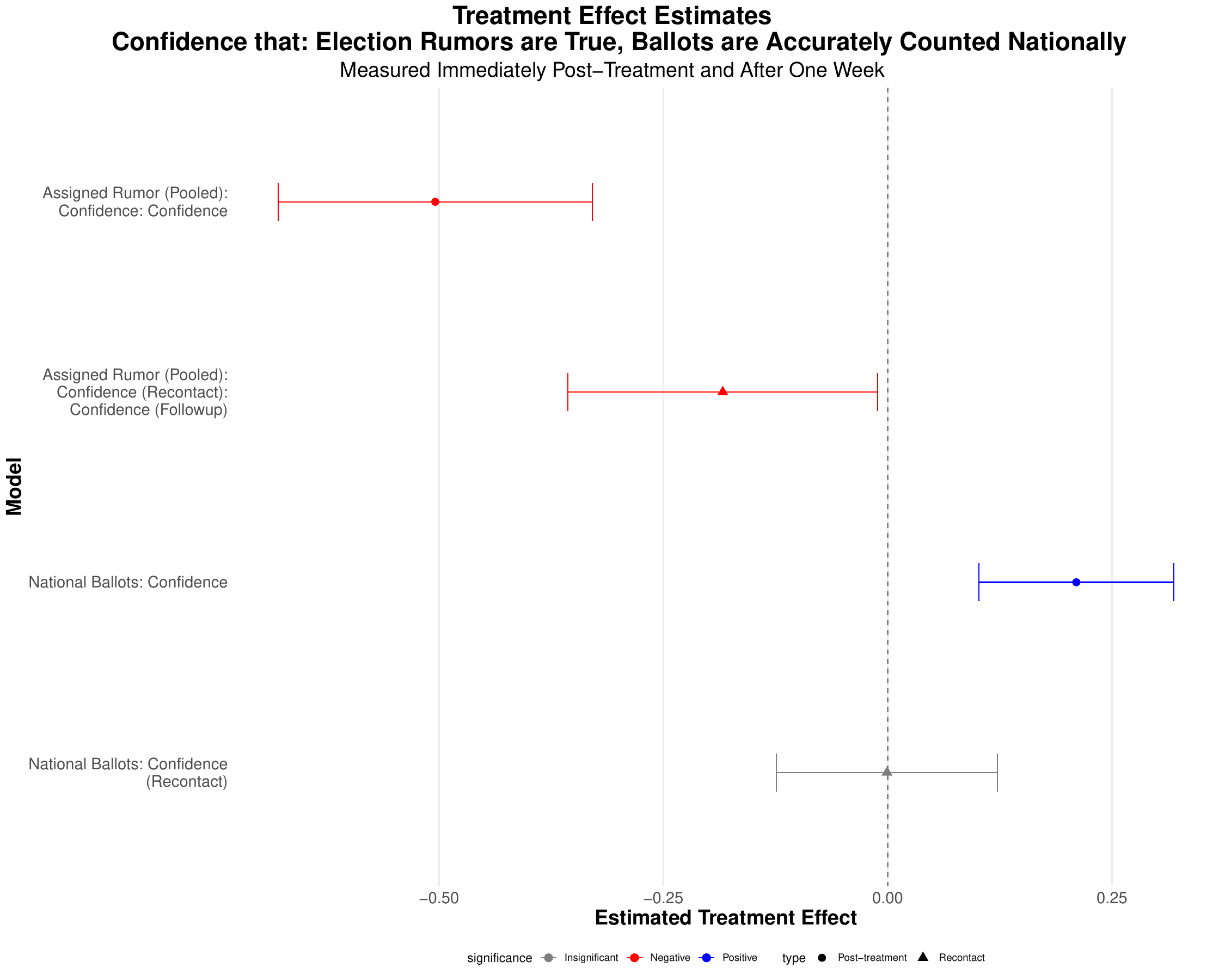}
    \caption{\textbf{Estimated treatment effects.} Results are averaged over all participants, regardless of treatment arm (that is, pooling across all five assigned rumors). The ``National Ballots" questions represent respondent confidence that ballots will be counted accurately at the national level (again, including participants in all five treatment arms). The ``Assigned Rumor" questions represent respondent confidence that the false election rumor they were assigned to is true. All questions are measured on a 0-10 scale. Recontact measures are taken one week after treatment. Error bars represent 95\% confidence intervals.}
    \label{fig:main}
\end{figure}

Prebunking is able to push back against the specific pieces of election misinformation included in the ``full exposure" articles shown to all participants, to increase knowledge of true election-related facts, and, in the short term, to increase confidence in national-level elections. Figure \ref{fig:main} shows the average treatment effect (ATE) estimates from each regression\footnote{Equivalent regressions can be found at Table \ref{tab:main}. Similar plots for confidence that one's own ballot or ballots in one's county will be counted accurately are contained in Figures \ref{fig:treatment_effects_election} and \ref{fig:treatment_effects_rumor}, with equivalents in Tables \ref{tab:rumor} and \ref{tab:other_election_integrity}.}. We focus on the pooled estimates, which aggregate across rumor assignments. All regressions include age, gender, race, education, party, ideology, urban status, level of political interest, degree of endorsement of populist and conspiratorial beliefs, and level of susceptibility to misinformation (as measured by the MIST-8 \citep{maertens2024misinformation}).

\subsection*{Confidence in the Truth of False Election Rumors}

\begin{figure}[!hp]
    \centering
    \includegraphics[width=0.95\linewidth]{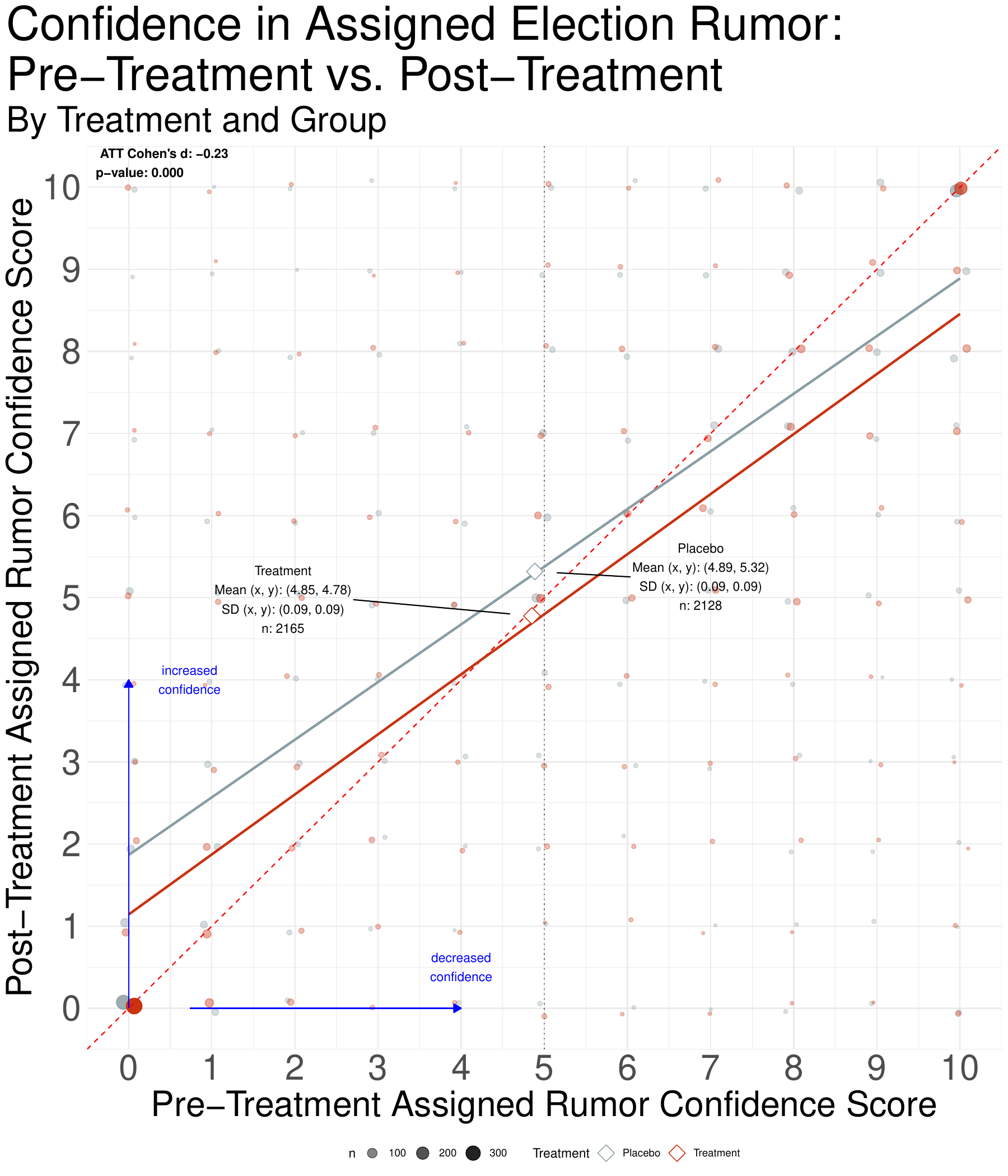}
    \caption{\textbf{Pre-treatment confidence in election rumors vs. post-treatment confidence in the rumor.} Diamonds indicate average pre- and post-treatment levels of confidence on a 0 to 10 scale, by treatment status. The red dotted 45-degree line indicates the same pre- and post-treatment measured confidence; above the line indicates increased confidence post-treatment. Respondents in the control condition (who read the ``full exposure" articles with no inoculation) show higher post-treatment beliefs in election rumors. Inoculation mitigates or reverses these increases.}
    \label{fig:rumor_conf}
\end{figure}

In support of our pre-registered hypothesis H1, the LLM-written inoculation decreases confidence in the veracity of the related election rumor. The effect is large -- the pooled treatment effect is around 0.5 on a 10 point scale -- and is still statistically significant after a week, though it decreases in magnitude (see Figure \ref{fig:main}). As we show in Figure \ref{fig:rumor_conf}, this effect is driven both by treated participants having lower confidence in election rumors (on average going from 4.85 pre-treatment to 4.78 post-treatment, around one SD) as well as control participants having higher confidence (on average going from 4.89 to 5.32, around 5 SD).

The results are further driven both by respondents who already had low belief in the rumor (e.g. reducing a 1 to a 0), as well as by respondents who firmly believed in the rumor (e.g. reducing a 10 to an 8). We see little evidence of heterogenous treatment effects overall: Figure \ref{fig:pre_post_cisa_party} (see also Table \ref{tab:party}) show that neither conditioning on party nor interacting party with treatment status has a significant effect; estimated treatment effects are significantly different than zero in each case. Importantly, we do not see evidence that inoculation articles written with human assistance are more effective than those using only AI: the interaction between treatment and an indicator for whether the article is written with human assistance is not distinguishable from zero (see Table \ref{tab:human_in_the_loop}). 



\subsection*{Confidence in Election Administration}\label{sec:national_ballots}

\begin{figure}[!hp]
    \centering
    \includegraphics[width=0.95\linewidth]{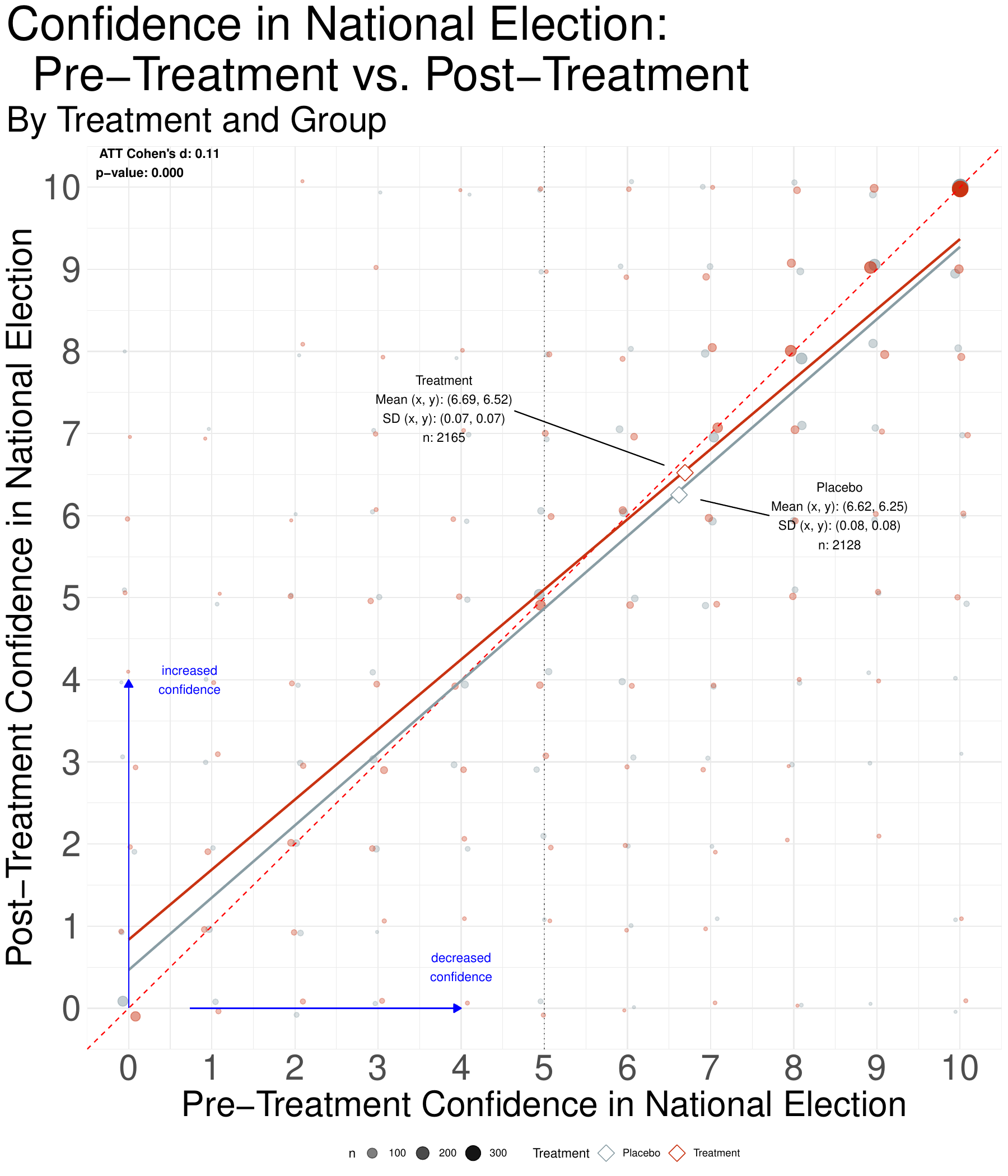}
    \caption{\textbf{Pre-treatment confidence in integrity of national election vs. post-treatment confidence.}  Diamonds indicate average pre- and post-treatment levels of confidence on a 0 to 10 scale, by treatment status. The red dotted 45-degree line indicates the same pre- and post-treatment measured confidence; above the line indicates increased confidence post-treatment. Respondents in the control condition (who read the ``full exposure" articles with no inoculation) show lower post-treatment confidence in the integrity of the national election. Inoculation mitigates or reverses these decreases.}
    \label{fig:election_conf}
\end{figure}

In support of pre-registered hypothesis H2 (that prebunking will increase confidence in the administration of the election), we see in Figure \ref{fig:main} that treatment increases confidence in the administration of the national election. As we show in Figure \ref{fig:election_conf}, this effect is driven both by participants in the placebo condition demonstrating lower confidence in the national election (going from an average of 6.62 to 6.25; a difference of around 4 SD, while treated participants go from an average of 6.69 to 6.52, a smaller decrease of around 2 SD). These results are replicated in Table \ref{tab:main}, which shows an average treatment effect of 0.189 (SE: 0.046) on a ten-point scale. 

We see little evidence of heterogenous treatment effects overall: Figure \ref{fig:election_conf} (see also Table \ref{tab:party}) shows that neither conditioning on party nor interacting party with treatment status has a significant effect: in either case the magnitude and direction of the effect remains unchanged. We see no evidence that the effects of human-assisted and purely AI-written articles are distinguishable (see Table \ref{tab:human_in_the_loop}). Counter to our pre-registered hypothesis concerning the durability of these effects, these results (increased confidence in the national election) are not significant when measured one week post-treatment, as shown in Figure \ref{fig:main}.

\subsection{Party-Level Heterogeneity}

One concern with prebunking interventions is the potential for backlash effects, where exposure to counter-attitudinal information strengthens rather than weakens pre-existing false beliefs. In our context, backlash would manifest as participants who initially believed election rumors exhibiting heightened confidence in those false narratives after exposure to our prebunking treatment, or alternatively, showing decreased confidence in the broader electoral system. Such effects would be particularly concerning given that prebunking aims to inoculate against misinformation rather than inadvertently reinforce it.

\begin{figure}[ht]
    \centering
    \includegraphics[width=0.9\linewidth]{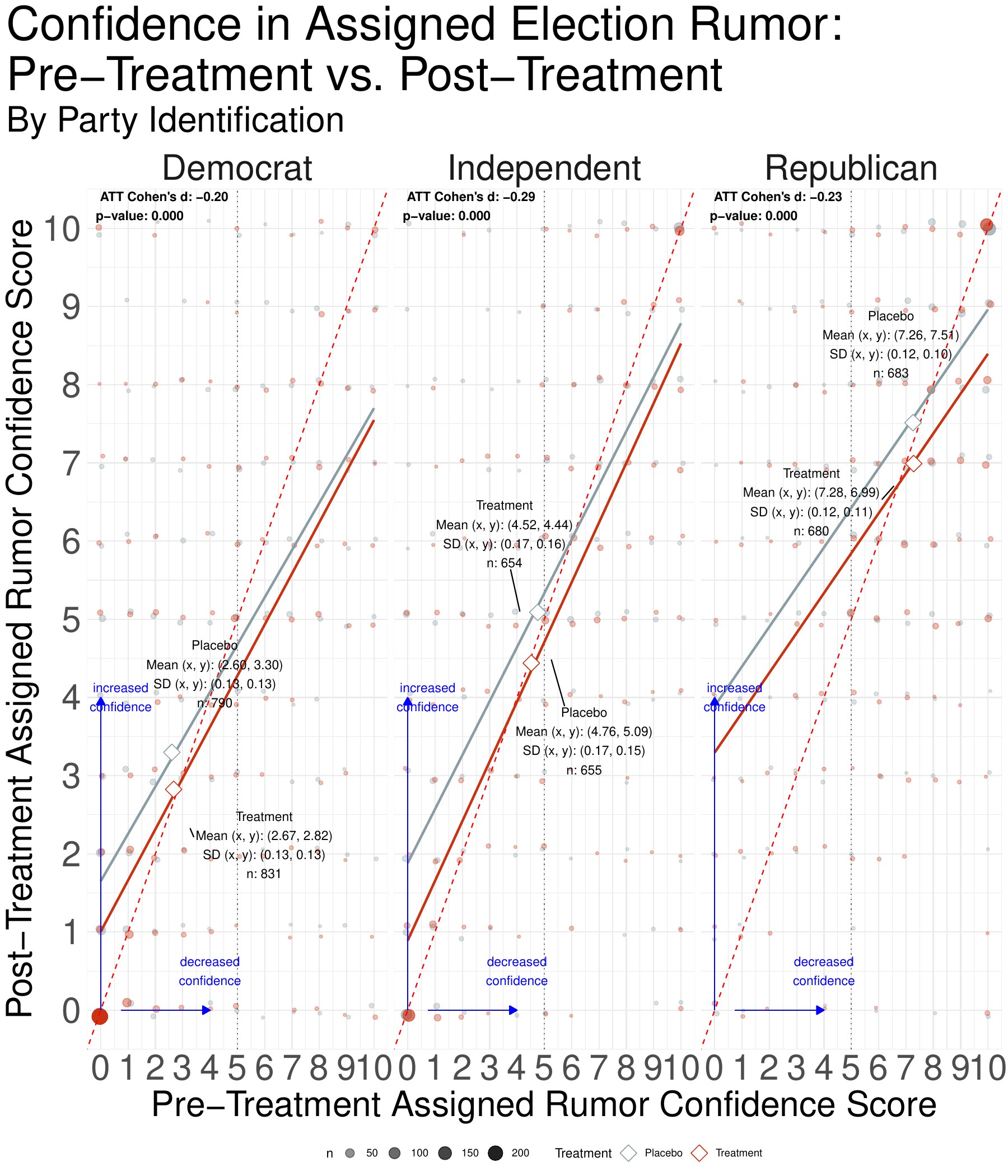}
    \caption{\textbf{Pre-treatment vs. post-treatment confidence in assigned election rumor, all rumors pooled together, by party and treatment status.}  Diamonds indicate average pre- and post-treatment levels of confidence on a 0 to 10 scale, by treatment status. The red dotted 45-degree line indicates the same pre- and post-treatment measured confidence; above the line indicates increased confidence post-treatment. Across parties, respondents in the control condition (who read the ``full exposure" articles with no inoculation) show higher post-treatment beliefs in election rumors. Inoculation mitigates or reverses these increases.}
    \label{fig:pre_post_cisa_party}
\end{figure}

\begin{figure}[ht]
    \centering
    \includegraphics[width=0.9\linewidth]{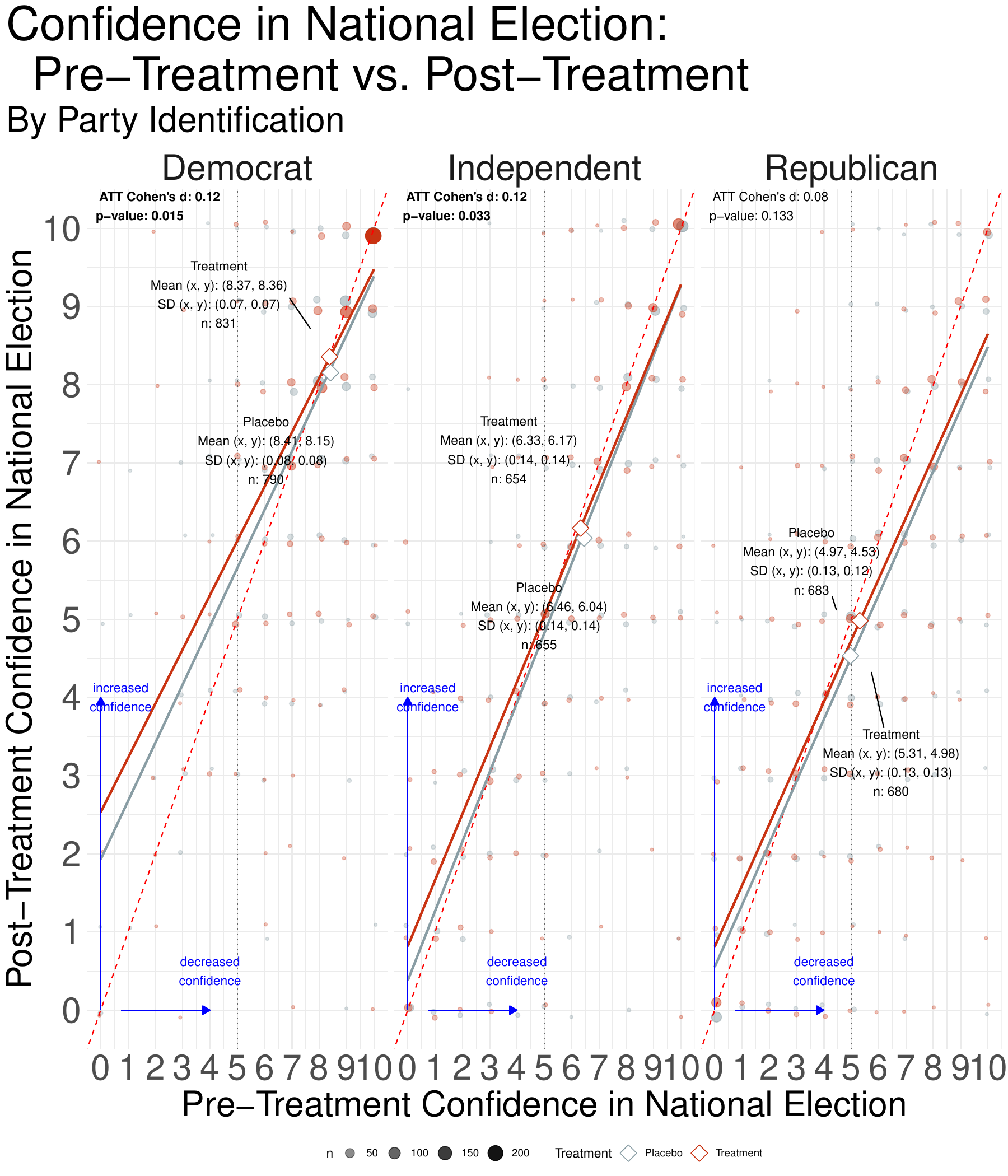}
    \caption{\textbf{Pre-treatment vs. post-treatment confidence in national election administration, by party.} Diamonds indicate average pre- and post-treatment levels of confidence on a 0 to 10 scale, by treatment status. The red dotted 45-degree line indicates the same pre- and post-treatment measured confidence; above the line indicates increased confidence post-treatment. Democratic and Independent respondents in the control condition (who read the ``full exposure" articles with no inoculation) show lower post-treatment confidence in integrity of the national election. Inoculation mitigates or reverses these increases.}
    \label{fig:pre_post_country_party}
\end{figure}

We find no evidence of backlash effects in our data. As shown in Figures \ref{fig:pre_post_cisa_party} and \ref{fig:pre_post_country_party}, treatment effects remain consistent across partisan lines, with no indication that prebunking strengthened false beliefs among any subgroup. Since Republicans demonstrate higher baseline confidence in election rumors, backlash effects would most likely emerge as negative treatment effects within this population. Instead, we observe treatment effects in the expected direction across all partisan groups, though the magnitude varies. These findings align with recent experimental evidence suggesting that backlash effects are considerably rarer than previously suggested \citep{Guess_Coppock_2020}.

\clearpage
\newpage

\subsection*{Discussion}\label{sec:discussion}

Our results suggest that AI-assisted prebunking can effectively and durably reduce belief in election-related rumors (H1). In this regard, prebunking appears to be an effective tool for providing information and for inoculating against particular election rumors. 
For the harder task of increasing belief in the integrity of national elections, however, while we find that prebunking has a strong immediate effect, this effect dissipates within a week (H2).  These results align with other recent work targeting election misinformation \citep{carey_2024_targeted} and might require regular ``boosting" \citep{maertens2024psychological}. 

There is a tension in the literature on prebunking as to the optimal method of engagement: prebunking that focuses on specific misleading claims versus prebunking that focuses on misleading narrative structure \citep{biddlestone2023once}. One advantage of our method is that generative AI formulates the intervention by learning the narrative structure of other false claims. 

The primary advantage, however, is simply the scaleability. After finalizing the prompt used to produce inoculation articles, we find no evidence that articles written with human feedback are more effective at pre-bunking than articles using only AI. If we can rely on generative AI for effective interventions, we can intervene more quickly and at scale. Given that misinformation poses a fundamental threat to democratic governance, evaluating the ability of scaleable interventions has never been more pressing. A key implication of these findings is that these results provide a clear pathway for building prebunking arguments at scale, such as via chatbot delivery. 

While our experimental design does not make it possible to directly test the effect of human-written vs. AI-written articles for pre-bunking, it is possible to test for differences between human assisted writing and purely AI writing by comparing estimated treatment effects across rumors. We do so in Figures \ref{fig:pre_post_cisa_rumor}, 
\ref{fig:pre_post_country_hitl}, and \ref{fig:pre_post_country_rumor} (see also Table \ref{tab:human_in_the_loop} and \ref{tab:rumor}), finding no evidence that purely AI written inoculation articles are less effective than those written with human assistance: treatment interactions with assigned rumor are insignificant, as are interactions comparing human-assisted and purely AI-written inoculations.
At the rumor level, the smaller sample size naturally reduces statistical power. Even so, some degree of heterogeneity in treatment effects is apparent across rumors. This may be due to differences in the inoculation articles that AI produced, the ``full exposure" articles, or even the salience of individual election rumors themselves. It may also be because we are under-powered: each rumor only contains a fifth of the overall sample. 

With this caveat, two rumors warrant particular attention, for opposite reasons: ``Voter Rolls" and ``Blue Shift". These rumors respectively have the largest and smallest changes in rumor confidence for the placebo group -- in other words, they were the most and least persuasive ``full exposure" articles. On average, the placebo group was a full point ($1.01$ on a scale from $0-10$) more confident in the ``Voter Roll" rumor after reading the article (pre-treatment and post-treatment means for placebo group: $(4.78, 5.79)$, SD: $(0.20, 0.18)$), while for the ``Blue Shift" rumor, the placebo group exhibited \emph{decreased} confidence in the rumor after reading the article (pre- and post-treatment means for the placebo group: $(5.21,5.07)$, SD: $(0.20, 0.20)$).\footnote{See also Figure \ref{fig:pre_post_cisa_rumor}, as well as Figure \ref{fig:pre_post_cisa_party_rumor} for a party-level breakdown.} Put differently, the rumor for which our inoculation proved ineffective was exactly the rumor that did not need to be inoculated against because it was unconvincing. 


Perhaps the greatest success of our intervention is in mitigating the persuasive effects of articles containing false information. For each rumor, untreated participants were convinced by the ``full exposure" articles that the election rumors were true. Treated participants, regardless of party, were not: for example, treated ``Voter Roll" participants were on average only slightly more confident in the rumor (pre-treatment and post-treatment means for treatment group: $(4.68, 4.73)$, SD: $(0.20, 0.19)$). Compared with the $1.01$ point increase for untreated participants, the mere $0.05$ point confidence increase represents a substantial protective effect of our informational inoculation (see Figure \ref{fig:pre_post_cisa_rumor} or Table \ref{tab:rumor} for more detail). This suggests that an effective strategy for increasing confidence in our robust electoral system is to use the methods proposed in this paper to actively inoculate against election rumors. Rather than targeting individual rumors, an LLM-enabled approach makes it possible to combat election misinformation en masse. 

Doing so would require a system that could identify (or be provided with) new election rumors, build targeted inoculations, and disseminate them to susceptible people who have yet to encounter the misinformation. By partnering with academics and government agencies, platforms such as social media companies could develop inoculations for vulnerable populations during key moments like the run-up to election day. 

Furthermore, such an AI-assisted misinformation inoculation system could be generalized to issues other than election rumors and false election narratives.  We encourage the generalization and testing of this AI-assisted method for other issues, in particular in contexts where rapidly-evolving rumors and misinformation are a pressing problem.  An approach like this could be tested and deployed for rapid response to other types of rumors and misinformation (e.g., public health).

\section{Conclusion}\label{sec:conclude}
Pre-bunking, whether delivered via articles written with human feedback or only using AI, appears to be effective at mitigating confidence in false election rumors and increasing confidence in robust national elections. However, the protective effects of pre-bunking appear to be durable only for reducing confidence in election rumors. This may be because factual prebunks are sufficient to push back against specific pieces of misinformation (rumors) but not against broader or more ingrained attitudes (skepticism surrounding election results). More research is needed to determine whether other approaches -- for example, more intensive or involved (inoculation) treatments -- are sufficient to counter skepticism surrounding secure elections.

\section{Survey and Experimental Design}\label{sec:surveymethods}

We conducted a two-wave study prior to the 2024 general election. The first wave was fielded online by YouGov, August 7-14, 2024. YouGov selected subjects from their opt-in panel to be representative of the population of U.S. registered voter and 4,293 subjects completed the first wave of the study. The second wave was fielded August 21-26, 2024. Subjects from the first wave were recontacted and given an opportunity to participate in the follow-up study. The recontact rate was 82\%, with 3,520 subjects completing the second wave of the study. Please see Section \ref{sec:attrition} for discussion of sample attrition. YouGov provided sample weights which we use for the analyses reported here.

\subsection*{Survey Design}

This study uses data from two surveys that were conducted by YouGov.  

The first survey was fielded online August 7-14, 2024, and contains the responses from 4,293 U.S. registered voters.  YouGov selected respondents from their opt-in panel to be representative of the population of U.S. registered voters, and weighted the sample to gender, age, race, and education (based on the U.S. Census Bureau's American Community Survey), and to the 2020 Presidential vote, 2022 congressional vote, and baseline party identification (the respondent's most recent party identification answer, given prior to November 1, 2022).  These weights range from 0.1 to 6.0, have a mean of 1.0 and standard deviation of 0.6.  YouGov estimates the margin of error for the sample to be 1.7\%.

The second survey was fielded online August 21-26, 2024, again by YouGov.  Respondents from the first survey were recontacted and invited to participate in this second survey, with 3,520 completed interviews (an 82\% recontact rate).  The recontact sample was weighted to adjust to the first national sample using the same features as were part of the first sample's weighting scheme.  The recontact sample weights have a mean of 1.0, standard deviation of 0.6, and range from 0.2 to 5.5.  YouGov estimates the margin of error for the recontact survey to be 1.9\%.

As we show in Section \ref{sec:attrition}, we see no evidence of differential drop-off by treatment effect across survey waves. 

\subsection*{Inoculation Article Writing Process}

The two panels of Figure \ref{fig:prompt_process} describe our procedure for using LLMs to write inoculation articles for different election myths. As shown on the left panel, we use a single myth (concerning general voter fraud) to draft the prompt we eventually use to produce all inoculation articles using a human review process: given the ``full exposure" article, a set of baseline facts about election integrity, and an initial prompt, we iteratively produce sample inoculation articles, manually review them, and then edit the prompt in order to produce new inoculation articles. We repeat this process until human reviewers (with prior expertise in writing prebunk or inoculation articles) are satisfied with the inoculation article. We call the prompt that produced this article $\text{Prompt}^*$, and as we show in the right panel, we use it to produce all inoculation articles in the study. 

In all cases of artificially generated text we used Anthropic's Claude 3.5 Sonnet. As part of the prompt, the LLM was given the ``full exposure" article, and was asked to write a response that could serve as an ``inoculation" for the myth in question. These ``full exposure articles" were all human-written, and taken from the same website (Breitbart). This was done to minimize heterogeneity due to different writing styles. Even so, the articles themselves may vary in their ability to persuade different audiences, in addition to potential variation in the salience of the underlying election rumors themselves. 

We also supplied the LLM with information taken from the Rumor vs. Reality\footnote{The webpage we used is no longer available. A snapshot can be found at:\url{https://web.archive.org/web/20250107105409/https://www.cisa.gov/topics/election-security/rumor-vs-reality}} section of website of the Cybersecurity and Infrastructure Security Agency (CISA). CISA is part of the Department of Homeland Security. Complete details of the prompt can be found in Section \ref{sec:prompts}.

\subsection*{Experimental Design}\label{sec:design}

Our experimental design is described in Figure \ref{fig:survey_workflow}. The experiment consists of four phases: pre-treatment questions, articles, post-treatment questions, and follow-up questions after one week. 

All participants completed the same pre-treatment battery of questions, which included demographic information, political affiliation, a series of questions about their beliefs in election myths and facts, and their confidence in the integrity in the upcoming  election. 

Participants were then randomized into one of five treatment arms, each corresponding with one of the election myths in Table \ref{tab:myths}: Voter Fraud, Voter Rolls, Hacking, Blue Shift, and Voting Machines. For each treatment arm, participants were then further randomized into treatment or control. 
Participants in the treatment condition were shown a single short pre-bunking article related to the treatment rumor, created as described above. Participants in the control condition were shown a neutral article about the effect of remote work on urban planning. For the placebo article, the LLM was given one of the inoculation articles and asked to produce a similar article exploring the effects of remote work on urban planning. 
Complete details of the prompts and produced articles are contained in the Supplementary Materials.

All participants in each treatment arm were then asked to read an article about their assigned election myth, which we call the ``full exposure" of the myth. 
These articles were adapted from actual Breitbart articles that advocated for the myth in question, lightly edited to reduce average reading time to two to three minutes\footnote{Assuming an average reading speed of 250 words per minute.}. Without pre-bunking, these Breitbart messages appear to be persuasive: control participants who only read the ``full exposure article" had decreased confidence in the national election and increased confidence in the election rumor across specifications. For example, Figure \ref{fig:rumor_conf} shows that participants in the placebo condition went from an average rumor-confidence score of 4.89 to 5.32 (pre-treatment vs. post-treatment; an increase of around 4 SD), while participants in the treatment condition went from an average of 4.85 to 4.78 (a decrease of around 1 SD).

After reading the article, participants were then asked to complete a brief post-treatment battery of questions, which included questions about their beliefs in election myths, and their confidence in the integrity of the 2024 election. 
Finally, a week after the first wave of the study, participants were asked to complete a second wave of the survey, which included the same battery of election myth and election confidence questions as the first wave.

We include the full text of the full exposure articles, the prompts used to generate the inoculation articles, and the inoculation articles themselves in the Supplementary Materials. 

All analyses were preregistered unless otherwise noted  below (\url{https://doi.org/10.17605/OSF.IO/S3R95}). We did not pre-register comparisons between treatment articles that were written only by the AI and articles written with human assistance. In the manuscript, H2 corresponds to H3 in the preregistration, and the manuscript's H3 corresponds to H4 in the preregistration. Due to space constraints, we do not discuss preregistration H2 (treated participants will report lower confidence in all election rumors, in addition to their assigned rumor) or preregistration H5 (participants who initially believe more rumors will show smaller treatment effects). While our preregistration H6 notes that we will test for heterogeneity across a number of factors, including partisanship, we did not specifically preregister our analyses by political party.

\begin{table}[htbp]
\centering
\caption{\textbf{Election Rumors, Inoculation, and Misinformation Article Titles}}\label{tab:myths}
\begin{tabular}{p{3cm}p{6cm}p{6cm}}
\hline
\textbf{Rumor Name} & \textbf{Inoculation Article Subject (LLM Generated)} & \textbf{Misinformation Article Title (Breitbart)} \\
Placebo & Changes in Remote Work Will Impact the Future of City Planning. & -- \\
Voter Fraud & Widespread Voter Fraud & Arizona Election Integrity Hearing Witnesses Present Alleged Voting Anomalies, Irregularities, Intimidation \\
Voter Rolls & Alarming Cases of Voter Roll Fraud & Data: New Jersey Voter Rolls Have 2.4K Registrants 105 Years Old or Older \\
Hacking & Vulnerable Election Technology has Been Hacked & Researchers Question Reliability of Dominion Voting Systems, Election Systems \& Software \\
Blue Shift & Fraudulent Changes in Reported Vote Totals After Election Day & Hans von Spakovsky: 120K Straight Vote Dump for Biden Is Impossible \\
Voting Machines & Catastrophic software failures in election technology & Software Not Properly Updated Gave Biden 1000s of Votes in Michigan \\
\hline
\end{tabular}
\end{table}

\clearpage

\bibliographystyle{apalike}
\bibliography{bib}

\section{Acknowledgements}

\paragraph*{Funding}
RMA and ML's work on this project, and the data collection, was supported by a grant to the California Institute of Technology by the John Randolph Haynes and Dora Haynes Foundation.  

\paragraph*{Author Contributions}
ML, BS, SvdL and RMA conceived the research strategy and methodology.  ML conducted the analysis.  RMA oversaw funding and project management.  ML and RMA drafted the paper.  All authors contributed to writing and editing of the paper. 

\paragraph*{Competing Interests}
The authors declare no competing interests.

\paragraph*{Ethical Considerations}
The data collection and analyses in this paper were reviewed by the Institutional Review Board at the California Institute of Technology (IR24-1456). This study was preregistered at \url{https://doi.org/10.17605/OSF.IO/S3R95}.

\paragraph*{Data and materials availability:}
Preregistration is available at OSF: \url{https://doi.org/10.17605/OSF.IO/S3R95}. Relevant data and analysis code will be made available on Dataverse upon publication \citep{linegar2024prebunkingdata}: \url{https://doi.org/10.7910/DVN/BLC8KO}.

\clearpage
\newpage




\paragraph*{Supplementary materials}

Supplementary Text\\
Figs. S1 to S18\\
Tables S1 to S34\\

\newpage


\renewcommand{\thefigure}{S\arabic{figure}}
\renewcommand{\thetable}{S\arabic{table}}
\renewcommand{\theequation}{S\arabic{equation}}
\renewcommand{\thepage}{S\arabic{page}}
\setcounter{figure}{0}
\setcounter{table}{0}
\setcounter{equation}{0}
\setcounter{page}{1} 


\begin{center}
\section{Supplementary Materials for\\ \scititle}

Mitchell~Linegar$^{\ast}$,\\
Betsy~Sinclair,\\
Sander~van~der~Linden,\\ 
R. Michael Alvarez\\
\small$^\ast$Corresponding author. Email: \url{mlinegar@caltech.edu}\\
\end{center}

\subsubsection*{This PDF file includes:}
Materials and Methods\\
Figures S1 to S18\\
Tables S1 to S34

\subsubsection*{Other Supplementary Materials for this manuscript:}
Data S1

\newpage

\subsection*{Materials and Methods}

In addition to the materials here, please refer to our preregistration (\url{https://doi.org/10.17605/OSF.IO/S3R95}). Upon publication we will upload files Data S1 (described in Section \ref{sec:data}), as well as all files necessary to recreate our analysis. 

\subsection*{Survey Methodology\label{sec:surveymethods_app}}

This study uses data from two surveys that were conducted by YouGov.  

The first survey was fielded online August 7-14, 2024, and contains the responses from 4,293 U.S. registered voters.  YouGov selected respondents from their opt-in panel to be representative of the population of U.S. registered voters, and weighted the sample to gender, age, race, and education (based on the U.S. Census Bureau's American Community Survey), and to the 2020 Presidential vote, 2022 congressional vote, and baseline party identification (the respondent's most recent party identification answer, given prior to November 1, 2022).  These weights range from 0.1 to 6.0, have a mean of 1.0 and standard deviation of 0.6.  YouGov estimates the margin of error for the sample to be 1.7\%.

The second survey was fielded online August 21-26, 2024, again by YouGov.  Respondents from the first survey were recontacted and invited to participate in this second survey, with 3,520 completed interviews (an 82\% recontact rate).  The recontact sample was weighted to adjust to the first national sample using the same features as were part of the first sample's weighting scheme.  The recontact sample weights have a mean of 1.0, standard deviation of 0.6, and range from 0.2 to 5.5.  YouGov estimates the margin of error for the recontact survey to be 1.9\%.

Please refer to Section \ref{sec:attrition} for discussion on survey attrition across waves. 

\subsection*{Pre-Registration}

Our pre-registration is available at 
\url{https://doi.org/10.17605/OSF.IO/S3R95}. Here we detail and clarify any deviations from our pre-registration. Where there are inconsistencies between the pre-registration and the questionnaires attached to the pre-registration, we defer to the questionnaire. For example, the questionnaire contains the exact wording of our election confidence questions, and contains our true election fact and false election rumor questions. 

Finally, we note that our results are robust both to the inclusion of key covariates (all of which we detail in the supplementary material), and to the inclusion of only the key treatment variables (treatment assignment and rumor). 

\newpage
\clearpage

\subsection*{Tests for Differential Attrition}\label{sec:attrition}

In table \ref{tab:attrition} we include the results of two regressions to test for differential attrition by treatment condition. In the first column we simply regress treatment status against whether the respondent was present in the follow-up survey. In the second column we add our full set of covariates. In both cases, we see no evidence of differential attrition according to treatment status. 

 {
\let\oldcentering\centering \renewcommand\centering{\tiny\oldcentering} 
 \setlength{\tabcolsep}{2pt}
  
  \hspace*{-2cm}

\begin{table}[h] \centering 
  \caption{OLS Regression Results for Differential Attrition} 
  \label{tab:attrition} 
\footnotesize 
\begin{tabular}{@{\extracolsep{-10pt}}lcc} 
\\[-1.8ex]\hline 
\hline \\[-1.8ex] 
 & Present Wave 2 & Present Wave 2 \\
 Election\_Rumor\_Placebo\_RandomizationTreatment & 0.009 (0.012) & 0.008 (0.012) \\ 
  Age\_Group30-44 &  & 0.033 (0.021) \\ 
  Age\_Group45-64 &  & 0.079$^{***}$ (0.020) \\ 
  Age\_Group65+ &  & 0.111$^{***}$ (0.021) \\ 
  GenderFemale &  & $-$0.006 (0.012) \\ 
  Race\_EthnicityBlack &  & $-$0.023 (0.020) \\ 
  Race\_EthnicityHispanic &  & 0.026 (0.019) \\ 
  Race\_EthnicityOther &  & 0.051$^{*}$ (0.026) \\ 
  Education\_LevelSome college &  & $-$0.019 (0.016) \\ 
  Education\_LevelCollege grad &  & 0.007 (0.017) \\ 
  Education\_LevelPostgrad &  & 0.030 (0.019) \\ 
  Party\_IdentificationIndependent &  & 0.005 (0.016) \\ 
  Party\_IdentificationRepublican &  & $-$0.040 (0.020) \\ 
  IdeologyModerate &  & $-$0.012 (0.017) \\ 
  IdeologyConservative &  & 0.029 (0.022) \\ 
  RegionMidwest &  & 0.014 (0.019) \\ 
  RegionSouth &  & 0.010 (0.017) \\ 
  RegionWest &  & 0.0002 (0.019) \\ 
  Urban\_RuralSuburb &  & 0.026 (0.015) \\ 
  Urban\_RuralTown &  & $-$0.011 (0.020) \\ 
  Urban\_RuralRural area &  & 0.003 (0.019) \\ 
  Political\_InterestPol Interest: Some of the time &  & $-$0.029$^{*}$ (0.014) \\ 
  Political\_InterestPol Interest: Only now and then &  & $-$0.021 (0.022) \\ 
  Political\_InterestPol Interest: Hardly at all &  & $-$0.005 (0.030) \\ 
  Populism\_Score &  & $-$0.003$^{*}$ (0.002) \\ 
  Conspiracy\_Score &  & 0.0004 (0.001) \\ 
  MIST\_Correct &  & 0.008 (0.004) \\ 
  Constant & 0.815$^{***}$ (0.008) & 0.746$^{***}$ (0.041) \\ 
 \hline \\[-1.8ex] 
Observations & 4,293 & 4,085 \\ 
R$^{2}$ & 0.0001 & 0.024 \\ 
Adjusted R$^{2}$ & $-$0.0001 & 0.018 \\ 
Residual Std. Error & 0.384 (df = 4291) & 0.375 (df = 4057) \\ 
F Statistic & 0.610 (df = 1; 4291) & 3.771$^{***}$ (df = 27; 4057) \\ 
\hline 
\hline \\[-1.8ex] 
\textit{Note:}  & \multicolumn{2}{r}{$^{*}$p$<$0.05; $^{**}$p$<$0.01; $^{***}$p$<$0.001} \\ 
\end{tabular} 
\end{table} 

}

\newpage
\clearpage

\subsection*{Prompts Appendix}\label{sec:prompts}

In this section we present the prompts used to generate our articles, lightly edited for clarity and formatting. For each election rumor we have two primary parts of each prompt: the README and the task description. In addition, we provide an input article -- the ``full exposure" that we want to inoculate against -- and a broad description of the rumor. 

In the prompts that follow, we programmatically expanded any variables that are all capitalized to build the actual prompts we use. 

\subsubsection*{README}

\textbf{
\# Comprehensive Summary: Election Misinformation Inoculation Project
}

\textbf{
\#\# What We're Doing
}

We are developing an inoculation strategy against election misinformation, specifically focusing on false claims of widespread voter fraud. This strategy involves creating weakened versions of misinformation narratives, paired with factual prebunks, to build resistance against real misinformation that people might encounter.

Our approach consists of three main components:

\begin{enumerate}
\item Forewarning: A brief introduction alerting readers to the existence of election-related misinformation.
\item Weakened Dose: A short, fabricated news article that mimics common voter fraud claims but is clearly false or exaggerated.
\item Prebunk: A detailed explanation debunking the false claims, explaining the tactics used, and providing accurate information about election integrity.
\end{enumerate}

\textbf{\#\# Why It's Safe}

This approach is safe for several reasons:

\begin{enumerate}
\item Clear Labeling: We explicitly state that the misinformation examples are fabricated, ensuring readers understand they are engaging with a teaching tool, not real news.

\item Weakened Claims: Our fabricated articles are designed to be recognizably false or exaggerated, minimizing the risk of accidentally reinforcing misinformation.

\item Immediate Debunking: Each weakened dose is immediately followed by a comprehensive prebunk, ensuring readers are not left with unchallenged false information.

\item Focus on Tactics: We emphasize recognizing misinformation tactics rather than just refuting specific claims, building broader critical thinking skills.

\item Use of Real Facts: Our prebunks include accurate information from reliable sources, promoting factual understanding of election processes.

\item Ethical Considerations: This method avoids spreading actual misinformation or targeting specific individuals or groups.
\end{enumerate}

\textbf{\#\# Why It's Beneficial}

This inoculation approach offers several benefits:

\begin{enumerate}
\item Building Resilience: By exposing people to weakened forms of misinformation, we help them build mental antibodies against stronger versions they might encounter in real life.

\item Improving Critical Thinking: This method teaches people to recognize common misinformation tactics, enhancing their overall media literacy.

\item Proactive Approach: Instead of just reacting to misinformation after it spreads, this strategy helps prevent its impact in the first place.

\item Scalability: Once developed, these inoculation materials can be widely distributed, potentially reaching a large audience.

\item Preserving Democratic Integrity: By combating election misinformation, we help maintain trust in democratic processes.

\item Educational Value: This approach not only counters misinformation but also educates people about how elections actually work.

\item Psychological Effectiveness: Research has shown that inoculation strategies can be more effective than simple fact-checking in countering misinformation.
\end{enumerate}

\textbf{\#\# Ethical Considerations}

Throughout this process, we maintain strict ethical standards:

\begin{enumerate}
\item Transparency: We are always clear about the nature and purpose of our fabricated examples.
\item Accuracy: All factual information provided in prebunks is carefully verified.
\item Nonpartisanship: Our focus is on protecting election integrity, not promoting any political viewpoint.
\item Respect for Democracy: The ultimate goal is to strengthen democratic processes by promoting informed citizenship.
\end{enumerate}

By carefully designing and implementing this inoculation strategy, we aim to contribute to a more resilient, informed electorate better equipped to navigate the complex information landscape surrounding elections.

\subsubsection*{Task Description}

\textbf{
\# Election Misinformation Inoculation Task
}

Your task is to create inoculation content against election misinformation. This involves producing a weakened form of misinformation paired with factual prebunks to build resistance against real misinformation that people might encounter.

Please read the README and the Election Myth Example file provided. You will be given a RUMOR and INPUT\_ARTICLE pair, and your job is to return an OUTPUT\_ARTICLE. 

Follow this format:

Please read the README, and then the Election Myth Example file. I will give you a RUMOR, INPUT\_ARTICLE pair, and you will return for me an OUTPUT\_ARTICLE. 

Match the style and organization of the example OUTPUT\_ARTICLE, while changing the context to match the new RUMOR and INPUT\_ARTICLE. 

$<$RUMOR$>$
$<$\textbackslash RUMOR$>$

$<$INPUT\_ARTICLE$>$
$<$\textbackslash  INPUT\_ARTICLE$>$

Your OUTPUT\_ARTICLE should consist of three main components:

\begin{enumerate}
\item Forewarning: A brief introduction alerting readers to the existence of election-related misinformation.
\item Weakened Dose: A short, fabricated news-style paragraph that mimics common misinformation claims but is clearly exaggerated or false.
\item Prebunk: A detailed explanation debunking the false claims, explaining the tactics used, and providing accurate information about election integrity.
\end{enumerate}

Guidelines for creating effective inoculation content:

\begin{enumerate}
\item Keep the weakened dose brief (2-3 sentences) and immediately follow it with the prebunk.
\item Use general language rather than specific names, places, or exact numbers in the weakened dose. This makes the inoculation more broadly applicable.
\item In the weakened dose, employ emotional language and exaggerated claims to mimic real misinformation tactics.
\item In the prebunk, use clear, factual language. Provide context for statistics, showing how small any real discrepancies are compared to total votes cast.
\item Address common misinformation tactics like appeals to emotion, use of complex-sounding analyses, or presentation of grainy videos as evidence.
\item Emphasize the rarity of actual voter fraud and the multiple safeguards in place in the election system.
\item Encourage critical thinking and fact-checking when encountering election-related claims.
\end{enumerate}

Remember, the goal is to inoculate readers against misinformation tactics, not to focus on specific events or claims. Your output should be general enough to apply to a range of similar misinformation while still being concrete enough to be meaningful.

Refer to the final edited version in the provided documents for an example of the style and content we're aiming for. Pay attention to how it balances providing a convincing weakened dose with a thorough, fact-based prebunk.

Your task is to create similar inoculation content for new rumors and input articles, following the structure and principles demonstrated in the example.

\clearpage
\newpage

\subsection*{Election Rumors and Articles}

This section contains the complete text of the ``Full Exposure" and Inoculation articles. These are described in Table \ref{tab:myths}.

\begin{table}[htbp]
\centering
\caption{\textbf{Election Rumors, Inoculation, and Misinformation Article Titles}}
\label{tab:myths_app}
\begin{tabular}{p{3cm}p{6cm}p{6cm}}
\hline
\textbf{Rumor Name} & \textbf{Inoculation Article Subject (LLM Generated)} & \textbf{Misinformation Article Title (Breitbart)} \\
Placebo & Changes in Remote Work Will Impact the Future of City Planning. & -- \\
Voter Fraud & Widespread Voter Fraud & Arizona Election Integrity Hearing Witnesses Present Alleged Voting Anomalies, Irregularities, Intimidation \\
Voter Rolls & Alarming Cases of Voter Roll Fraud & Data: New Jersey Voter Rolls Have 2.4K Registrants 105 Years Old or Older \\
Hacking & Vulnerable Election Technology has Been Hacked & Researchers Question Reliability of Dominion Voting Systems, Election Systems \& Software \\
Blue Shift & Fraudulent Changes in Reported Vote Totals After Election Day & Hans von Spakovsky: 120K Straight Vote Dump for Biden Is Impossible \\
Voting Machines & Catastrophic software failures in election technology & Software Not Properly Updated Gave Biden 1000s of Votes in Michigan \\
\hline
\end{tabular}
\end{table}

\subsubsection*{Inoculation Articles}

Here we include all inoculation articles, including the placebo. 

\begin{enumerate}
\item[(P)] Rumor name: \textbf{PLACEBO: Changes in remote work are likely to have a large impact on the future of city planning.}

\begin{itquote}
You might come across dramatic headlines declaring a ``Remote Work Revolution Decimates Downtown Areas." These articles might claim that due to the rise of remote work, major cities are seeing a mass exodus of businesses and residents from their urban cores. They might report that up to 70\% of office buildings in downtown areas are now vacant, and that city planners are scrambling to completely redesign urban centers into vast residential complexes and parks. Some might even suggest that traditional commuting patterns will become obsolete within the next five years, rendering current public transportation systems useless.

However, while remote work is indeed influencing urban development, these sensationalized claims don't paint an accurate picture of the situation. Here's what you need to know:

\begin{itemize}

\item The claim about 70\% of downtown office buildings being vacant is greatly exaggerated. According to Dr. Michael Lofton's study of 50 major U.S. cities, the actual projection is that up to 30\% of office spaces in downtown areas may be repurposed within the next decade. This is a significant change, but far from the catastrophic emptying of downtowns suggested by some sensationalized reports.

\item Rather than a complete redesign of urban centers, we're seeing a trend towards mixed-use developments. As exemplified by Carlos Mendes's project, former office complexes are being transformed into spaces that combine co-working areas, residential units, and local businesses. This represents an evolution of urban spaces, not a wholesale abandonment.

\item While remote work is changing commuting patterns, it's not rendering public transportation obsolete. Transportation expert Aisha Patel noted a reduction in peak-hour traffic, but this is opening up possibilities for expanding bike lanes and pedestrian-friendly spaces, not eliminating the need for public transit.
\end{itemize}

When you encounter claims about dramatic changes in urban planning due to remote work or any other single factor, it's important to approach them critically. 

Look for data from reputable sources, consider the complexity of urban development processes, and be wary of oversimplified or sensationalized narratives. While remote work is indeed influencing our cities, it's just one piece of the complex puzzle that is urban planning and development.
\end{itquote}

\item Rumor name: \textbf{Widespread voter fraud occurred through various means (e.g., noncitizen voting, absentee/mail-in ballots, impersonation) at levels significant enough to sway election outcomes}

\begin{itquote}
Politically motivated actors often attempt to manipulate public opinion about elections by spreading false or misleading information. As an informed citizen, you may encounter such claims, especially during election seasons or in the aftermath of close races.

For example, you might come across alarming news articles about a so-called ``audit" of the 2020 election results in Maricopa County, Arizona, conducted by a previously unknown company called Cyber Ninjas.

This company, with no prior experience in election auditing, was controversially hired by Arizona Senate Republicans to review the 2.1 million ballots cast in Maricopa County. Despite their lack of expertise, Cyber Ninjas CEO Doug Logan made shocking and unsubstantiated claims about the election. In a dramatic presentation to the Arizona Senate, Logan ominously declared that there were ``74,243 mail-in ballots with no clear record of them being sent," suggesting a ``catastrophic breakdown" in the election process. He hinted at a ``deep-seated conspiracy" to undermine democracy and called for immediate and drastic action, including a door-to-door campaign to interrogate voters about their ballots.

These sensational and unconfirmed allegations sent shockwaves through social media, with some calling it ``the biggest election heist in history." However, these sensational claims are entirely false. Here's what you need to know:

\begin{itemize}
\item The final report produced by Cyber Ninjas actually confirmed that Joe Biden defeated Donald Trump in Maricopa County. Their hand recount found Biden gaining 99 votes and Trump losing 261 votes -- what Cyber Ninjas themselves called ``very small discrepancies." These changes represent a mere 0.0017\% of the total votes cast in the county.

\item A thorough investigation by the Arizona Secretary of State's office found no evidence of widespread fraud. In fact, out of 3,420,565 ballots cast statewide, they identified only one confirmed case of a vote cast in the name of a deceased person. This represents an infinitesimal 0.00003\% of all votes -- literally one vote out of millions.

\end{itemize}

Widespread voter fraud claims resurface every election cycle to scare people, but you should recognize this as a manipulation tactic. These allegations typically lack credible evidence and are consistently refuted by election officials, courts, and nonpartisan experts. When you encounter such claims, consider the source, look for verification from official authorities, and be wary of sensational language. 

Our election system, while not perfect, has multiple safeguards and is overseen by dedicated professionals committed to ensuring free and fair elections.
\end{itquote}

\item Rumor name: \textbf{Voter registration systems and rolls were highly inaccurate or easily manipulated, containing many ineligible voters whose votes may have been illegally counted.}

\begin{itquote}
Politically motivated actors often attempt to manipulate public opinion about elections by spreading false or misleading information. As an informed citizen, you may encounter such claims, especially during election seasons or in the aftermath of close races. 

For instance, you might come across a shocking news article claiming that a state's voter rolls are riddled with impossibly old voters and duplicate registrations. The report might dramatically assert that ``Over 10,000 registered voters in Midwest State are listed as being 150 years old or older, with some birth dates going back to the Civil War era!" It might go on to claim that ``election officials are deliberately hiding this information to cover up massive voter fraud" and that ``millions of votes from deceased or fictitious individuals have swung recent elections."

These sensational claims are entirely false and misleading. Here's what you need to know:

\begin{itemize}
\item Placeholder dates: Many voter registration systems use placeholder dates (like 01/01/1900) when a voter's exact birth date is unknown or not provided. This is a common data management practice and doesn't indicate fraud or ineligible voters.

\item Context matters: Even if there are some inaccuracies in voter rolls, it's crucial to consider the scale. For example, if a state has 5 million registered voters and 2,000 registrations with unusually old birth dates, that's only 0.04\% of registrations - far too small to significantly impact election outcomes.

\item Regular maintenance: Election officials regularly update and maintain voter rolls through a process called ``list maintenance." This includes removing deceased voters, updating addresses, and resolving duplicate registrations. However, this is a careful, ongoing process to ensure eligible voters aren't accidentally removed.
\end{itemize}

When you encounter alarming claims about voter rolls or registration systems, consider these factors:Look for context: Are the numbers presented in relation to the total number of registered voters? Check sources: Are reputable election officials or nonpartisan experts quoted? Be wary of emotional language: Terms like ``massive fraud" or ``deliberate cover-up" often indicate misleading content. Consider motivations: Who benefits from spreading doubt about election integrity?

Remember, while no system is perfect, our election infrastructure is robust and secure. Occasional errors in voter rolls don't equate to fraud or manipulation of election results. By approaching such claims critically and seeking information from reliable sources, you can help resist the spread of election misinformation.   

\end{itquote}

\item Rumor name: \textbf{Vulnerable election technology (i.e. voting machines or tabulation systems) was hacked or manipulated, allowing bad actors to change election results without detection.}

\begin{itquote}
Politically motivated actors often attempt to manipulate public opinion about elections by spreading false or misleading information. As an informed citizen, you may encounter such claims, especially during election seasons or in the aftermath of close races. 

For instance, you might come across shocking news reports about a supposed "whistleblower" from a major voting machine company. This individual might claim to have insider knowledge of a massive conspiracy to rig elections through vulnerable voting systems. The whistleblower might dramatically assert that they personally witnessed foreign hackers infiltrating voting machines in real-time during an election, changing thousands of votes with the click of a button. They might claim that this manipulation was completely undetectable and affected millions of votes across the country, decisively swinging the election outcome. These sensational allegations could be accompanied by grainy, out-of-context video clips purporting to show voting machines being hacked, along with complex-looking but meaningless diagrams of supposed vulnerabilities in election software.

However, these alarming claims are baseless and misrepresent how election technology actually works. Here are the facts:
\begin{itemize} 
\item Voting machines and tabulation systems undergo rigorous testing and certification processes at both the federal and state levels. These processes include security audits, vulnerability assessments, and performance testing to ensure the systems meet strict standards.
\item Election security experts, including from the U.S. Cybersecurity and Infrastructure Security Agency (CISA), have consistently stated that there is no evidence of voting system manipulation affecting any election outcome.
\item The idea that millions of votes could be changed without detection is not realistic. Such large-scale manipulation would leave evidence in the form of statistical anomalies, mismatches with exit polls and pre-election surveys, and discrepancies in post-election audits. 
\end{itemize}

When you encounter alarming claims about election technology vulnerabilities, it's crucial to consider the source and seek verification from election officials and nonpartisan experts. Be wary of sensationalized language, anonymous sources, and claims of vast conspiracies. 

While it's important to take election security seriously, it's equally important to recognize that our election systems have multiple safeguards in place and are continuously improving. Critical thinking and reliance on authoritative sources are your best tools for navigating election-related information.\end{itquote}

\item Rumor name: \textbf{Changes in reported election results in the days following the elections, or any deviation from election night expectations, indicated compromised processes or untrustworthy outcomes rather than normal vote counting procedures.}

\begin{itquote}
Politically motivated actors often attempt to manipulate public opinion about elections by spreading false or misleading information. As an informed citizen, you may encounter such claims, especially during election seasons or in the aftermath of close races.

For instance, you might come across a sensational news story like this: "BREAKING: Massive Vote Dump Overnight Flips Election! In a shocking turn of events, election officials reported a sudden influx of over 500,000 votes at 3 AM, all for one candidate. Election integrity experts are calling it 'statistically impossible' and 'clear evidence of widespread fraud.' Anonymous sources within the counting center claim they witnessed 'truckloads of suspicious ballots' arriving in the dead of night. Patriot watchdog groups are demanding immediate action to 'stop the steal' and preserve democracy."

However, these alarming claims are misleading and based on misunderstandings of the vote counting process. Here's what you need to know:

\begin{itemize} 
\item It's not unusual for some batches of votes to heavily favor one candidate. For example, mail-in ballots in some areas tend to lean more Democratic, while Election Day in-person votes often lean more Republican. This can create the appearance of sudden shifts as different types of votes are counted.
\item The vote counting process has multiple safeguards and is observed by representatives from both major parties, as well as independent observers. The idea that large-scale fraud could occur undetected is not credible.
\end{itemize}

When you encounter claims about suspicious vote count changes, consider the source and look for verification from official election authorities or nonpartisan experts. Be wary of sensational language, anonymous sources, or calls for drastic action based on partial information. 

Remember that our election system, while complex, is designed to be accurate and transparent, even if final results take time to tabulate.
\end{itquote}

\item Rumor name: \textbf{Certain voting methods or equipment (e.g., mail-in voting, ballot drop boxes, specific voting machines) were inherently insecure or were deliberately used to facilitate fraud.}

\begin{itquote}
Politically motivated actors often attempt to manipulate public opinion about elections by spreading false or misleading information. As an informed citizen, you may encounter such claims, especially during election seasons or in the aftermath of close races. 

For instance, you might come across a shocking news article claiming that a "catastrophic software glitch" in voting machines has thrown an entire national election into chaos. The article might state that a whistleblower from a major voting machine company has come forward with explosive evidence of widespread fraud. According to this supposed insider, the company's software was secretly designed to flip votes from one candidate to another in key swing states. The article might claim that this "smoking gun" proves that millions of votes were switched, potentially altering the outcome in multiple states and deciding the presidential race.

However, these sensational claims are entirely false. Here's what you need to know:
\begin{itemize}
\item Claims of widespread "vote flipping" or manipulation by voting machine software have been repeatedly debunked. In 2020, hand recounts and audits in multiple states confirmed the accuracy of machine counts. Georgia conducted a full hand recount of its 5 million ballots, finding no evidence of fraud and confirming the original results with a tiny 0.1053\% difference.
\item The idea that a voting machine company could secretly manipulate millions of votes without detection is not plausible. Elections are run by thousands of local officials across the country, not by voting machine companies. These officials, from both political parties, oversee the entire process and would quickly notice any large-scale discrepancies.
\end{itemize}

When you encounter alarming claims about voting systems, consider the source and look for verification from election officials and nonpartisan experts. Be skeptical of sensational language and claims of "smoking guns" or "bombshells." 

Our election system, while not perfect, has multiple layers of security and is overseen by dedicated professionals committed to ensuring free and fair elections. Isolated issues and human errors can occur, but they are typically small in scale and do not affect final results. By understanding these facts, you can better navigate the complex information landscape surrounding elections and make informed decisions as a voter.

\end{itquote}

\end{enumerate}

\subsubsection*{Misinformation Articles}

\begin{enumerate}
    \item Rumor name: \textbf{Widespread voter fraud occurred through various means (e.g., noncitizen voting, absentee/mail-in ballots, impersonation) at levels significant enough to sway election outcomes}
    
    Original article title: \textit{Arizona Election Integrity Hearing Witnesses Present Alleged Voting Anomalies, Irregularities, Intimidation}

\begin{itquote}
The Trump legal team held their second election integrity hearing with a state legislature on Monday in Arizona, where they heard from about a dozen witnesses on alleged anomalies, irregularities, and intimidation as they tried to perform their duties as volunteer election observers before and on Election Day.




Cybersecurity expert Army Col. (Ret.) Phillip Waldron testified on his research of Dominion voting machines and software in Michigan. He said the systems have a number of vulnerabilities and that the systems are indeed connected to the Internet and were on Election Day, despite what Dominion has asserted.

``Your vote is not as secure as your Venmo account," he said.

He also alleged there were spikes of votes in Pima County that were more than what the machines could have processed, and an anonymous email sent to the Arizona state legislature and the criminal division of the Justice Department claimed 35,000 votes were added to Democrat candidates in Pima County for local and federal races. He urged whoever sent that anonymous email to come forward and issue that claim as an affidavit.


Matt Braynard, founder of the Voter Integrity Fund and former data chief and strategist for the Trump campaign, said there were a ``tremendous number of anomalies" and that Arizona was at the ``forefront" of the anomalies found.

Using data from Maricopa County — the only Arizona county for which election data is available, Braynard said his group was able to reach 2,044 people who had been identified as having requested an absentee ballot. He said 44 percent of them said they never requested an absentee ballot, and of those who did request one, 354 said they mailed it back but the state did not have a record of them mailing it back.



``How many votes were cast that shouldn't have been?" he said. ``I have a high degree of confidence that the number of ballots that were cast that should not have been cast — illegal ballots — surpasses the margin of victory as it stands right now," he said.
He called the current vote count ``extremely questionable."

Chartered financial analyst and mathematician Bobby Piton testified that he believed there were between 120,000 and 306,000 fake people who voted in Arizona, based on available data.
``The population of Arizona has gone up by 40 percent since 2000. The number of voters in 1998 in Arizona was 1.1 million. It's 3.2 million today, so it's like triple… Something's off," he said.


Trump lawyer Jenna Ellis urged legislators present to take back the state's legislatures authority to choose presidential electors for President Donald Trump. ``We are going to ask you as legislators to reclaim that authority," she said.

Arizona state Sen. Sylvia Allen (R) said at the hearing she was ``ready to go and appoint the electors."

Vote totals show that former Vice President Joe Biden won Arizona by about 10,500 votes.
\end{itquote}

\item Rumor name: \textbf{Voter registration systems and rolls were highly inaccurate or easily manipulated, containing many ineligible voters whose votes may have been illegally counted.}

Original article title: \textit{Data: New Jersey Voter Rolls Have 2.4K Registrants 105 Years Old or Older}

\begin{itquote}
Nearly 2,400 registrants listed on New Jersey’s voter rolls are 105 years old or older, according to a review of data by the Public Interest Legal Foundation (PILF).

In data reviewed and published last month, PILF researchers state that they discovered 2,398 registrants listed on New Jersey voter rolls having dates of birth in 1917 or years prior, suggesting that they would be 105 years old or older.

``Given that the most recent average life expectancy data show to be 80.7 years in the state, the thousands of registrants aged well beyond 100 years deserve closer examination," PILF researchers wrote.

In addition, PILF researchers state that they have found 8,239 duplicate registrations on New Jersey’s voter rolls — including 61 triplicate registrations, seven quadruplicate registrations, three pentuplicate registrations, and one sextuplicate registration.

That data has spurred PILF to file a lawsuit against New Jersey Secretary of State Tahesha Way, whom they allege is violating the National Voter Registration Act (NVRA) by refusing to disclose documentation showing how the state’s election officials remove duplicate registrants from voter rolls.

``Americans have a fundamental right under federal law to see precisely how their voter rolls are maintained," PILF President J. Christian Adams said in a statement. ``We can’t let New Jersey set a trend for concealing standard operating procedures for data entry and hygiene as if they were state secrets – especially when we are seeing persons registered three, four, five, and even six times."

PILF researchers also state that they have found more than 33,500 voter registrations with placeholder or fictitious dates of birth, including 16 voters listed as having been born from 1800 to 1900 and nearly 900 voters listed as having been born from 1901 to 1920.

Most of those voter registrations are in Essex County and Middlesex County which are strong Democrat strongholds that went for President Joe Biden over former President Trump in the 2020 presidential election.

As of 2018, there are nearly 250 counties across the United States with more registered voters on the voter rolls than eligible citizen voters. There are also nearly three million individuals who are registered to vote in more than one state. In Illinois alone, close to 580 noncitizen voters were improperly registered to vote in the 2018 election.
    
\end{itquote}

\item Rumor name: \textbf{Vulnerable election technology (i.e. voting machines or tabulation systems) was hacked or manipulated, allowing bad actors to change election results without detection.}


Original article title: \textit{Researchers Question Reliability of Dominion Voting Systems, Election Systems \& Software}

\begin{itquote}
Researchers have questioned the reliability of new voting machines that state and local officials have rushed to implement at their polling locations ahead of the 2020 presidential election.

``Some of the most popular ballot-marking machines, made by Election Systems \& Software and Dominion Voting Systems, register votes in bar codes that the human eye cannot decipher," according to a February report by Associated Press.

But according to researchers, that’s a problem, as ``voters could end up with printouts that accurately spell out the names of the candidates they picked, but, because of a hack, the bar codes do not reflect those choices."

``Because the bar codes are what’s tabulated, voters would never know that their ballots benefited another candidate," the report adds.

State and local officials had reportedly rushed to replace old voting systems with the new software ahead of the 2020 presidential election out of fear of ``unreliable electronic voting machines" in the wake of so-called ``Russia’s interference in the 2016 U.S. presidential race."

But instead of using hand-marked paper ballots — which are most resistant to tampering due to the fact that paper cannot be hacked — many have opted out for technology that computer security experts believe to be nearly as risky as the older electronic systems.

Election Systems \& Software disagrees, insisting that the security and accuracy of the company’s ballot-marking machines ``have been proven through thousands of hours of testing and tens of thousands of successful elections," according to a company spokesperson, Katina Granger.

Nonetheless, critics see the machines as vulnerable to hacking and noted that tinkerers at last year’s DefCon hacker convention in Las Vegas were able to ``hack two older ballot-marking devices" in less than eight hours.




Dominion Voting Systems election software was implemented in all of Georgia’s counties for the first time this year.

``Georgia’s new electronic voting system is vulnerable to cyberattacks that could undermine public confidence, create chaos at the polls or even manipulate the results on Election Day," reported the Atlanta Journal-Constitution (AJC) in October.



The report continued:

    Officials tell voters to verify their selections on a paper ballot before feeding it into an optical scanner. But the scanner doesn’t record the text that voters see; rather, it reads an unencrypted quick response, or QR, barcode that is indecipherable to the human eye. Either by tampering with individual voting machines or by infiltrating the state’s central elections server, hackers could systematically alter the barcodes to change votes.

    Such a manipulation could not be detected without an audit after the election.

    The new voting system ``presents serious security vulnerability and operational issues" caused by ``fundamental deficits and exposure," U.S. District Judge Amy Totenberg wrote in a recent order, in which she criticized state officials for not taking the problems more seriously.

    ``These risks," Totenberg wrote, ``are neither hypothetical nor remote under the current circumstances."


Dominion disagrees with these findings, stating that multiple large local governments across the country — such Cook County, Illinois, which includes Chicago, and San Francisco and San Diego counties in California — have purchased their system.

U.S. intelligence agencies, however, have warned that such systems can be targets of foreign governments trying to disrupt elections.

\end{itquote}

\item Rumor name: \textbf{Changes in reported election results in the days following the elections, or any deviation from election night expectations, indicated compromised processes or untrustworthy outcomes rather than normal vote counting procedures.}

Original article title: \textit{Hans von Spakovsky: 120K Straight Vote Dump for Biden Is Impossible
}

\begin{itquote}
Vote dumps entirely for former Vice President Joe Biden are not credible, assessed Hans von Spakovsky, manager of the Heritage Foundation’s Election Law Reform Initiative and a senior legal fellow of the Meese Center for Legal and Judicial Studies, offering his analysis on Thursday’s edition of SiriusXM’s Breitbart News Daily with host Alex Marlow.

Marlow asked about reports of drastic spikes in vote counts for Joe Biden in the early hours of Wednesday morning.

``If those reports are correct, I don’t understand it. The way you do counting is you simply count all of the ballots," Von Spakovsky said. ``You don’t divide. They’re not divided up between the candidates. So the [precinct] reporting that’s coming in ought to be reporting of the total vote count, regardless of who it’s for. So again, if it’s confirmed that there are these weird reports coming out of votes only for one candidate and not the other, you’ve got to question, what exactly is going on?"

    The only thing we did on Election Day was tell them how many votes they needed on Election Night. pic.twitter.com/lOG2iV4l2e

    — Andy Swan (@AndySwan) November 4, 2020

Reports of Republican poll watchers being denied observation of vote counting raise questions about electoral misconduct, stated von Spakovsky.

``It does raise concerns when you know that all the people that are working there are clearly Democrats," remarked von Spakovsky. ``That’s why it’s so important that those places comply with state poll-watching laws. All the campaigns [and] all the political parties are legally entitled to have poll watchers watching every aspect of the election process, including the counting process."

``For places like Detroit to chase out and not allow legally appointed poll watchers in there to watch them processing these absentee ballots raises serious questions about possible misconduct going on," Von Spakovsky said. ``Now, I don’t have evidence of misconduct. But the point is, if the poll watchers were there we would know what was exactly going on in the vote counting process."

Poll watchers are needed to reduce the risk of ineligible ballots being included in vote count, explained von Spakovsky.

``What I worry about is absentee ballots being accepted, processed, and counted that don’t comply with state law requirements, because they know it’s going to be the vote they like," von Spakovsky warned. ``What I mean by that is an absentee ballot comes in, and the signature doesn’t match, so clearly it may be fraudulent. Or it came in late, but they counted anyway. That’s the kind of thing you don’t want to have happening, because that is simply illegal."

Marlow asked von Spakovsky what advice he would offer the president with respect to protecting electoral integrity.

Von Spakovsky replied, ``The only that’s going to help [Trump] now is lawful means, court orders ordering election officials to comply with the law [and] to not count absentee ballots that have been received in violation of state law. That’s where my resources would be concentrated if I was doing this."
    
\end{itquote}

\item Rumor name: \textbf{Certain voting methods or equipment (e.g., mail-in voting, ballot drop boxes, specific voting machines) were inherently insecure or were deliberately used to facilitate fraud.}


Original article title: \textit{Software Not Properly Updated Gave Biden 1000s of Votes in Michigan}

\begin{itquote}
The election software that ``glitched" in both Georgia and Michigan — which in Michigan’s case, incorrectly gave Joe Biden thousands of votes — is being used in 28 states, according to the software company’s website.

The software company, Dominion Voting Systems, ``glitched" in Michigan, causing thousands of ballots that were meant for Republican candidates to be wrongly counted for Democrats in the state’s Antrim County. Antrim is also one of 47 counties in Michigan that uses the same software that experienced this ``glitch."

The presidential election results for Antrim County were later corrected, flipping the county from Joe Biden to President Donald Trump after the ``glitch" was fixed.

Two Georgia counties — which use the same electronic voting software — also reported encountering glitches during the 2020 election, which caused their voting machines to crash.

A Georgia election official said that a technical glitch that halted voting in the state’s Spalding and Morgan counties was caused by a vendor uploading an update to their election machines the night before the election.

``That is something that they don’t ever do. I’ve never seen them update anything the day before the election," said Marcia Ridley, elections supervisor at Spalding County Board of Election.

A third county in Georgia — Gwinnett County — which uses the same software, also experienced a glitch. This glitch, however, had caused the delay of counting thousands of votes in the 2020 presidential election.

Election officials estimate that roughly 80,000 absentee ballots were impacted by this glitch, yet decided to push the impacted votes through, knowing some of the votes would likely change.

The software was implemented in all of Georgia’s counties for the first time this year. Last year, the Dominion Democracy Suite 5.5A was certified by the Pennsylvania Department of State.

Dominion Voting Systems boasts on its website of having ``customers in 28 states," including ``9 of the top 20 counties" and ``4 of the top 10 counties" throughout the United States.

``Dominion is ready to make a difference in your next election," the company advertises on its website.
\end{itquote}

\end{enumerate}

\subsection*{Data Description}

Here we include summary statistics, balance tables, and other descriptions of the experimental and survey data. 

\subsubsection*{Variable Definitions}

Here we briefly define any variables or abbreviations used in the following analysis. 

\begin{itemize}
    \item X Ballot Confidence: on a 0-10 scale, how confident are you that X will be accurately in the upcoming election? (X: Own Ballot, Ballots in County, Ballots around the Country)
    \item Pol Interest: How often do you pay attention to politics? 
    \item Rumor Diff: Difference between pre-treatment and post-treatment 0-10 measures of confidence in \textbf{assigned} rumor
    \item Rumor Diff Recontact: Difference between pre-treatment and follow-up 0-10 measures of confidence in \textbf{assigned} rumor
    \item All Rumors Diff Recontact: Difference between pre-treatment and follow-up 0-10 measures of confidence in \textbf{all} five election rumors, divided by five to normalize to a 0-10 scale
    \item All Facts Diff Recontact: Difference between pre-treatment and follow-up 0-10 measures of confidence in all five true election facts, divided by five to normalize to a 0-10 scale
    \item MIST Correct: total number of Misinformation Susceptibility Test (MIST-8) scores correct, from 0-8.

\end{itemize}

\subsubsection*{Binary Respondent Confidence} \label{sec:binary_confidence}



\begin{table}[h]
\centering
\caption{Proportion of participants confident in the election integrity (score $>$ 5 out of 10) before, immediately after, and at recontact} 
\label{tab:prop_election}
\begingroup\small
\begin{tabular}{llll}
  \hline
Treatment & Pre-Election & Post-Election & Recontact-Election \\ 
  \hline
Placebo & 68.4\% & 63.9\% & 69.3\% \\ 
  Treatment & 69.0\% & 66.2\% & 69.5\% \\ 
   \hline
\end{tabular}
\endgroup
\end{table}

\begin{table}[h]
\centering
\caption{Proportion of participants confident in the election integrity (score $>$ 5 out of 10) before, immediately after, and at recontact, by party identification} 
\label{tab:prop_election_party}
\begingroup\small
\begin{tabular}{lllll}
  \hline
Treatment & Party & Pre-Election & Post-Election & Recontact-Election \\ 
  \hline
Placebo & Democrat & 92.5\% & 89.4\% & 91.8\% \\ 
  Placebo & Independent & 64.7\% & 60.6\% & 66.8\% \\ 
  Placebo & Republican & 44.1\% & 37.5\% & 45.5\% \\ 
  Treatment & Democrat & 91.3\% & 90.1\% & 93.7\% \\ 
  Treatment & Independent & 61.2\% & 58.0\% & 61.9\% \\ 
  Treatment & Republican & 49.3\% & 44.8\% & 47.3\% \\ 
   \hline
\end{tabular}
\endgroup
\end{table}

\begin{table}[h]
\centering
\caption{Proportion of participants confident in the election integrity (score $>$ 5 out of 10) before, immediately after, and at recontact, by assigned rumor} 
\label{tab:prop_election_rumor}
\begingroup\small
\begin{tabular}{lllll}
  \hline
Rumor & Treatment & Pre-Election & Post-Election & Recontact-Election \\ 
  \hline
Voter Fraud & Placebo & 65.7\% & 62.4\% & 67.9\% \\ 
  Voter Fraud & Treatment & 70.0\% & 67.5\% & 71.3\% \\ 
  Voter Rolls & Placebo & 71.1\% & 65.1\% & 72.0\% \\ 
  Voter Rolls & Treatment & 70.7\% & 69.0\% & 72.3\% \\ 
  Hacking & Placebo & 67.9\% & 63.1\% & 68.5\% \\ 
  Hacking & Treatment & 66.3\% & 63.1\% & 65.1\% \\ 
  Blue Shift & Placebo & 67.1\% & 64.3\% & 69.7\% \\ 
  Blue Shift & Treatment & 68.0\% & 63.1\% & 68.4\% \\ 
  Voting Machines & Placebo & 70.6\% & 64.6\% & 68.7\% \\ 
  Voting Machines & Treatment & 70.0\% & 68.1\% & 70.4\% \\ 
   \hline
\end{tabular}
\endgroup
\end{table}

\newpage
\clearpage

\begin{table}[h]
\centering
\caption{Proportion of participants confident in the election integrity (score $>$ 5 out of 10) before, immediately after, and at recontact, by assigned rumor and party identification} 
\label{tab:prop_election_indiv_rumor_party}
\begingroup\small
\begin{tabular}{llllll}
  \hline
Rumor & Treatment & Party & Pre-Election & Post-Election & Recontact-Election \\ 
  \hline
Voter Fraud & Placebo & Democrat & 89.6\% & 89.6\% & 93.4\% \\ 
  Voter Fraud & Treatment & Democrat & 93.8\% & 95.0\% & 96.2\% \\ 
  Voter Fraud & Placebo & Independent & 61.8\% & 58.1\% & 65.0\% \\ 
  Voter Fraud & Treatment & Independent & 59.0\% & 54.1\% & 59.8\% \\ 
  Voter Fraud & Placebo & Republican & 43.3\% & 36.9\% & 43.4\% \\ 
  Voter Fraud & Treatment & Republican & 49.6\% & 44.6\% & 50.0\% \\ 
  Voter Rolls & Placebo & Democrat & 91.1\% & 86.1\% & 89.5\% \\ 
  Voter Rolls & Treatment & Democrat & 93.5\% & 90.3\% & 93.7\% \\ 
  Voter Rolls & Placebo & Independent & 65.9\% & 62.1\% & 70.3\% \\ 
  Voter Rolls & Treatment & Independent & 60.4\% & 59.7\% & 60.4\% \\ 
  Voter Rolls & Placebo & Republican & 51.6\% & 42.2\% & 51.4\% \\ 
  Voter Rolls & Treatment & Republican & 50.0\% & 49.3\% & 53.8\% \\ 
  Hacking & Placebo & Democrat & 92.6\% & 90.1\% & 91.6\% \\ 
  Hacking & Treatment & Democrat & 88.8\% & 88.1\% & 91.3\% \\ 
  Hacking & Placebo & Independent & 62.8\% & 57.4\% & 59.6\% \\ 
  Hacking & Treatment & Independent & 56.6\% & 52.5\% & 57.3\% \\ 
  Hacking & Placebo & Republican & 44.8\% & 37.9\% & 50.0\% \\ 
  Hacking & Treatment & Republican & 47.7\% & 42.3\% & 42.3\% \\ 
  Blue Shift & Placebo & Democrat & 93.3\% & 91.4\% & 92.0\% \\ 
  Blue Shift & Treatment & Democrat & 91.1\% & 85.4\% & 90.8\% \\ 
  Blue Shift & Placebo & Independent & 65.2\% & 62.9\% & 72.7\% \\ 
  Blue Shift & Treatment & Independent & 61.8\% & 60.4\% & 64.7\% \\ 
  Blue Shift & Placebo & Republican & 37.5\% & 33.1\% & 39.3\% \\ 
  Blue Shift & Treatment & Republican & 50.3\% & 42.9\% & 48.8\% \\ 
  Voting Machines & Placebo & Democrat & 96.1\% & 89.5\% & 92.5\% \\ 
  Voting Machines & Treatment & Democrat & 89.3\% & 91.7\% & 96.3\% \\ 
  Voting Machines & Placebo & Independent & 68.3\% & 62.7\% & 66.7\% \\ 
  Voting Machines & Treatment & Independent & 67.4\% & 62.1\% & 66.7\% \\ 
  Voting Machines & Placebo & Republican & 43.6\% & 37.6\% & 43.9\% \\ 
  Voting Machines & Treatment & Republican & 48.5\% & 44.9\% & 41.0\% \\ 
   \hline
\end{tabular}
\endgroup
\end{table}

\newpage
\clearpage

\begin{table}[h]
\centering
\caption{Proportion of participants confident election rumors is true (score $>$ 5 out of 10) before, immediately after, and at recontact} 
\label{tab:prop_rumor}
\begingroup\small
\begin{tabular}{llll}
  \hline
Treatment & Pre-Rumor & Post-Rumor & Recontact-Rumor \\ 
  \hline
Placebo & 44.2\% & 48.8\% & 43.4\% \\ 
  Treatment & 44.6\% & 43.3\% & 42.3\% \\ 
   \hline
\end{tabular}
\endgroup
\end{table}

\begin{table}[h]
\centering
\caption{Proportion of participants confident election rumors is true (score $>$ 5 out of 10) before, immediately after, and at recontact, by party identification} 
\label{tab:prop_rumor_party}
\begingroup\small
\begin{tabular}{lllll}
  \hline
Treatment & Party & Pre-Rumor & Post-Rumor & Recontact-Rumor \\ 
  \hline
Placebo & Democrat & 20.6\% & 24.2\% & 18.8\% \\ 
  Placebo & Independent & 39.7\% & 44.3\% & 40.0\% \\ 
  Placebo & Republican & 75.8\% & 81.6\% & 75.6\% \\ 
  Treatment & Democrat & 20.7\% & 20.8\% & 19.7\% \\ 
  Treatment & Independent & 42.7\% & 40.4\% & 37.6\% \\ 
  Treatment & Republican & 75.7\% & 73.5\% & 74.5\% \\ 
   \hline
\end{tabular}
\endgroup
\end{table}

\begin{table}[h]
\centering
\caption{Proportion of participants confident election rumors is true (score $>$ 5 out of 10) before, immediately after, and at recontact, by assigned rumor} 
\label{tab:prop_indiv_rumor}
\begingroup\small
\begin{tabular}{lllll}
  \hline
Rumor & Treatment & Pre-Rumor & Post-Rumor & Recontact-Rumor \\ 
  \hline
Voter Fraud & Placebo & 47.8\% & 52.0\% & 47.4\% \\ 
  Voter Fraud & Treatment & 43.9\% & 42.9\% & 41.7\% \\ 
  Voter Rolls & Placebo & 42.1\% & 52.2\% & 43.7\% \\ 
  Voter Rolls & Treatment & 44.2\% & 45.4\% & 39.6\% \\ 
  Hacking & Placebo & 40.6\% & 46.6\% & 40.6\% \\ 
  Hacking & Treatment & 45.4\% & 40.0\% & 43.1\% \\ 
  Blue Shift & Placebo & 46.9\% & 48.0\% & 42.2\% \\ 
  Blue Shift & Treatment & 44.3\% & 46.1\% & 43.1\% \\ 
  Voting Machines & Placebo & 43.7\% & 45.1\% & 43.2\% \\ 
  Voting Machines & Treatment & 45.2\% & 41.5\% & 44.3\% \\ 
   \hline
\end{tabular}
\endgroup
\end{table}

\clearpage
\newpage

\begin{table}[h]
\centering
\caption{Proportion of participants confident election rumors is true (score $>$ 5 out of 10) before, immediately after, and at recontact, by assigned rumor and party identification} 
\label{tab:prop_rumor_indiv_rumor_party}
\begingroup\small
\begin{tabular}{llllll}
  \hline
Rumor & Treatment & Party & Pre-Rumor & Post-Rumor & Recontact-Rumor \\ 
  \hline
Voter Fraud & Placebo & Democrat & 17.5\% & 23.4\% & 19.7\% \\ 
  Voter Fraud & Treatment & Democrat & 19.4\% & 15.0\% & 18.0\% \\ 
  Voter Fraud & Placebo & Independent & 43.4\% & 48.5\% & 42.7\% \\ 
  Voter Fraud & Treatment & Independent & 43.4\% & 44.3\% & 36.4\% \\ 
  Voter Fraud & Placebo & Republican & 85.1\% & 86.5\% & 82.3\% \\ 
  Voter Fraud & Treatment & Republican & 76.9\% & 78.5\% & 79.6\% \\ 
  Voter Rolls & Placebo & Democrat & 19.6\% & 31.0\% & 17.3\% \\ 
  Voter Rolls & Treatment & Democrat & 20.4\% & 22.6\% & 19.0\% \\ 
  Voter Rolls & Placebo & Independent & 38.6\% & 46.2\% & 43.6\% \\ 
  Voter Rolls & Treatment & Independent & 42.5\% & 42.5\% & 37.7\% \\ 
  Voter Rolls & Placebo & Republican & 73.4\% & 84.4\% & 77.1\% \\ 
  Voter Rolls & Treatment & Republican & 78.1\% & 79.0\% & 69.2\% \\ 
  Hacking & Placebo & Democrat & 18.5\% & 22.2\% & 19.8\% \\ 
  Hacking & Treatment & Democrat & 23.8\% & 21.2\% & 22.8\% \\ 
  Hacking & Placebo & Independent & 37.2\% & 41.1\% & 33.9\% \\ 
  Hacking & Treatment & Independent & 43.4\% & 36.9\% & 37.9\% \\ 
  Hacking & Placebo & Republican & 68.3\% & 78.6\% & 71.4\% \\ 
  Hacking & Treatment & Republican & 73.8\% & 66.2\% & 71.2\% \\ 
  Blue Shift & Placebo & Democrat & 30.7\% & 27.0\% & 21.0\% \\ 
  Blue Shift & Treatment & Democrat & 22.9\% & 27.4\% & 20.6\% \\ 
  Blue Shift & Placebo & Independent & 38.6\% & 43.2\% & 41.8\% \\ 
  Blue Shift & Treatment & Independent & 38.9\% & 38.9\% & 33.6\% \\ 
  Blue Shift & Placebo & Republican & 74.3\% & 77.9\% & 68.8\% \\ 
  Blue Shift & Treatment & Republican & 71.0\% & 71.6\% & 74.4\% \\ 
  Voting Machines & Placebo & Democrat & 16.3\% & 17.0\% & 16.0\% \\ 
  Voting Machines & Treatment & Democrat & 17.3\% & 17.9\% & 18.5\% \\ 
  Voting Machines & Placebo & Independent & 40.5\% & 42.1\% & 38.1\% \\ 
  Voting Machines & Treatment & Independent & 45.5\% & 39.4\% & 42.6\% \\ 
  Voting Machines & Placebo & Republican & 78.2\% & 80.5\% & 78.5\% \\ 
  Voting Machines & Treatment & Republican & 79.4\% & 72.8\% & 79.0\% \\ 
   \hline
\end{tabular}
\endgroup
\end{table}

\clearpage
\newpage

\subsubsection*{Summary Statistics and Balance Tables}


 {
\let\oldcentering\centering \renewcommand\centering{\tiny\oldcentering} 
 \setlength{\tabcolsep}{2pt}

\begin{table}[!h]
\label{tab:tab:summary_statistics_all}
\centering
\caption{Summary Statistics - All}
\centering
\begin{tabular}[t]{lllllllllllll}

\textbf{Group} & \textbf{Placebo} & \textbf{Variable} & \textbf{N} & \textbf{Mean} & \textbf{Var} & \textbf{Min} & \textbf{Q10} & \textbf{Q25} & \textbf{Median} & \textbf{Q75} & \textbf{Q90} & \textbf{Max}\\

All & Placebo & Confidence Country Ballots Diff & 2,128 & -0.33 & 2.53 & -7.40 & -2.00 & -1.00 & 0.00 & 0.00 & 1.00 & 6.60\\
All & Treatment & Confidence Country Ballots Diff & 2,164 & -0.16 & 2.80 & -7.80 & -2.00 & -1.00 & 0.00 & 0.20 & 1.50 & 6.40\\
All & Placebo & Confidence County Ballots Diff & 2,128 & -0.41 & 2.78 & -9.20 & -2.20 & -1.00 & 0.00 & 0.00 & 1.00 & 7.40\\
All & Treatment & Confidence County Ballots Diff & 2,165 & -0.34 & 2.72 & -9.60 & -2.00 & -0.95 & 0.00 & 0.00 & 1.00 & 7.80\\
All & Placebo & Confidence Own Ballot Diff & 2,128 & -0.52 & 2.97 & -9.60 & -2.40 & -1.00 & 0.00 & 0.00 & 0.90 & 8.40\\

All & Treatment & Confidence Own Ballot Diff & 2,165 & -0.49 & 2.85 & -10.00 & -2.40 & -1.00 & 0.00 & 0.00 & 1.00 & 6.20\\
All & Placebo & Conspiracy Score & 2,128 & 20.33 & 59.82 & 5.60 & 9.94 & 13.95 & 19.50 & 28.75 & 30.00 & 30.00\\
All & Treatment & Conspiracy Score & 2,165 & 20.45 & 58.28 & 6.00 & 10.00 & 14.00 & 20.50 & 28.00 & 30.00 & 30.00\\
All & Placebo & MIST Correct & 2,128 & 5.57 & 3.90 & 0.00 & 3.00 & 4.00 & 6.00 & 7.00 & 8.00 & 8.00\\
All & Treatment & MIST Correct & 2,165 & 5.63 & 3.72 & 0.20 & 3.00 & 4.00 & 6.00 & 7.00 & 8.00 & 8.00\\

All & Placebo & Populism Score & 2,128 & 12.70 & 16.73 & 6.00 & 7.02 & 9.60 & 12.40 & 15.40 & 18.00 & 27.20\\
All & Treatment & Populism Score & 2,165 & 12.63 & 16.53 & 6.00 & 6.90 & 9.60 & 12.40 & 15.80 & 18.00 & 26.20\\
All & Placebo & Post Confidence Country Ballots & 2,128 & 6.37 & 9.40 & 0.00 & 1.70 & 4.40 & 7.00 & 9.00 & 10.00 & 10.00\\
All & Treatment & Post Confidence Country Ballots & 2,164 & 6.61 & 9.11 & 0.00 & 1.60 & 4.90 & 7.40 & 9.00 & 10.00 & 10.00\\
All & Placebo & Post Confidence County Ballots & 2,128 & 7.35 & 7.41 & 0.00 & 3.20 & 5.75 & 8.00 & 9.85 & 10.00 & 10.00\\

All & Treatment & Post Confidence County Ballots & 2,165 & 7.48 & 7.03 & 0.00 & 3.60 & 5.95 & 8.00 & 10.00 & 10.00 & 10.00\\
All & Placebo & Post Confidence Own Ballot & 2,128 & 7.29 & 7.86 & 0.00 & 2.94 & 5.40 & 8.00 & 9.80 & 10.00 & 10.00\\
All & Treatment & Post Confidence Own Ballot & 2,165 & 7.44 & 7.27 & 0.00 & 3.70 & 5.80 & 8.00 & 10.00 & 10.00 & 10.00\\
All & Placebo & Pre Confidence Country Ballots & 2,128 & 6.70 & 9.01 & 0.00 & 2.20 & 5.00 & 7.20 & 9.00 & 10.00 & 10.00\\
All & Treatment & Pre Confidence Country Ballots & 2,165 & 6.76 & 8.95 & 0.00 & 2.14 & 5.00 & 7.80 & 9.00 & 10.00 & 10.00\\

All & Placebo & Pre Confidence County Ballots & 2,128 & 7.76 & 6.85 & 0.00 & 4.00 & 6.80 & 8.80 & 10.00 & 10.00 & 10.00\\
All & Treatment & Pre Confidence County Ballots & 2,165 & 7.83 & 6.34 & 0.00 & 4.60 & 7.00 & 8.80 & 10.00 & 10.00 & 10.00\\
All & Placebo & Pre Confidence Own Ballot & 2,128 & 7.81 & 6.89 & 0.00 & 4.34 & 6.80 & 9.00 & 10.00 & 10.00 & 10.00\\
All & Treatment & Pre Confidence Own Ballot & 2,165 & 7.93 & 5.93 & 0.00 & 5.00 & 7.00 & 9.00 & 10.00 & 10.00 & 10.00\\

\multicolumn{13}{l}{\rule{0pt}{1em}\textit{Note:} }\\
\end{tabular}
\end{table}

}

 {
\let\oldcentering\centering \renewcommand\centering{\tiny\oldcentering} 
 \setlength{\tabcolsep}{2pt}
  
  \hspace*{-2cm}
\begin{table}[!h]
\centering
\label{tab:tab:summary_statistics_voter_fraud}
\caption{Summary Statistics - Voter Fraud}
\centering
\begin{tabular}[t]{lllllllllllll}

\textbf{Group} & \textbf{Placebo} & \textbf{Variable} & \textbf{N} & \textbf{Mean} & \textbf{Var} & \textbf{Min} & \textbf{Q10} & \textbf{Q25} & \textbf{Median} & \textbf{Q75} & \textbf{Q90} & \textbf{Max}\\

Voter Fraud & Placebo & Pre Confidence Own Ballot & 431 & 7.67 & 7.39 & 0.00 & 4.00 & 6.00 & 9.00 & 10.00 & 10.00 & 10.00\\
Voter Fraud & Treatment & Pre Confidence Own Ballot & 403 & 8.00 & 5.54 & 0.00 & 5.00 & 7.00 & 9.00 & 10.00 & 10.00 & 10.00\\
Voter Fraud & Placebo & Pre Confidence County Ballots & 431 & 7.63 & 7.17 & 0.00 & 3.00 & 6.00 & 8.00 & 10.00 & 10.00 & 10.00\\
Voter Fraud & Treatment & Pre Confidence County Ballots & 403 & 7.87 & 5.92 & 0.00 & 5.00 & 7.00 & 9.00 & 10.00 & 10.00 & 10.00\\
Voter Fraud & Placebo & Pre Confidence Country Ballots & 431 & 6.50 & 9.22 & 0.00 & 2.00 & 5.00 & 7.00 & 9.00 & 10.00 & 10.00\\

Voter Fraud & Treatment & Pre Confidence Country Ballots & 403 & 6.77 & 8.71 & 0.00 & 2.00 & 5.00 & 8.00 & 9.00 & 10.00 & 10.00\\
Voter Fraud & Placebo & Post Confidence Own Ballot & 431 & 7.24 & 8.48 & 0.00 & 2.00 & 5.00 & 8.00 & 10.00 & 10.00 & 10.00\\
Voter Fraud & Treatment & Post Confidence Own Ballot & 403 & 7.45 & 7.32 & 0.00 & 4.00 & 6.00 & 8.00 & 10.00 & 10.00 & 10.00\\
Voter Fraud & Placebo & Post Confidence County Ballots & 431 & 7.27 & 7.76 & 0.00 & 3.00 & 5.00 & 8.00 & 10.00 & 10.00 & 10.00\\
Voter Fraud & Treatment & Post Confidence County Ballots & 403 & 7.63 & 6.52 & 0.00 & 4.00 & 6.00 & 8.00 & 10.00 & 10.00 & 10.00\\

Voter Fraud & Placebo & Post Confidence Country Ballots & 431 & 6.17 & 9.68 & 0.00 & 1.00 & 4.00 & 7.00 & 9.00 & 10.00 & 10.00\\
Voter Fraud & Treatment & Post Confidence Country Ballots & 403 & 6.74 & 8.86 & 0.00 & 2.00 & 5.00 & 8.00 & 9.00 & 10.00 & 10.00\\
Voter Fraud & Placebo & Confidence Own Ballot Diff & 431 & -0.43 & 2.55 & -9.00 & -2.00 & -1.00 & 0.00 & 0.00 & 1.00 & 10.00\\
Voter Fraud & Treatment & Confidence Own Ballot Diff & 403 & -0.54 & 2.74 & -10.00 & -3.00 & -1.00 & 0.00 & 0.00 & 1.00 & 5.00\\
Voter Fraud & Placebo & Confidence County Ballots Diff & 431 & -0.37 & 2.26 & -7.00 & -2.00 & -1.00 & 0.00 & 0.00 & 1.00 & 7.00\\

Voter Fraud & Treatment & Confidence County Ballots Diff & 403 & -0.25 & 2.99 & -10.00 & -2.00 & -1.00 & 0.00 & 0.00 & 1.00 & 9.00\\
Voter Fraud & Placebo & Confidence Country Ballots Diff & 431 & -0.32 & 2.36 & -7.00 & -2.00 & -1.00 & 0.00 & 0.00 & 1.00 & 6.00\\
Voter Fraud & Treatment & Confidence Country Ballots Diff & 403 & -0.03 & 2.65 & -5.00 & -2.00 & -1.00 & 0.00 & 1.00 & 2.00 & 6.00\\
Voter Fraud & Placebo & Populism Score & 431 & 12.55 & 15.87 & 6.00 & 7.00 & 10.00 & 12.00 & 15.00 & 18.00 & 25.00\\
Voter Fraud & Treatment & Populism Score & 403 & 12.53 & 17.68 & 6.00 & 6.00 & 9.00 & 13.00 & 16.00 & 18.00 & 26.00\\

Voter Fraud & Placebo & Conspiracy Score & 431 & 20.24 & 59.59 & 4.00 & 9.00 & 14.00 & 19.00 & 29.00 & 30.00 & 30.00\\
Voter Fraud & Treatment & Conspiracy Score & 403 & 20.76 & 58.76 & 6.00 & 10.00 & 15.00 & 21.00 & 28.00 & 30.00 & 30.00\\
Voter Fraud & Placebo & MIST Correct & 431 & 5.64 & 3.73 & 0.00 & 3.00 & 4.00 & 6.00 & 7.00 & 8.00 & 8.00\\
Voter Fraud & Treatment & MIST Correct & 403 & 5.67 & 3.82 & 0.00 & 3.00 & 4.00 & 6.00 & 7.00 & 8.00 & 8.00\\

\multicolumn{13}{l}{\rule{0pt}{1em}\textit{Note:} }\\
\end{tabular}
\end{table}

}

\clearpage
\newpage

 {
\let\oldcentering\centering \renewcommand\centering{\tiny\oldcentering} 
 \setlength{\tabcolsep}{2pt}
  
  \hspace*{-2cm}
\begin{table}[!h]
\centering
\label{tab:tab:summary_statistics_voter_rolls}
\caption{Summary Statistics - Voter Rolls}
\centering
\begin{tabular}[t]{lllllllllllll}

\textbf{Group} & \textbf{Placebo} & \textbf{Variable} & \textbf{N} & \textbf{Mean} & \textbf{Var} & \textbf{Min} & \textbf{Q10} & \textbf{Q25} & \textbf{Median} & \textbf{Q75} & \textbf{Q90} & \textbf{Max}\\

Voter Rolls & Placebo & Pre Confidence Own Ballot & 418 & 7.83 & 6.87 & 0.00 & 4.70 & 7.00 & 9.00 & 10.00 & 10.00 & 10.00\\
Voter Rolls & Treatment & Pre Confidence Own Ballot & 458 & 7.90 & 6.03 & 0.00 & 5.00 & 7.00 & 9.00 & 10.00 & 10.00 & 10.00\\
Voter Rolls & Placebo & Pre Confidence County Ballots & 418 & 7.85 & 6.41 & 0.00 & 5.00 & 7.00 & 9.00 & 10.00 & 10.00 & 10.00\\
Voter Rolls & Treatment & Pre Confidence County Ballots & 458 & 7.92 & 6.01 & 0.00 & 5.00 & 7.00 & 9.00 & 10.00 & 10.00 & 10.00\\
Voter Rolls & Placebo & Pre Confidence Country Ballots & 418 & 6.98 & 8.51 & 0.00 & 3.00 & 5.00 & 8.00 & 9.00 & 10.00 & 10.00\\

Voter Rolls & Treatment & Pre Confidence Country Ballots & 458 & 6.90 & 8.10 & 0.00 & 2.70 & 5.00 & 8.00 & 9.00 & 10.00 & 10.00\\
Voter Rolls & Placebo & Post Confidence Own Ballot & 418 & 7.37 & 7.40 & 0.00 & 3.70 & 6.00 & 8.00 & 10.00 & 10.00 & 10.00\\
Voter Rolls & Treatment & Post Confidence Own Ballot & 458 & 7.66 & 6.64 & 0.00 & 4.00 & 6.00 & 8.00 & 10.00 & 10.00 & 10.00\\
Voter Rolls & Placebo & Post Confidence County Ballots & 418 & 7.45 & 6.54 & 0.00 & 4.00 & 6.00 & 8.00 & 10.00 & 10.00 & 10.00\\
Voter Rolls & Treatment & Post Confidence County Ballots & 458 & 7.67 & 6.67 & 0.00 & 4.00 & 6.00 & 8.00 & 10.00 & 10.00 & 10.00\\

Voter Rolls & Placebo & Post Confidence Country Ballots & 418 & 6.49 & 8.94 & 0.00 & 2.00 & 5.00 & 7.00 & 9.00 & 10.00 & 10.00\\
Voter Rolls & Treatment & Post Confidence Country Ballots & 458 & 6.75 & 8.27 & 0.00 & 2.00 & 5.00 & 7.00 & 9.00 & 10.00 & 10.00\\
Voter Rolls & Placebo & Confidence Own Ballot Diff & 418 & -0.46 & 2.86 & -10.00 & -2.00 & -1.00 & 0.00 & 0.00 & 1.00 & 10.00\\
Voter Rolls & Treatment & Confidence Own Ballot Diff & 458 & -0.24 & 2.54 & -10.00 & -2.00 & -1.00 & 0.00 & 0.00 & 1.00 & 6.00\\
Voter Rolls & Placebo & Confidence County Ballots Diff & 418 & -0.39 & 3.18 & -10.00 & -2.00 & -1.00 & 0.00 & 0.00 & 1.00 & 10.00\\

Voter Rolls & Treatment & Confidence County Ballots Diff & 458 & -0.25 & 2.43 & -10.00 & -2.00 & -0.75 & 0.00 & 0.00 & 1.00 & 8.00\\
Voter Rolls & Placebo & Confidence Country Ballots Diff & 418 & -0.49 & 2.92 & -10.00 & -2.00 & -1.00 & 0.00 & 0.00 & 1.00 & 8.00\\
Voter Rolls & Treatment & Confidence Country Ballots Diff & 458 & -0.16 & 3.08 & -10.00 & -2.00 & -1.00 & 0.00 & 0.00 & 2.00 & 6.00\\
Voter Rolls & Placebo & Populism Score & 418 & 12.70 & 15.86 & 6.00 & 7.00 & 10.00 & 13.00 & 15.00 & 18.00 & 27.00\\
Voter Rolls & Treatment & Populism Score & 458 & 12.93 & 17.18 & 6.00 & 7.00 & 10.00 & 13.00 & 16.00 & 18.00 & 25.00\\

Voter Rolls & Placebo & Conspiracy Score & 418 & 20.61 & 59.91 & 6.00 & 9.70 & 15.00 & 20.00 & 29.00 & 30.00 & 30.00\\
Voter Rolls & Treatment & Conspiracy Score & 458 & 20.78 & 58.50 & 6.00 & 10.00 & 14.00 & 21.50 & 29.00 & 30.00 & 30.00\\
Voter Rolls & Placebo & MIST Correct & 418 & 5.65 & 3.92 & 0.00 & 3.00 & 4.00 & 6.00 & 7.00 & 8.00 & 8.00\\
Voter Rolls & Treatment & MIST Correct & 458 & 5.83 & 3.42 & 1.00 & 3.00 & 4.00 & 6.00 & 7.00 & 8.00 & 8.00\\

\multicolumn{13}{l}{\rule{0pt}{1em}\textit{Note:} }\\
\end{tabular}
\end{table}

}

 {
\let\oldcentering\centering \renewcommand\centering{\tiny\oldcentering} 
 \setlength{\tabcolsep}{2pt}
  
  \hspace*{-2cm}
\begin{table}[!h]
\centering
\label{tab:tab:summary_statistics_hacking}
\caption{Summary Statistics - Hacking}
\centering
\begin{tabular}[t]{lllllllllllll}

\textbf{Group} & \textbf{Placebo} & \textbf{Variable} & \textbf{N} & \textbf{Mean} & \textbf{Var} & \textbf{Min} & \textbf{Q10} & \textbf{Q25} & \textbf{Median} & \textbf{Q75} & \textbf{Q90} & \textbf{Max}\\

Hacking & Placebo & Pre Confidence Own Ballot & 436 & 7.80 & 7.19 & 0.00 & 4.00 & 7.00 & 9.00 & 10.00 & 10.00 & 10.00\\
Hacking & Treatment & Pre Confidence Own Ballot & 412 & 7.89 & 6.37 & 0.00 & 5.00 & 7.00 & 9.00 & 10.00 & 10.00 & 10.00\\
Hacking & Placebo & Pre Confidence County Ballots & 436 & 7.71 & 7.33 & 0.00 & 3.00 & 7.00 & 9.00 & 10.00 & 10.00 & 10.00\\
Hacking & Treatment & Pre Confidence County Ballots & 412 & 7.69 & 6.93 & 0.00 & 4.00 & 7.00 & 8.00 & 10.00 & 10.00 & 10.00\\
Hacking & Placebo & Pre Confidence Country Ballots & 436 & 6.59 & 9.23 & 0.00 & 2.00 & 5.00 & 7.00 & 9.00 & 10.00 & 10.00\\

Hacking & Treatment & Pre Confidence Country Ballots & 412 & 6.62 & 9.38 & 0.00 & 2.00 & 5.00 & 7.00 & 9.00 & 10.00 & 10.00\\
Hacking & Placebo & Post Confidence Own Ballot & 436 & 7.08 & 7.75 & 0.00 & 3.00 & 5.00 & 8.00 & 9.00 & 10.00 & 10.00\\
Hacking & Treatment & Post Confidence Own Ballot & 412 & 7.14 & 8.03 & 0.00 & 3.00 & 5.00 & 8.00 & 10.00 & 10.00 & 10.00\\
Hacking & Placebo & Post Confidence County Ballots & 436 & 7.24 & 7.36 & 0.00 & 3.00 & 6.00 & 8.00 & 9.25 & 10.00 & 10.00\\
Hacking & Treatment & Post Confidence County Ballots & 412 & 7.24 & 7.76 & 0.00 & 3.00 & 5.75 & 8.00 & 10.00 & 10.00 & 10.00\\

Hacking & Placebo & Post Confidence Country Ballots & 436 & 6.22 & 9.17 & 0.00 & 1.50 & 4.00 & 7.00 & 9.00 & 10.00 & 10.00\\
Hacking & Treatment & Post Confidence Country Ballots & 412 & 6.36 & 9.89 & 0.00 & 1.00 & 5.00 & 7.00 & 9.00 & 10.00 & 10.00\\
Hacking & Placebo & Confidence Own Ballot Diff & 436 & -0.72 & 3.42 & -10.00 & -3.00 & -1.00 & 0.00 & 0.00 & 0.50 & 7.00\\
Hacking & Treatment & Confidence Own Ballot Diff & 412 & -0.75 & 3.27 & -10.00 & -3.00 & -1.00 & 0.00 & 0.00 & 1.00 & 4.00\\
Hacking & Placebo & Confidence County Ballots Diff & 436 & -0.48 & 3.47 & -10.00 & -3.00 & -1.00 & 0.00 & 0.00 & 1.00 & 10.00\\

Hacking & Treatment & Confidence County Ballots Diff & 412 & -0.45 & 2.85 & -10.00 & -2.00 & -1.00 & 0.00 & 0.00 & 1.00 & 6.00\\
Hacking & Placebo & Confidence Country Ballots Diff & 436 & -0.37 & 2.46 & -6.00 & -2.00 & -1.00 & 0.00 & 0.00 & 1.00 & 7.00\\
Hacking & Treatment & Confidence Country Ballots Diff & 412 & -0.26 & 3.02 & -10.00 & -2.00 & -1.00 & 0.00 & 0.00 & 1.00 & 6.00\\
Hacking & Placebo & Populism Score & 436 & 12.52 & 16.49 & 6.00 & 7.00 & 9.00 & 12.00 & 15.00 & 18.00 & 28.00\\
Hacking & Treatment & Populism Score & 412 & 12.54 & 15.78 & 6.00 & 7.00 & 10.00 & 12.00 & 16.00 & 18.00 & 27.00\\

Hacking & Placebo & Conspiracy Score & 436 & 20.11 & 60.40 & 6.00 & 10.00 & 13.00 & 19.50 & 28.25 & 30.00 & 30.00\\
Hacking & Treatment & Conspiracy Score & 412 & 20.05 & 54.35 & 6.00 & 10.00 & 14.00 & 19.00 & 27.00 & 30.00 & 30.00\\
Hacking & Placebo & MIST Correct & 436 & 5.56 & 4.02 & 0.00 & 3.00 & 4.00 & 6.00 & 7.00 & 8.00 & 8.00\\
Hacking & Treatment & MIST Correct & 412 & 5.52 & 3.45 & 0.00 & 3.00 & 4.00 & 6.00 & 7.00 & 8.00 & 8.00\\

\multicolumn{13}{l}{\rule{0pt}{1em}\textit{Note:} }\\
\end{tabular}
\end{table}

}

\clearpage
\newpage

 {
\let\oldcentering\centering \renewcommand\centering{\tiny\oldcentering} 
 \setlength{\tabcolsep}{2pt}
  
  \hspace*{-2cm}
\begin{table}[!h]
\centering
\label{tab:tab:summary_statistics_blue_shift}
\caption{Summary Statistics - Blue Shift}
\centering
\begin{tabular}[t]{lllllllllllll}

\textbf{Group} & \textbf{Placebo} & \textbf{Variable} & \textbf{N} & \textbf{Mean} & \textbf{Var} & \textbf{Min} & \textbf{Q10} & \textbf{Q25} & \textbf{Median} & \textbf{Q75} & \textbf{Q90} & \textbf{Max}\\

Blue Shift & Placebo & Pre Confidence Own Ballot & 431 & 7.82 & 6.90 & 0.00 & 4.00 & 7.00 & 9.00 & 10.00 & 10.00 & 10.00\\
Blue Shift & Treatment & Pre Confidence Own Ballot & 456 & 7.93 & 6.11 & 0.00 & 5.00 & 7.00 & 9.00 & 10.00 & 10.00 & 10.00\\
Blue Shift & Placebo & Pre Confidence County Ballots & 431 & 7.77 & 6.75 & 0.00 & 4.00 & 7.00 & 9.00 & 10.00 & 10.00 & 10.00\\
Blue Shift & Treatment & Pre Confidence County Ballots & 456 & 7.75 & 7.08 & 0.00 & 4.00 & 7.00 & 9.00 & 10.00 & 10.00 & 10.00\\
Blue Shift & Placebo & Pre Confidence Country Ballots & 431 & 6.64 & 9.09 & 0.00 & 2.00 & 5.00 & 7.00 & 9.00 & 10.00 & 10.00\\

Blue Shift & Treatment & Pre Confidence Country Ballots & 456 & 6.70 & 9.57 & 0.00 & 2.00 & 5.00 & 8.00 & 9.00 & 10.00 & 10.00\\
Blue Shift & Placebo & Post Confidence Own Ballot & 431 & 7.42 & 7.87 & 0.00 & 3.00 & 6.00 & 8.00 & 10.00 & 10.00 & 10.00\\
Blue Shift & Treatment & Post Confidence Own Ballot & 456 & 7.43 & 7.46 & 0.00 & 3.50 & 6.00 & 8.00 & 10.00 & 10.00 & 10.00\\
Blue Shift & Placebo & Post Confidence County Ballots & 431 & 7.36 & 7.82 & 0.00 & 3.00 & 6.00 & 8.00 & 10.00 & 10.00 & 10.00\\
Blue Shift & Treatment & Post Confidence County Ballots & 456 & 7.31 & 7.48 & 0.00 & 3.00 & 6.00 & 8.00 & 10.00 & 10.00 & 10.00\\

Blue Shift & Placebo & Post Confidence Country Ballots & 431 & 6.43 & 9.62 & 0.00 & 2.00 & 4.00 & 7.00 & 9.00 & 10.00 & 10.00\\
Blue Shift & Treatment & Post Confidence Country Ballots & 455 & 6.45 & 9.73 & 0.00 & 1.00 & 4.50 & 7.00 & 9.00 & 10.00 & 10.00\\
Blue Shift & Placebo & Confidence Own Ballot Diff & 431 & -0.40 & 3.40 & -9.00 & -2.00 & -1.00 & 0.00 & 0.00 & 1.00 & 10.00\\
Blue Shift & Treatment & Confidence Own Ballot Diff & 456 & -0.50 & 2.51 & -10.00 & -2.00 & -1.00 & 0.00 & 0.00 & 1.00 & 6.00\\
Blue Shift & Placebo & Confidence County Ballots Diff & 431 & -0.42 & 2.75 & -10.00 & -2.00 & -1.00 & 0.00 & 0.00 & 1.00 & 5.00\\

Blue Shift & Treatment & Confidence County Ballots Diff & 456 & -0.44 & 2.46 & -10.00 & -2.00 & -1.00 & 0.00 & 0.00 & 1.00 & 6.00\\
Blue Shift & Placebo & Confidence Country Ballots Diff & 431 & -0.20 & 2.27 & -9.00 & -2.00 & -1.00 & 0.00 & 0.00 & 1.00 & 6.00\\
Blue Shift & Treatment & Confidence Country Ballots Diff & 455 & -0.25 & 2.59 & -9.00 & -2.00 & -1.00 & 0.00 & 0.00 & 1.00 & 6.00\\
Blue Shift & Placebo & Populism Score & 431 & 12.70 & 19.47 & 6.00 & 7.00 & 9.00 & 12.00 & 16.00 & 18.00 & 30.00\\
Blue Shift & Treatment & Populism Score & 456 & 12.70 & 17.28 & 6.00 & 7.50 & 10.00 & 12.00 & 16.00 & 18.00 & 30.00\\

Blue Shift & Placebo & Conspiracy Score & 431 & 19.99 & 62.41 & 6.00 & 10.00 & 13.00 & 19.00 & 28.50 & 30.00 & 30.00\\
Blue Shift & Treatment & Conspiracy Score & 456 & 20.34 & 60.72 & 6.00 & 10.00 & 13.00 & 21.00 & 28.00 & 30.00 & 30.00\\
Blue Shift & Placebo & MIST Correct & 431 & 5.32 & 3.98 & 0.00 & 3.00 & 4.00 & 6.00 & 7.00 & 8.00 & 8.00\\
Blue Shift & Treatment & MIST Correct & 456 & 5.45 & 4.15 & 0.00 & 3.00 & 4.00 & 6.00 & 7.00 & 8.00 & 8.00\\

\multicolumn{13}{l}{\rule{0pt}{1em}\textit{Note:} }\\
\end{tabular}
\end{table}

}

\clearpage
\newpage

 {
\let\oldcentering\centering \renewcommand\centering{\tiny\oldcentering} 
 \setlength{\tabcolsep}{2pt}
  
  \hspace*{-2cm}

\begin{table}[!h]
\centering
\label{tab:tab:summary_statistics_voting_machines}
\caption{Summary Statistics - Voting Machines}
\centering
\begin{tabular}[t]{lllllllllllll}

\textbf{Group} & \textbf{Placebo} & \textbf{Variable} & \textbf{N} & \textbf{Mean} & \textbf{Var} & \textbf{Min} & \textbf{Q10} & \textbf{Q25} & \textbf{Median} & \textbf{Q75} & \textbf{Q90} & \textbf{Max}\\

Voting Machines & Placebo & Pre Confidence Own Ballot & 412 & 7.93 & 6.11 & 0.00 & 5.00 & 7.00 & 9.00 & 10.00 & 10.00 & 10.00\\
Voting Machines & Treatment & Pre Confidence Own Ballot & 436 & 7.94 & 5.61 & 0.00 & 5.00 & 7.00 & 9.00 & 10.00 & 10.00 & 10.00\\
Voting Machines & Placebo & Pre Confidence County Ballots & 412 & 7.85 & 6.59 & 0.00 & 5.00 & 7.00 & 9.00 & 10.00 & 10.00 & 10.00\\
Voting Machines & Treatment & Pre Confidence County Ballots & 436 & 7.89 & 5.74 & 0.00 & 5.00 & 7.00 & 9.00 & 10.00 & 10.00 & 10.00\\
Voting Machines & Placebo & Pre Confidence Country Ballots & 412 & 6.81 & 9.00 & 0.00 & 2.00 & 5.00 & 7.00 & 9.00 & 10.00 & 10.00\\

Voting Machines & Treatment & Pre Confidence Country Ballots & 436 & 6.83 & 8.96 & 0.00 & 2.00 & 5.00 & 8.00 & 9.00 & 10.00 & 10.00\\
Voting Machines & Placebo & Post Confidence Own Ballot & 412 & 7.36 & 7.78 & 0.00 & 3.00 & 5.00 & 8.00 & 10.00 & 10.00 & 10.00\\
Voting Machines & Treatment & Post Confidence Own Ballot & 436 & 7.53 & 6.91 & 0.00 & 4.00 & 6.00 & 8.00 & 10.00 & 10.00 & 10.00\\
Voting Machines & Placebo & Post Confidence County Ballots & 412 & 7.44 & 7.60 & 0.00 & 3.00 & 5.75 & 8.00 & 10.00 & 10.00 & 10.00\\
Voting Machines & Treatment & Post Confidence County Ballots & 436 & 7.55 & 6.73 & 0.00 & 4.00 & 6.00 & 8.00 & 10.00 & 10.00 & 10.00\\

Voting Machines & Placebo & Post Confidence Country Ballots & 412 & 6.52 & 9.57 & 0.00 & 2.00 & 5.00 & 7.00 & 9.00 & 10.00 & 10.00\\
Voting Machines & Treatment & Post Confidence Country Ballots & 436 & 6.74 & 8.82 & 0.00 & 2.00 & 5.00 & 8.00 & 9.00 & 10.00 & 10.00\\
Voting Machines & Placebo & Confidence Own Ballot Diff & 412 & -0.58 & 2.59 & -10.00 & -3.00 & -1.00 & 0.00 & 0.00 & 1.00 & 5.00\\
Voting Machines & Treatment & Confidence Own Ballot Diff & 436 & -0.41 & 3.22 & -10.00 & -2.00 & -1.00 & 0.00 & 0.00 & 1.00 & 10.00\\
Voting Machines & Placebo & Confidence County Ballots Diff & 412 & -0.41 & 2.26 & -9.00 & -2.00 & -1.00 & 0.00 & 0.00 & 1.00 & 5.00\\

Voting Machines & Treatment & Confidence County Ballots Diff & 436 & -0.34 & 2.90 & -8.00 & -2.00 & -1.00 & 0.00 & 0.00 & 1.00 & 10.00\\
Voting Machines & Placebo & Confidence Country Ballots Diff & 412 & -0.28 & 2.63 & -5.00 & -2.00 & -1.00 & 0.00 & 0.00 & 1.00 & 6.00\\
Voting Machines & Treatment & Confidence Country Ballots Diff & 436 & -0.09 & 2.65 & -5.00 & -2.00 & -1.00 & 0.00 & 0.00 & 1.50 & 8.00\\
Voting Machines & Placebo & Populism Score & 412 & 13.06 & 15.98 & 6.00 & 7.10 & 10.00 & 13.00 & 16.00 & 18.00 & 26.00\\
Voting Machines & Treatment & Populism Score & 436 & 12.44 & 14.74 & 6.00 & 7.00 & 9.00 & 12.00 & 15.00 & 18.00 & 23.00\\

Voting Machines & Placebo & Conspiracy Score & 412 & 20.69 & 56.77 & 6.00 & 11.00 & 14.75 & 20.00 & 29.00 & 30.00 & 30.00\\
Voting Machines & Treatment & Conspiracy Score & 436 & 20.34 & 59.09 & 6.00 & 10.00 & 14.00 & 20.00 & 28.00 & 30.00 & 30.00\\
Voting Machines & Placebo & MIST Correct & 412 & 5.69 & 3.86 & 0.00 & 3.00 & 4.00 & 6.00 & 7.00 & 8.00 & 8.00\\
Voting Machines & Treatment & MIST Correct & 436 & 5.67 & 3.75 & 0.00 & 3.00 & 4.00 & 6.00 & 7.00 & 8.00 & 8.00\\

\multicolumn{13}{l}{\rule{0pt}{1em}\textit{Note:} }\\
\end{tabular}
\end{table}

}

\clearpage
\newpage

\subsubsection*{Cross-Wave, Weighted and Unweighted Factor Variable Summary Statistics and Balance Tables}

 {
\let\oldcentering\centering \renewcommand\centering{\tiny\oldcentering} 
 \setlength{\tabcolsep}{2pt}
  
  \hspace*{-2cm}
\begin{table}[!h]
\centering
\label{tab:tab:cross_wave_factor_summary_statistics_age_group}
\caption{Cross-Wave Summary Statistics - Age Group}
\centering
\begin{tabular}[t]{lllllll}

\textbf{Treatment Status} & \textbf{Variable} & \textbf{Under 30} & \textbf{30-44} & \textbf{45-64} & \textbf{65+} & \textbf{Scenario}\\

Placebo & Age Group & 0.14 & 0.22 & 0.36 & 0.28 & Pre-treatment Unweighted\\
Treatment & Age Group & 0.13 & 0.23 & 0.37 & 0.27 & Pre-treatment Unweighted\\
Placebo & Age Group & 0.17 & 0.23 & 0.34 & 0.27 & Pre-treatment Weighted\\
Treatment & Age Group & 0.15 & 0.23 & 0.36 & 0.27 & Pre-treatment Weighted\\
Placebo & Age Group & 0.12 & 0.21 & 0.36 & 0.30 & Follow-up Unweighted\\

Treatment & Age Group & 0.12 & 0.21 & 0.38 & 0.29 & Follow-up Unweighted\\
Placebo & Age Group & 0.17 & 0.23 & 0.34 & 0.27 & Follow-up Weighted\\
Treatment & Age Group & 0.15 & 0.23 & 0.36 & 0.27 & Follow-up Weighted\\

\multicolumn{7}{l}{\rule{0pt}{1em}\textit{Note:} }\\
\end{tabular}
\end{table}

}

 {
\let\oldcentering\centering \renewcommand\centering{\tiny\oldcentering} 
 \setlength{\tabcolsep}{2pt}
  
  \hspace*{-2cm}
\begin{table}[!h]
\centering
\label{tab:tab:cross_wave_factor_summary_statistics_education_level}
\caption{Cross-Wave Summary Statistics - Education Level}
\centering
\begin{tabular}[t]{lllllll}

\textbf{Treatment Status} & \textbf{Variable} & \textbf{HS or less} & \textbf{Some college} & \textbf{College grad} & \textbf{Postgrad} & \textbf{Scenario}\\

Placebo & Education Level & 0.29 & 0.31 & 0.26 & 0.15 & Pre-treatment Unweighted\\
Treatment & Education Level & 0.29 & 0.30 & 0.25 & 0.15 & Pre-treatment Unweighted\\
Placebo & Education Level & 0.29 & 0.30 & 0.26 & 0.15 & Pre-treatment Weighted\\
Treatment & Education Level & 0.31 & 0.30 & 0.24 & 0.15 & Pre-treatment Weighted\\
Placebo & Education Level & 0.28 & 0.30 & 0.26 & 0.16 & Follow-up Unweighted\\

Treatment & Education Level & 0.29 & 0.30 & 0.25 & 0.16 & Follow-up Unweighted\\
Placebo & Education Level & 0.29 & 0.30 & 0.26 & 0.15 & Follow-up Weighted\\
Treatment & Education Level & 0.31 & 0.30 & 0.24 & 0.15 & Follow-up Weighted\\

\multicolumn{7}{l}{\rule{0pt}{1em}\textit{Note:} }\\
\end{tabular}
\end{table}

}

 {
\let\oldcentering\centering \renewcommand\centering{\tiny\oldcentering} 
 \setlength{\tabcolsep}{2pt}
  
  \hspace*{-2cm}
\begin{table}[!h]
\centering
\label{tab:tab:cross_wave_factor_summary_statistics_gender}
\caption{Cross-Wave Summary Statistics - Gender}
\centering
\begin{tabular}[t]{lllll}

\textbf{Treatment Status} & \textbf{Variable} & \textbf{Male} & \textbf{Female} & \textbf{Scenario}\\

Placebo & Gender & 0.48 & 0.52 & Pre-treatment Unweighted\\
Treatment & Gender & 0.46 & 0.54 & Pre-treatment Unweighted\\
Placebo & Gender & 0.47 & 0.53 & Pre-treatment Weighted\\
Treatment & Gender & 0.46 & 0.54 & Pre-treatment Weighted\\
Placebo & Gender & 0.48 & 0.52 & Follow-up Unweighted\\

Treatment & Gender & 0.46 & 0.54 & Follow-up Unweighted\\
Placebo & Gender & 0.48 & 0.52 & Follow-up Weighted\\
Treatment & Gender & 0.46 & 0.54 & Follow-up Weighted\\

\multicolumn{5}{l}{\rule{0pt}{1em}\textit{Note:} }\\
\end{tabular}
\end{table}

}

 {
\let\oldcentering\centering \renewcommand\centering{\tiny\oldcentering} 
 \setlength{\tabcolsep}{2pt}
  
  \hspace*{-2cm}
\begin{table}[!h]
\centering
\label{tab:tab:cross_wave_factor_summary_statistics_ideology}
\caption{Cross-Wave Summary Statistics - Ideology}
\centering
\begin{tabular}[t]{lllllll}

\textbf{Treatment Status} & \textbf{Variable} & \textbf{NA} & \textbf{Liberal} & \textbf{Moderate} & \textbf{Conservative} & \textbf{Scenario}\\

Placebo & Ideology & 0.04 & 0.30 & 0.30 & 0.36 & Pre-treatment Unweighted\\
Treatment & Ideology & 0.04 & 0.31 & 0.32 & 0.34 & Pre-treatment Unweighted\\
Placebo & Ideology & 0.04 & 0.29 & 0.27 & 0.40 & Pre-treatment Weighted\\
Treatment & Ideology & 0.04 & 0.29 & 0.31 & 0.36 & Pre-treatment Weighted\\
Placebo & Ideology & 0.03 & 0.31 & 0.29 & 0.37 & Follow-up Unweighted\\

Treatment & Ideology & 0.03 & 0.31 & 0.31 & 0.35 & Follow-up Unweighted\\
Placebo & Ideology & 0.03 & 0.30 & 0.27 & 0.41 & Follow-up Weighted\\
Treatment & Ideology & 0.04 & 0.30 & 0.30 & 0.37 & Follow-up Weighted\\

\multicolumn{7}{l}{\rule{0pt}{1em}\textit{Note:} }\\
\end{tabular}
\end{table}

}

 {
\let\oldcentering\centering \renewcommand\centering{\tiny\oldcentering} 
 \setlength{\tabcolsep}{2pt}
  
  \hspace*{-2cm}
\begin{table}[!h]
\centering
\label{tab:tab:cross_wave_factor_summary_statistics_party_identification}
\caption{Cross-Wave Summary Statistics - Party Identification}
\centering
\begin{tabular}[t]{llllll}

\textbf{Treatment Status} & \textbf{Variable} & \textbf{Democrat} & \textbf{Independent} & \textbf{Republican} & \textbf{Scenario}\\

Placebo & Party Identification & 0.37 & 0.31 & 0.32 & Pre-treatment Unweighted\\
Treatment & Party Identification & 0.38 & 0.30 & 0.31 & Pre-treatment Unweighted\\
Placebo & Party Identification & 0.36 & 0.28 & 0.36 & Pre-treatment Weighted\\
Treatment & Party Identification & 0.36 & 0.29 & 0.36 & Pre-treatment Weighted\\
Placebo & Party Identification & 0.37 & 0.31 & 0.32 & Follow-up Unweighted\\

Treatment & Party Identification & 0.38 & 0.30 & 0.31 & Follow-up Unweighted\\
Placebo & Party Identification & 0.35 & 0.29 & 0.36 & Follow-up Weighted\\
Treatment & Party Identification & 0.36 & 0.29 & 0.36 & Follow-up Weighted\\

\multicolumn{6}{l}{\rule{0pt}{1em}\textit{Note:} }\\
\end{tabular}
\end{table}

}

 {
\let\oldcentering\centering \renewcommand\centering{\tiny\oldcentering} 
 \setlength{\tabcolsep}{2pt}
  
  \hspace*{-2cm}
\begin{table}[!h]
\centering
\label{tab:tab:cross_wave_factor_summary_statistics_political_interest}
\caption{Cross-Wave Summary Statistics - Political Interest}
\centering
\begin{tabular}[t]{llllllll}

\textbf{Treatment Status} & \textbf{Variable} & \textbf{NA} & \textbf{Pol Interest: Most of the time} & \textbf{Pol Interest: Some of the time} & \textbf{Pol Interest: Only now and then} & \textbf{Pol Interest: Hardly at all} & \textbf{Scenario}\\

Placebo & Political Interest & 0.01 & 0.55 & 0.28 & 0.10 & 0.05 & Pre-treatment Unweighted\\
Treatment & Political Interest & 0.01 & 0.55 & 0.30 & 0.09 & 0.05 & Pre-treatment Unweighted\\
Placebo & Political Interest & 0.01 & 0.53 & 0.29 & 0.11 & 0.06 & Pre-treatment Weighted\\
Treatment & Political Interest & 0.01 & 0.52 & 0.32 & 0.09 & 0.06 & Pre-treatment Weighted\\
Placebo & Political Interest & 0.01 & 0.57 & 0.27 & 0.10 & 0.05 & Follow-up Unweighted\\

Treatment & Political Interest & 0.01 & 0.57 & 0.29 & 0.09 & 0.05 & Follow-up Unweighted\\
Placebo & Political Interest & 0.01 & 0.55 & 0.28 & 0.10 & 0.06 & Follow-up Weighted\\
Treatment & Political Interest & 0.01 & 0.53 & 0.32 & 0.09 & 0.05 & Follow-up Weighted\\

\multicolumn{8}{l}{\rule{0pt}{1em}\textit{Note:} }\\
\end{tabular}
\end{table}

}

 {
\let\oldcentering\centering \renewcommand\centering{\tiny\oldcentering} 
 \setlength{\tabcolsep}{2pt}
  
  \hspace*{-2cm}
\begin{table}[!h]
\centering
\label{tab:tab:cross_wave_factor_summary_statistics_race_ethnicity}
\caption{Cross-Wave Summary Statistics - Race Ethnicity}
\centering
\begin{tabular}[t]{lllllll}

\textbf{Treatment Status} & \textbf{Variable} & \textbf{White} & \textbf{Black} & \textbf{Hispanic} & \textbf{Other} & \textbf{Scenario}\\

Placebo & Race Ethnicity & 0.71 & 0.11 & 0.13 & 0.06 & Pre-treatment Unweighted\\
Treatment & Race Ethnicity & 0.70 & 0.12 & 0.12 & 0.06 & Pre-treatment Unweighted\\
Placebo & Race Ethnicity & 0.71 & 0.11 & 0.10 & 0.07 & Pre-treatment Weighted\\
Treatment & Race Ethnicity & 0.70 & 0.13 & 0.11 & 0.06 & Pre-treatment Weighted\\
Placebo & Race Ethnicity & 0.71 & 0.10 & 0.13 & 0.06 & Follow-up Unweighted\\

Treatment & Race Ethnicity & 0.71 & 0.12 & 0.11 & 0.06 & Follow-up Unweighted\\
Placebo & Race Ethnicity & 0.71 & 0.11 & 0.11 & 0.07 & Follow-up Weighted\\
Treatment & Race Ethnicity & 0.71 & 0.13 & 0.10 & 0.06 & Follow-up Weighted\\

\multicolumn{7}{l}{\rule{0pt}{1em}\textit{Note:} }\\
\end{tabular}
\end{table}

}

  






\clearpage
\newpage

\subsubsection*{Factor Variable Summary Statistics and Balance Tables}

 {
\let\oldcentering\centering \renewcommand\centering{\tiny\oldcentering} 
 \setlength{\tabcolsep}{2pt}
  
  \hspace*{-2cm}

\begin{table}[!h]
\centering
\label{tab:tab:cross_wave_factor_summary_statistics_urban_rural}
\caption{Cross-Wave Summary Statistics - Urban Rural}
\centering
\begin{tabular}[t]{llllllll}

\textbf{Treatment Status} & \textbf{Variable} & \textbf{NA} & \textbf{City} & \textbf{Suburb} & \textbf{Town} & \textbf{Rural area} & \textbf{Scenario}\\

Placebo & Urban Rural & 0.00 & 0.29 & 0.40 & 0.12 & 0.19 & Pre-treatment Unweighted\\
Treatment & Urban Rural & 0.00 & 0.30 & 0.39 & 0.13 & 0.18 & Pre-treatment Unweighted\\
Placebo & Urban Rural & 0.00 & 0.30 & 0.39 & 0.12 & 0.18 & Pre-treatment Weighted\\
Treatment & Urban Rural & 0.01 & 0.29 & 0.39 & 0.12 & 0.19 & Pre-treatment Weighted\\
Placebo & Urban Rural & 0.01 & 0.28 & 0.41 & 0.12 & 0.18 & Follow-up Unweighted\\

Treatment & Urban Rural & 0.00 & 0.29 & 0.40 & 0.13 & 0.18 & Follow-up Unweighted\\
Placebo & Urban Rural & 0.00 & 0.30 & 0.40 & 0.12 & 0.18 & Follow-up Weighted\\
Treatment & Urban Rural & 0.00 & 0.29 & 0.40 & 0.12 & 0.19 & Follow-up Weighted\\

\multicolumn{8}{l}{\rule{0pt}{1em}\textit{Note:} }\\
\end{tabular}
\end{table}

}

\clearpage
\newpage

 {
\let\oldcentering\centering \renewcommand\centering{\tiny\oldcentering} 
 \setlength{\tabcolsep}{2pt}
  
  \hspace*{-2cm}
\begin{table}[!h]
\centering
\label{tab:tab:cross_wave_factor_summary_statistics_region}
\caption{Cross-Wave Summary Statistics - Region}
\centering
\begin{tabular}[t]{lllllll}

\textbf{Treatment Status} & \textbf{Variable} & \textbf{Northeast} & \textbf{Midwest} & \textbf{South} & \textbf{West} & \textbf{Scenario}\\

Placebo & Region & 0.16 & 0.23 & 0.37 & 0.23 & Pre-treatment Unweighted\\
Treatment & Region & 0.19 & 0.22 & 0.37 & 0.22 & Pre-treatment Unweighted\\
Placebo & Region & 0.17 & 0.22 & 0.38 & 0.23 & Pre-treatment Weighted\\
Treatment & Region & 0.19 & 0.22 & 0.36 & 0.23 & Pre-treatment Weighted\\
Placebo & Region & 0.16 & 0.23 & 0.37 & 0.24 & Follow-up Unweighted\\

Treatment & Region & 0.19 & 0.22 & 0.37 & 0.22 & Follow-up Unweighted\\
Placebo & Region & 0.16 & 0.22 & 0.38 & 0.24 & Follow-up Weighted\\
Treatment & Region & 0.19 & 0.22 & 0.37 & 0.22 & Follow-up Weighted\\

\multicolumn{7}{l}{\rule{0pt}{1em}\textit{Note:} }\\
\end{tabular}
\end{table}

}

\clearpage
\newpage

 {
\let\oldcentering\centering \renewcommand\centering{\tiny\oldcentering} 
 \setlength{\tabcolsep}{2pt}
  
  \hspace*{-2cm}
\begin{table}[!h]
\centering
\label{tab:tab:summary_statistics_education_level}
\caption{Summary Statistics - Education Level}
\centering
\begin{tabular}[t]{lllllll}

\textbf{Treatment Status} & \textbf{Assigned Rumor} & \textbf{Variable} & \textbf{HS or less} & \textbf{Some college} & \textbf{College grad} & \textbf{Postgrad}\\

Placebo & Voter Fraud & Education Level & 0.31 & 0.30 & 0.23 & 0.15\\
Treatment & Voter Fraud & Education Level & 0.31 & 0.30 & 0.22 & 0.17\\
Placebo & Voter Rolls & Education Level & 0.28 & 0.31 & 0.25 & 0.17\\
Treatment & Voter Rolls & Education Level & 0.29 & 0.29 & 0.24 & 0.17\\
Placebo & Hacking & Education Level & 0.31 & 0.27 & 0.29 & 0.13\\

Treatment & Hacking & Education Level & 0.31 & 0.29 & 0.25 & 0.15\\
Placebo & Blue Shift & Education Level & 0.28 & 0.34 & 0.26 & 0.13\\
Treatment & Blue Shift & Education Level & 0.28 & 0.30 & 0.27 & 0.14\\
Placebo & Voting Machines & Education Level & 0.26 & 0.31 & 0.25 & 0.17\\
Treatment & Voting Machines & Education Level & 0.27 & 0.33 & 0.28 & 0.11\\

\multicolumn{7}{l}{\rule{0pt}{1em}\textit{Note:} }\\
\end{tabular}
\end{table}

}

 {
\let\oldcentering\centering \renewcommand\centering{\tiny\oldcentering} 
 \setlength{\tabcolsep}{2pt}
  
  \hspace*{-2cm}
\begin{table}[!h]
\centering
\label{tab:tab:summary_statistics_race_ethnicity}
\caption{Summary Statistics - Race Ethnicity}
\centering
\begin{tabular}[t]{lllllll}

\textbf{Treatment Status} & \textbf{Assigned Rumor} & \textbf{Variable} & \textbf{White} & \textbf{Black} & \textbf{Hispanic} & \textbf{Other}\\

Placebo & Voter Fraud & Race Ethnicity & 0.72 & 0.10 & 0.12 & 0.06\\
Treatment & Voter Fraud & Race Ethnicity & 0.71 & 0.13 & 0.09 & 0.07\\
Placebo & Voter Rolls & Race Ethnicity & 0.73 & 0.11 & 0.10 & 0.06\\
Treatment & Voter Rolls & Race Ethnicity & 0.70 & 0.12 & 0.12 & 0.05\\
Placebo & Hacking & Race Ethnicity & 0.71 & 0.11 & 0.13 & 0.05\\

Treatment & Hacking & Race Ethnicity & 0.69 & 0.13 & 0.12 & 0.06\\
Placebo & Blue Shift & Race Ethnicity & 0.65 & 0.12 & 0.15 & 0.08\\
Treatment & Blue Shift & Race Ethnicity & 0.68 & 0.13 & 0.14 & 0.05\\
Placebo & Voting Machines & Race Ethnicity & 0.71 & 0.11 & 0.12 & 0.06\\
Treatment & Voting Machines & Race Ethnicity & 0.72 & 0.10 & 0.12 & 0.06\\

\multicolumn{7}{l}{\rule{0pt}{1em}\textit{Note:} }\\
\end{tabular}
\end{table}

}

 {
\let\oldcentering\centering \renewcommand\centering{\tiny\oldcentering} 
 \setlength{\tabcolsep}{2pt}
  
  \hspace*{-2cm}
\begin{table}[!h]
\centering
\label{tab:tab:summary_statistics_gender}
\caption{Summary Statistics - Gender}
\centering
\begin{tabular}[t]{lllll}

\textbf{Treatment Status} & \textbf{Assigned Rumor} & \textbf{Variable} & \textbf{Male} & \textbf{Female}\\

Placebo & Voter Fraud & Gender & 0.47 & 0.53\\
Treatment & Voter Fraud & Gender & 0.43 & 0.57\\
Placebo & Voter Rolls & Gender & 0.44 & 0.56\\
Treatment & Voter Rolls & Gender & 0.46 & 0.54\\
Placebo & Hacking & Gender & 0.47 & 0.53\\

Treatment & Hacking & Gender & 0.47 & 0.53\\
Placebo & Blue Shift & Gender & 0.48 & 0.52\\
Treatment & Blue Shift & Gender & 0.49 & 0.51\\
Placebo & Voting Machines & Gender & 0.52 & 0.48\\
Treatment & Voting Machines & Gender & 0.45 & 0.55\\

\multicolumn{5}{l}{\rule{0pt}{1em}\textit{Note:} }\\
\end{tabular}
\end{table}

}

 {
\let\oldcentering\centering \renewcommand\centering{\tiny\oldcentering} 
 \setlength{\tabcolsep}{2pt}
  
  \hspace*{-2cm}
\begin{table}[!h]
\centering
\label{tab:tab:summary_statistics_age_group}
\caption{Summary Statistics - Age Group}
\centering
\begin{tabular}[t]{lllllll}

\textbf{Treatment Status} & \textbf{Assigned Rumor} & \textbf{Variable} & \textbf{Under 30} & \textbf{30-44} & \textbf{45-64} & \textbf{65+}\\

Placebo & Voter Fraud & Age Group & 0.15 & 0.18 & 0.35 & 0.32\\
Treatment & Voter Fraud & Age Group & 0.10 & 0.24 & 0.39 & 0.28\\
Placebo & Voter Rolls & Age Group & 0.10 & 0.26 & 0.36 & 0.28\\
Treatment & Voter Rolls & Age Group & 0.13 & 0.18 & 0.40 & 0.29\\
Placebo & Hacking & Age Group & 0.14 & 0.22 & 0.34 & 0.29\\

Treatment & Hacking & Age Group & 0.11 & 0.26 & 0.36 & 0.26\\
Placebo & Blue Shift & Age Group & 0.13 & 0.24 & 0.37 & 0.26\\
Treatment & Blue Shift & Age Group & 0.15 & 0.25 & 0.34 & 0.27\\
Placebo & Voting Machines & Age Group & 0.16 & 0.22 & 0.35 & 0.27\\
Treatment & Voting Machines & Age Group & 0.15 & 0.21 & 0.37 & 0.26\\

\multicolumn{7}{l}{\rule{0pt}{1em}\textit{Note:} }\\
\end{tabular}
\end{table}

}

\clearpage
\newpage

\subsection*{Regression Tables}

Here we report the results of our regressions. We also show the results graphically.

\begin{figure}[!hp]
    \centering
    \includegraphics[width=0.95\linewidth]{data/treatment_effects_plot_main.pdf}
    \caption{\textbf{Estimated treatment effects.} Results are pooled together (containing all participants, regardless of assigned rumor). All questions are measured on a 0-10 scale. Recontact measures are taken one week after treatment. Error bars represent 95\% confidence intervals.}
    \label{fig:main2}
\end{figure}

\begin{figure}[!hp]
    \centering
    \includegraphics[width=0.95\linewidth]{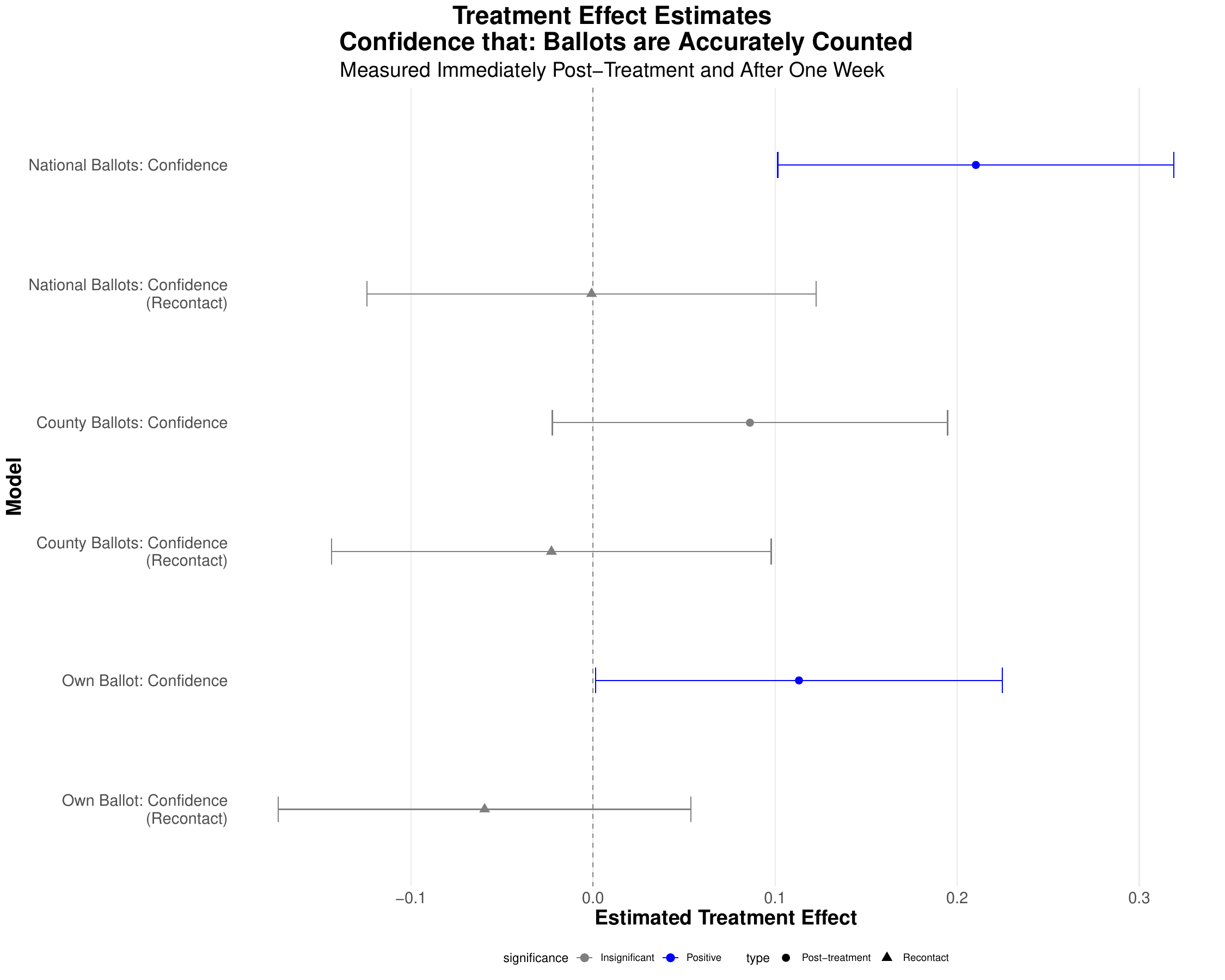}
    \caption{\textbf{Estimated treatment effects.} Results are pooled together (containing all participants, regardless of assigned rumor). All questions are measured on a 0-10 scale. Recontact measures are taken one week after treatment. Error bars represent 95\% confidence intervals.}
    \label{fig:treatment_effects_election}
\end{figure}

\begin{figure}[!hp]
    \centering
    \includegraphics[width=0.95\linewidth]{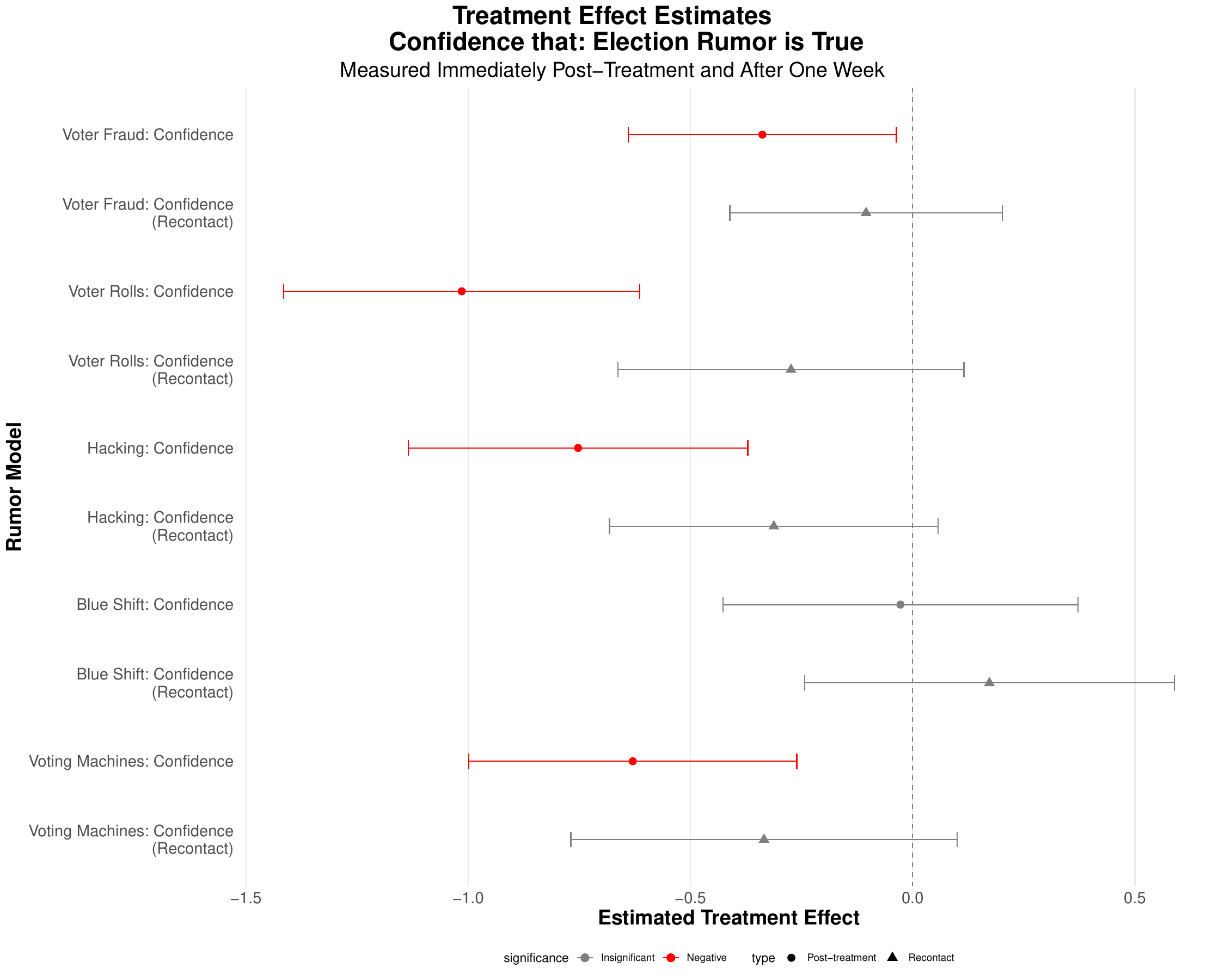}
    \caption{\textbf{Estimated treatment effects.} All questions are measured on a 0-10 scale. Recontact measures are taken one week after treatment. Error bars represent 95\% confidence intervals.}
    \label{fig:treatment_effects_rumor}
\end{figure}

\clearpage
\newpage
\hspace*{-6cm}
\input{data/Main_models}
\hspace*{-6cm}

\clearpage
\newpage

\clearpage
\newpage
\hspace*{-6cm}
\input{data/Human_in_the_Loop_models}
\hspace*{-6cm}

\clearpage
\newpage

\clearpage
\newpage
\hspace*{-6cm}
\input{data/Party_models}
\hspace*{-6cm}

\clearpage
\newpage

\clearpage
\newpage
\hspace*{-6cm}
\input{data/Rumor_models}
\hspace*{-6cm}

\clearpage
\newpage


\clearpage
\newpage

\hspace*{-6cm}
\input{data/Other_Election_Integrity_models}
\hspace*{-6cm}

\clearpage
\newpage


\subsection*{Figures}

\subsubsection*{Rumor Confidence}

\begin{figure}[ht]
    \centering
    \includegraphics[width=0.9\linewidth]{data/pre_post_cisa_confidence.pdf}
    \caption{\textbf{Pre-treatment vs. post-treatment confidence in assigned election rumor, all rumors pooled together.}}
    \label{fig:pre_post_cisa2}
\end{figure}

\newpage
\clearpage

\begin{figure}[ht]
    \centering
    \includegraphics[width=0.9\linewidth]{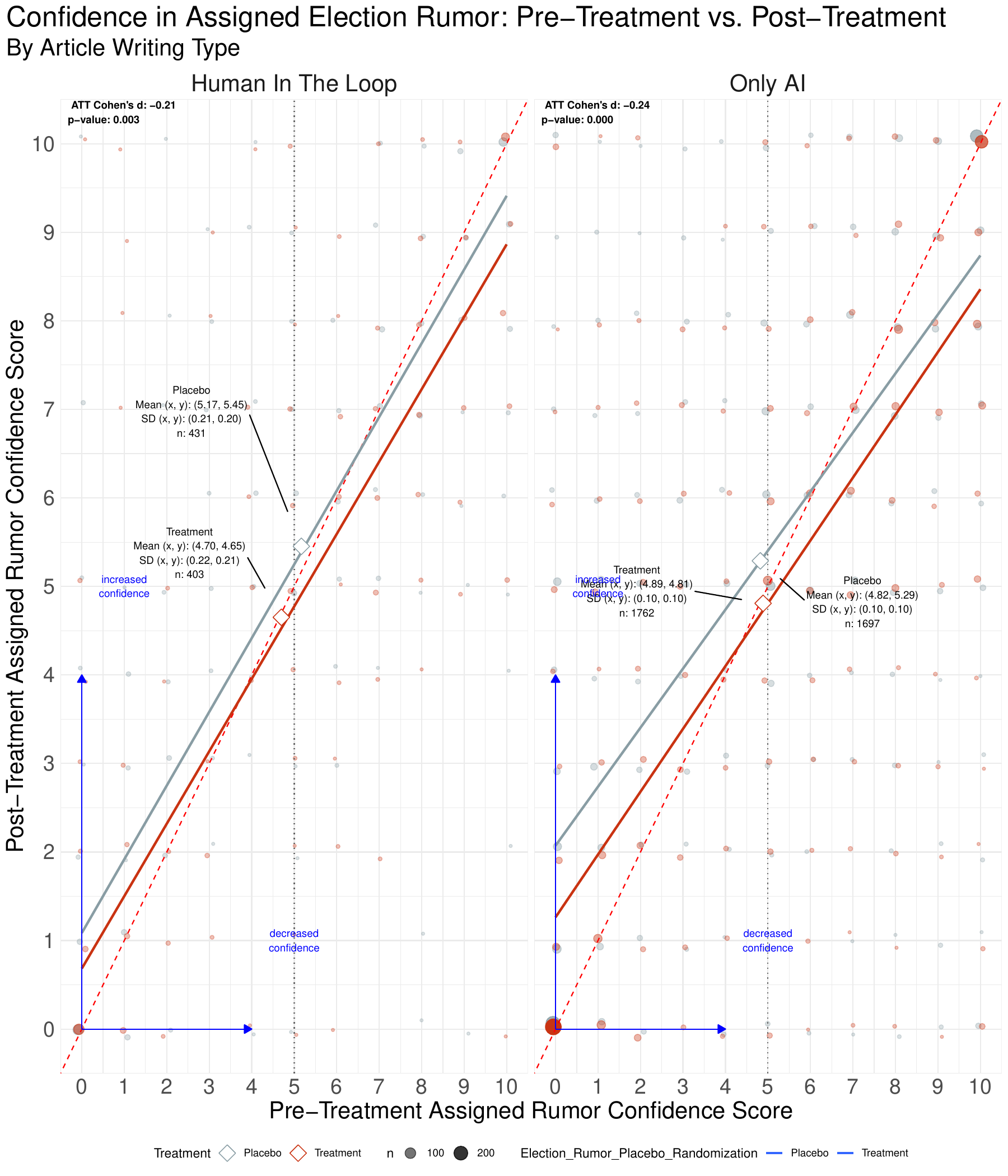}
    \caption{\textbf{Pre-treatment vs. post-treatment confidence in assigned election rumor, all rumors pooled together, by human assistance status.}}
    \label{fig:pre_post_cisa_hitl}
\end{figure}

\newpage
\clearpage

\begin{figure}[ht]
    \centering
    \includegraphics[width=0.9\linewidth]{data/pre_post_cisa_confidence_party.pdf}
    \caption{\textbf{Pre-treatment vs. post-treatment confidence in assigned election rumor, all rumors pooled together, by party.}}
    \label{fig:pre_post_cisa_party_app}
\end{figure}

\newpage
\clearpage

\begin{figure}[ht]
    \centering
    \includegraphics[width=0.9\linewidth]{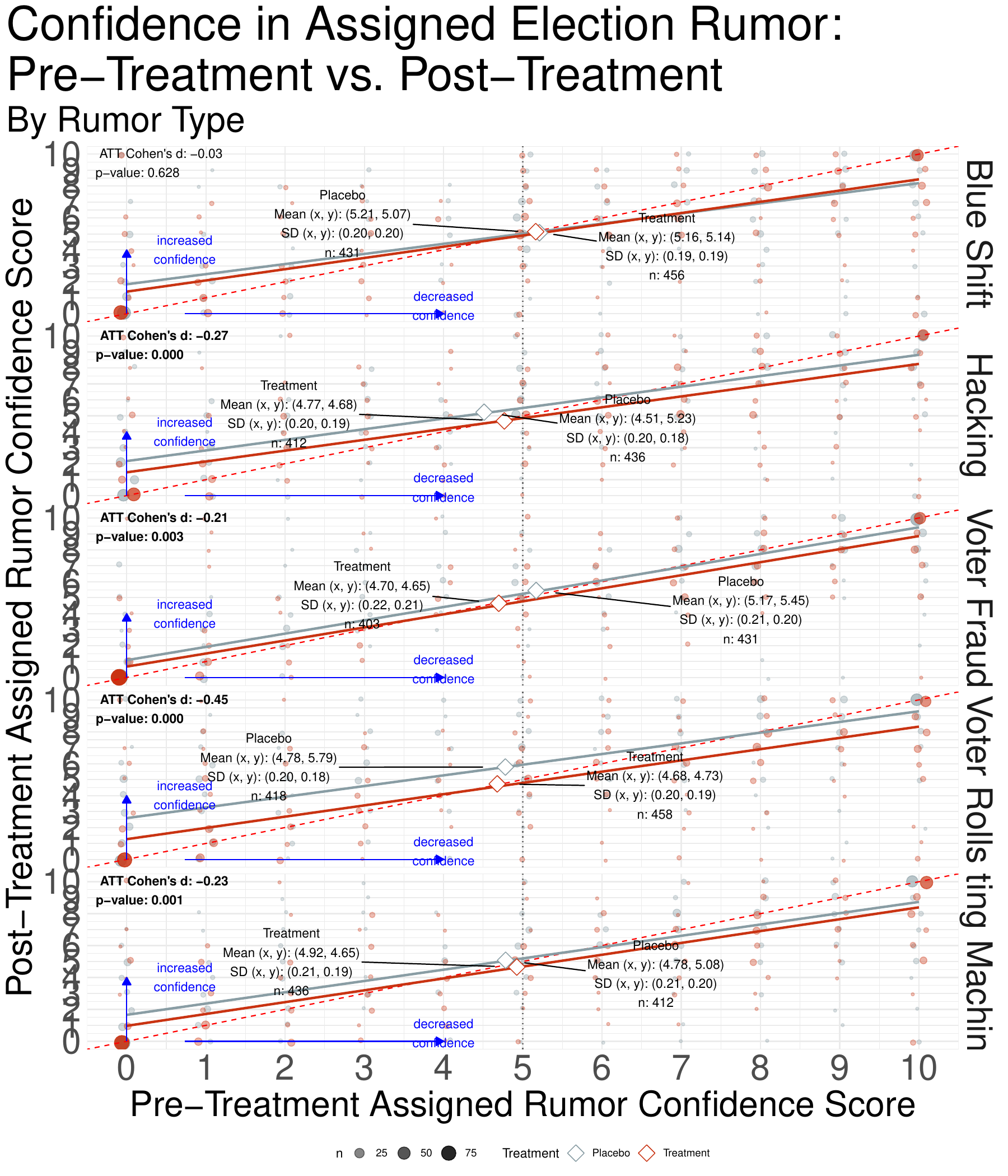}
    \caption{\textbf{Pre-treatment vs. post-treatment confidence in assigned election rumor, all rumors pooled together, by rumor.}}
    \label{fig:pre_post_cisa_rumor}
\end{figure}

\newpage
\clearpage

\begin{figure}[ht]
    \centering
    \includegraphics[width=0.9\linewidth]{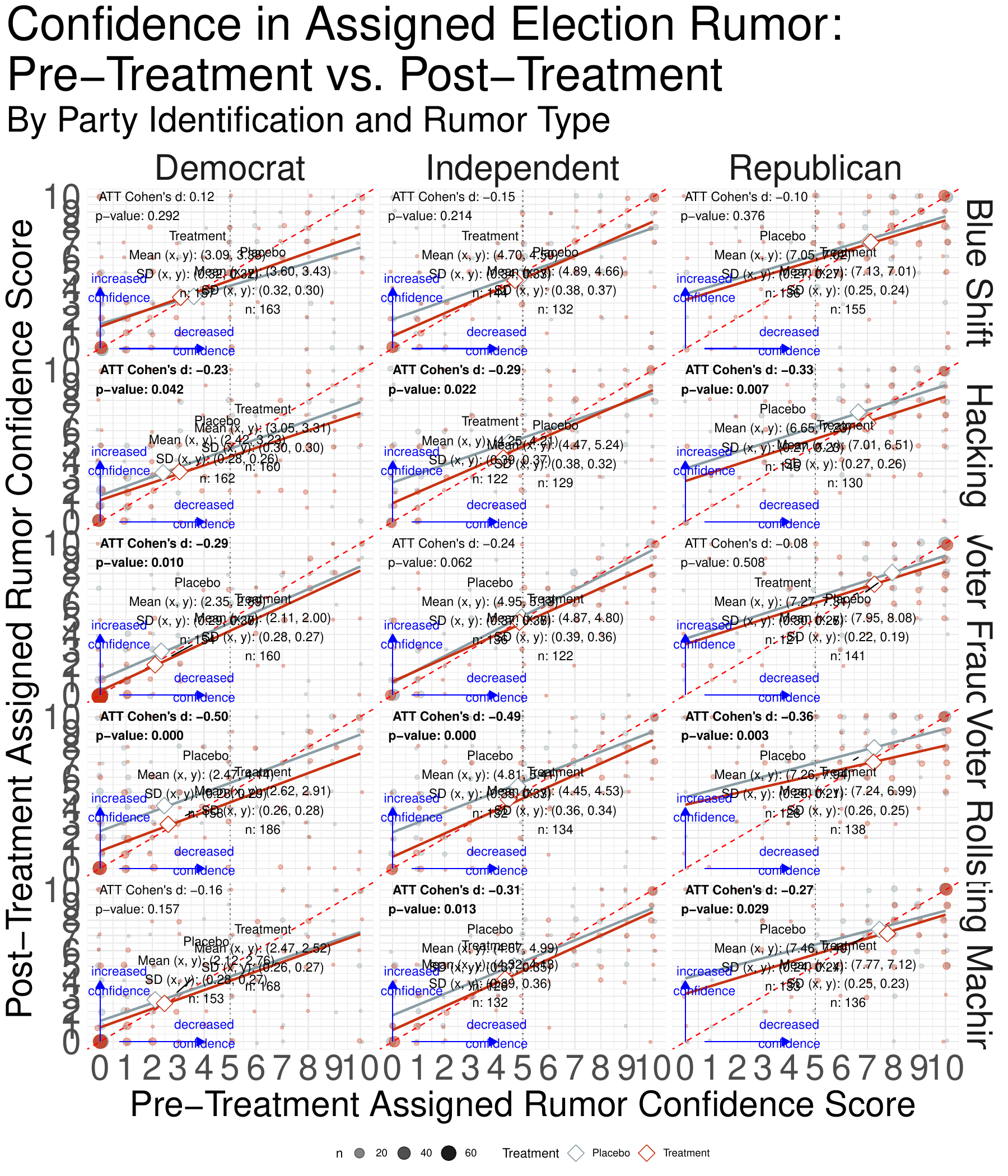}
    \caption{\textbf{Pre-treatment vs. post-treatment confidence in assigned election rumor, all rumors pooled together, by party and rumor.}}
    \label{fig:pre_post_cisa_party_rumor}
\end{figure}

\newpage
\clearpage

\subsubsection*{National Election Confidence (Initial Survey)} \label{sec:nat_conf}

\begin{figure}[ht]
    \centering
    \includegraphics[width=0.9\linewidth]{data/pre_post_confidence_country_ballots.pdf}
    \caption{\textbf{Pre-treatment vs. post-treatment confidence in national election administration.}}
    \label{fig:pre_post_country2}
\end{figure}

\newpage
\clearpage

\begin{figure}[ht]
    \centering
    \includegraphics[width=0.9\linewidth]{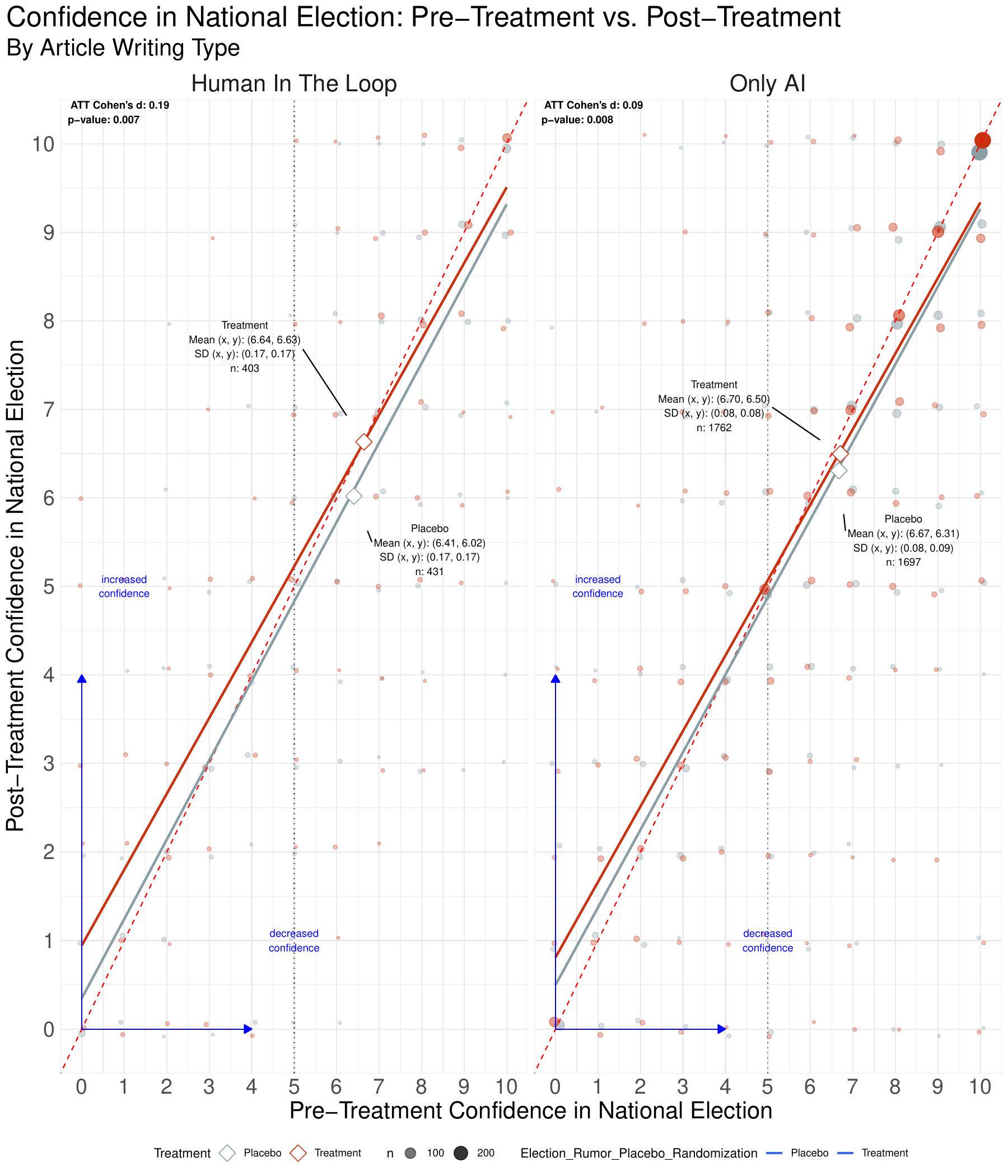}
    \caption{\textbf{Pre-treatment vs. post-treatment confidence in national election administration, by human assistance status.}}
    \label{fig:pre_post_country_hitl}
\end{figure}

\newpage
\clearpage

\begin{figure}[ht]
    \centering
    \includegraphics[width=0.9\linewidth]{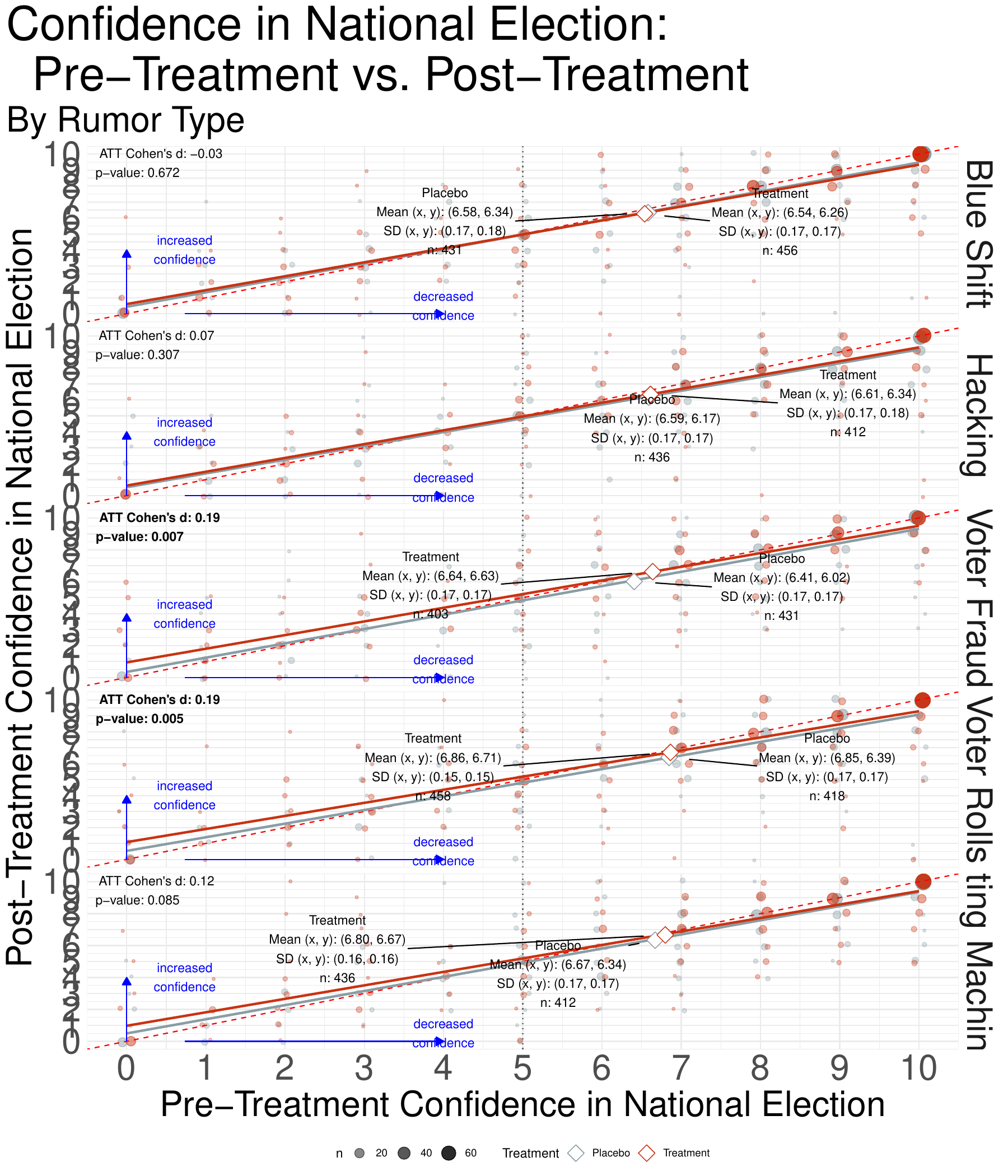}
    \caption{\textbf{Pre-treatment vs. post-treatment confidence in national election administration, by rumor.}}
    \label{fig:pre_post_country_rumor}
\end{figure}

\newpage
\clearpage

\begin{figure}[ht]
    \centering
    \includegraphics[width=0.9\linewidth]{data/pre_post_confidence_country_ballots_party.pdf}
    \caption{\textbf{Pre-treatment vs. post-treatment confidence in national election administration, by party.}}
    \label{fig:pre_post_country_party2}
\end{figure}

\newpage
\clearpage

\begin{figure}[ht]
    \centering
    \includegraphics[width=0.9\linewidth]{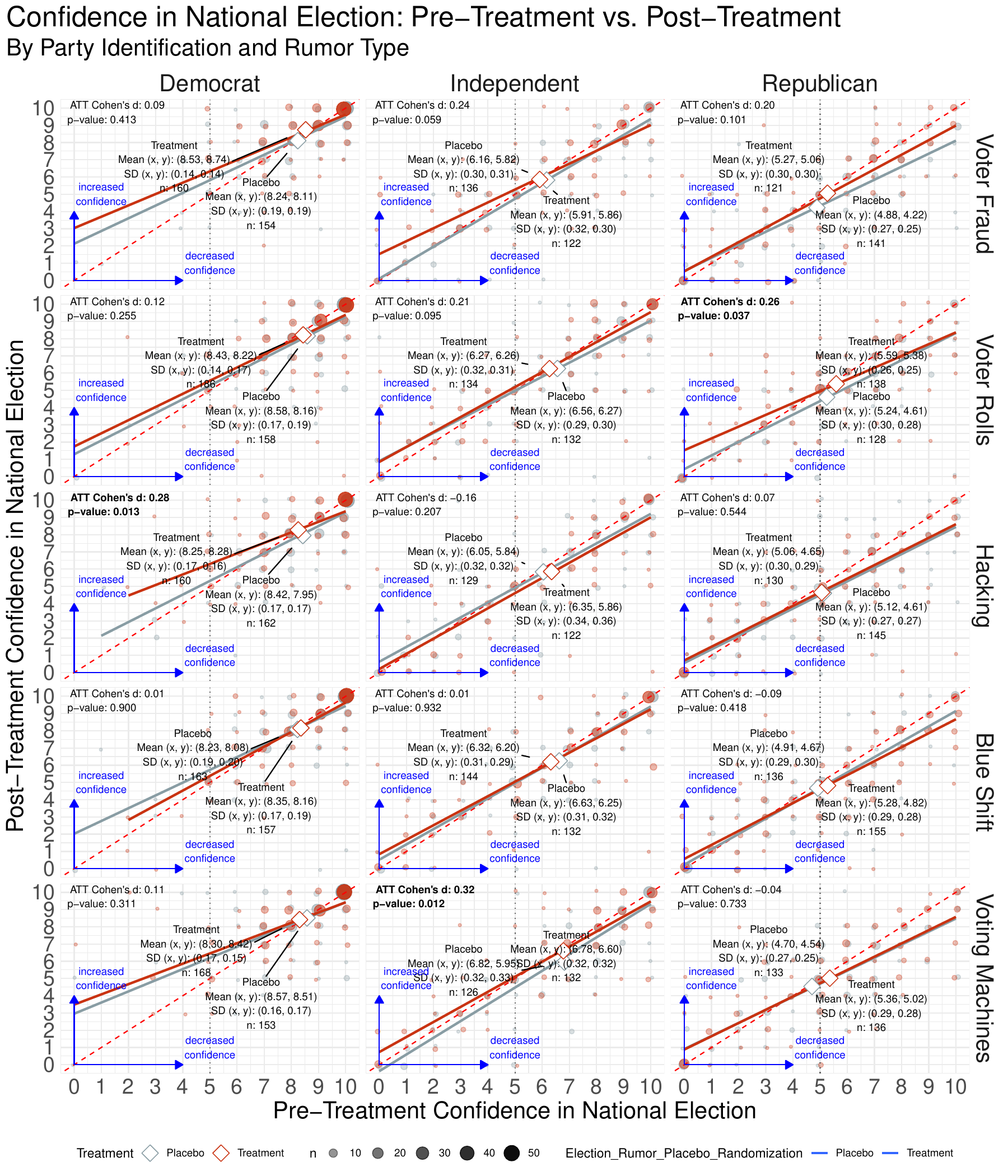}
    \caption{\textbf{Pre-treatment vs. post-treatment confidence in national election administration, by party and rumor.}}
    \label{fig:pre_post_country_party_rumor}
\end{figure}

\newpage
\clearpage

\subsubsection*{Rumor Confidence (Recontact Survey)}

\begin{figure}[ht]
    \centering
    \includegraphics[width=0.9\linewidth]{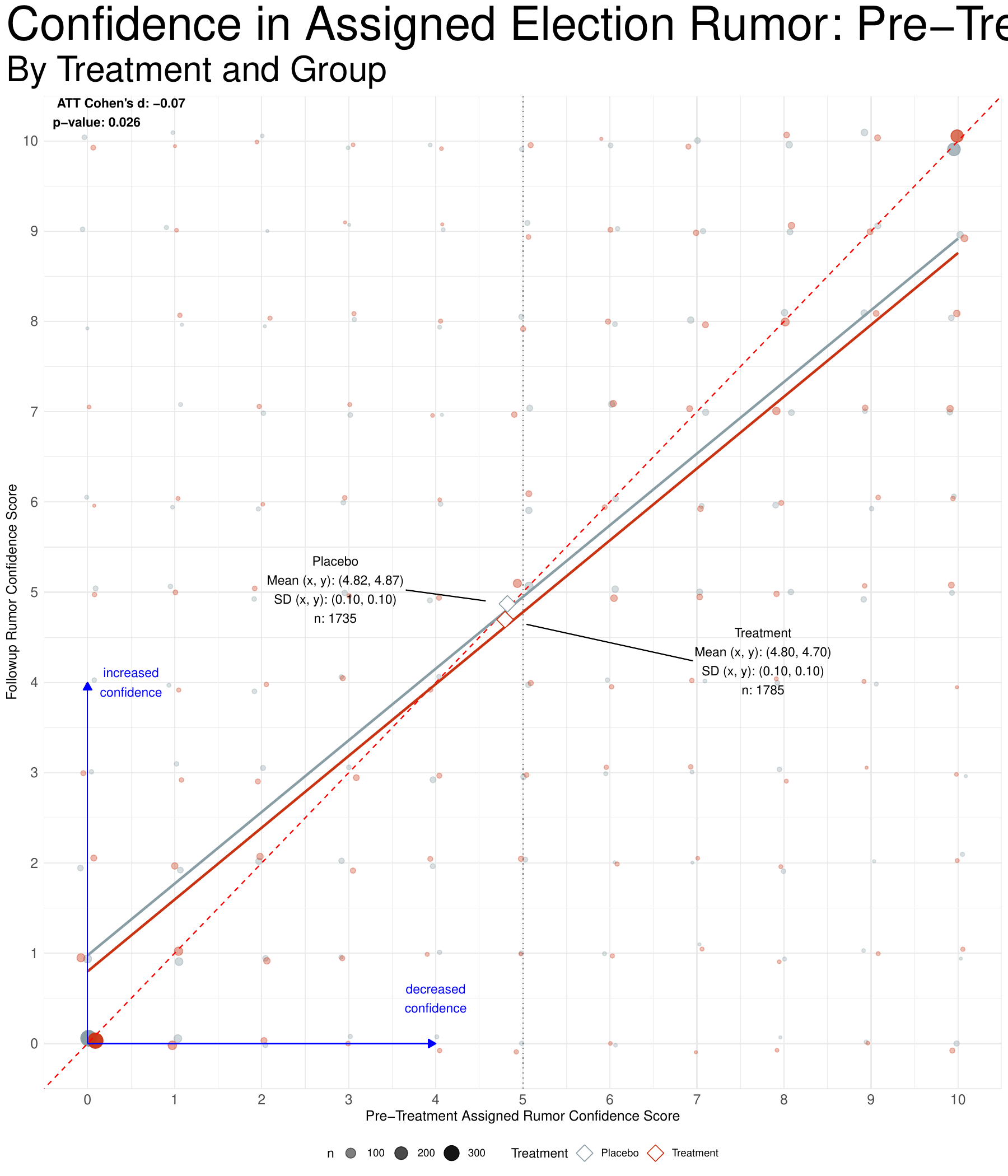}
    \caption{\textbf{Pre-treatment vs. recontact confidence in assigned election rumor, all rumors pooled together. }}
    \label{fig:pre_post_cisa_recontact}
\end{figure}

\newpage
\clearpage

\begin{figure}[ht]
    \centering
    \includegraphics[width=0.9\linewidth]{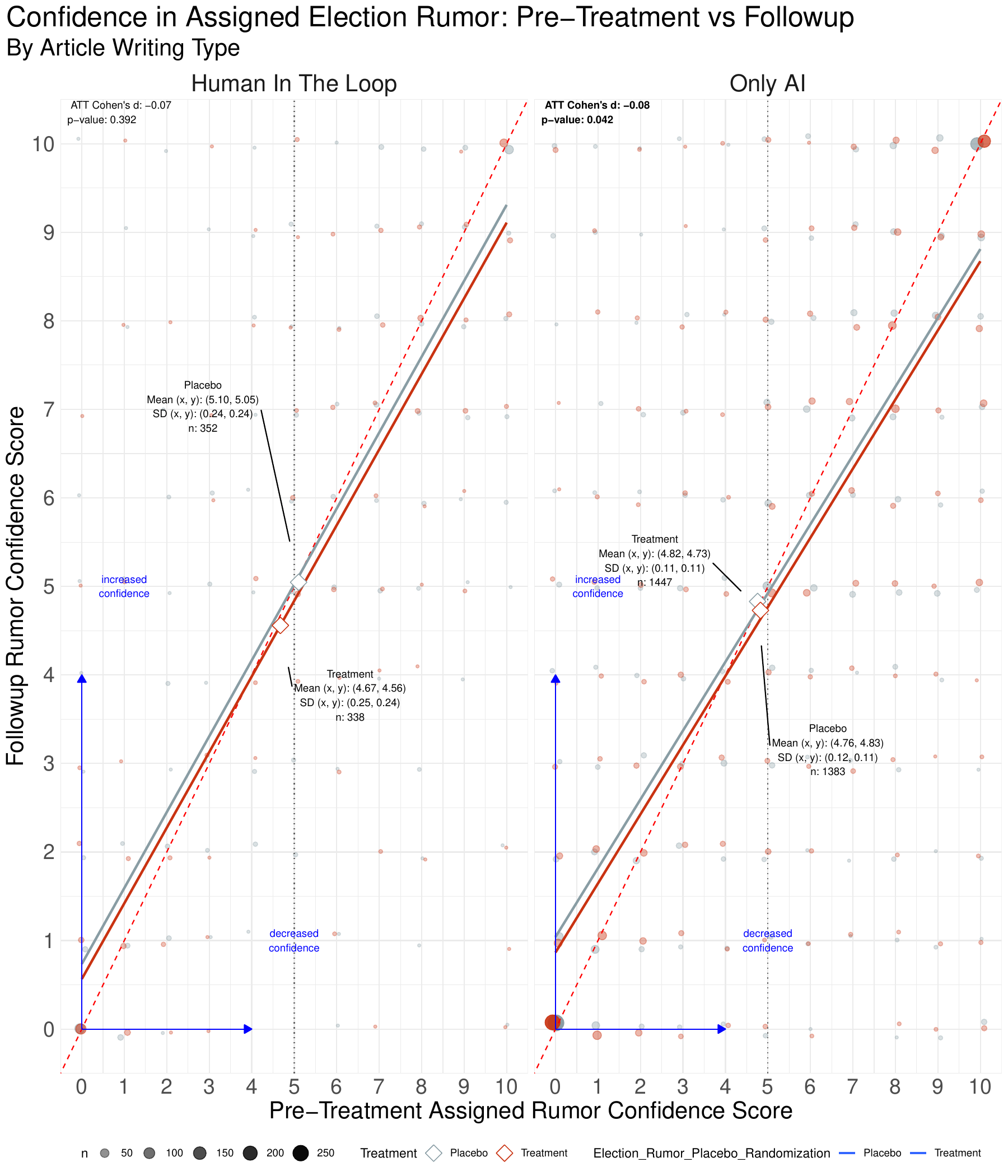}
    \caption{\textbf{Pre-treatment vs. recontact confidence in assigned election rumor, all rumors pooled together, by party. }}
    \label{fig:pre_post_cisa_party_recontact_hitl}
\end{figure}

\newpage
\clearpage

\begin{figure}[ht]
    \centering
    \includegraphics[width=0.9\linewidth]{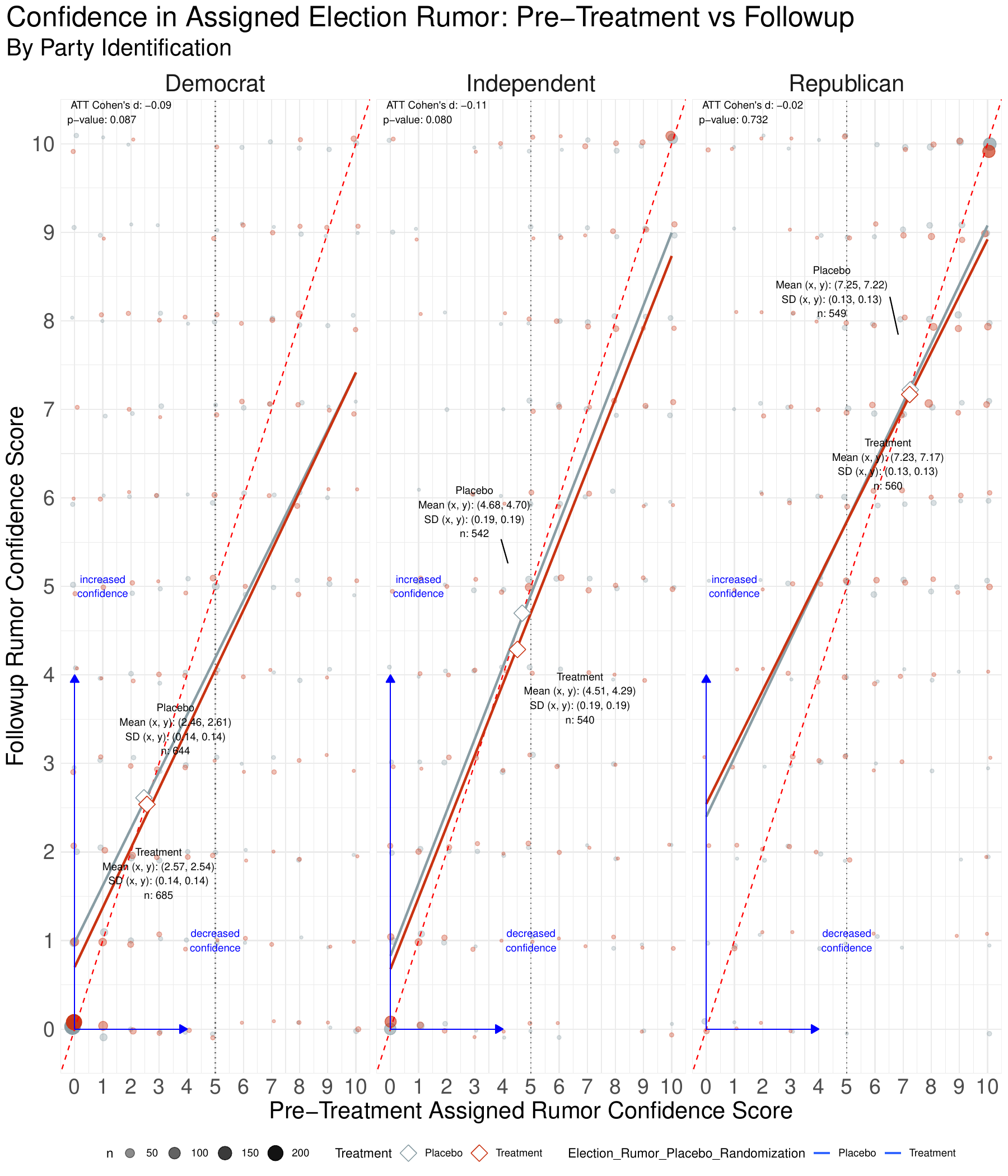}
    \caption{\textbf{Pre-treatment vs. recontact confidence in assigned election rumor, all rumors pooled together, by party. }}
    \label{fig:pre_post_cisa_party_recontact}
\end{figure}

\newpage
\clearpage

\begin{figure}[ht]
    \centering
    \includegraphics[width=0.9\linewidth]{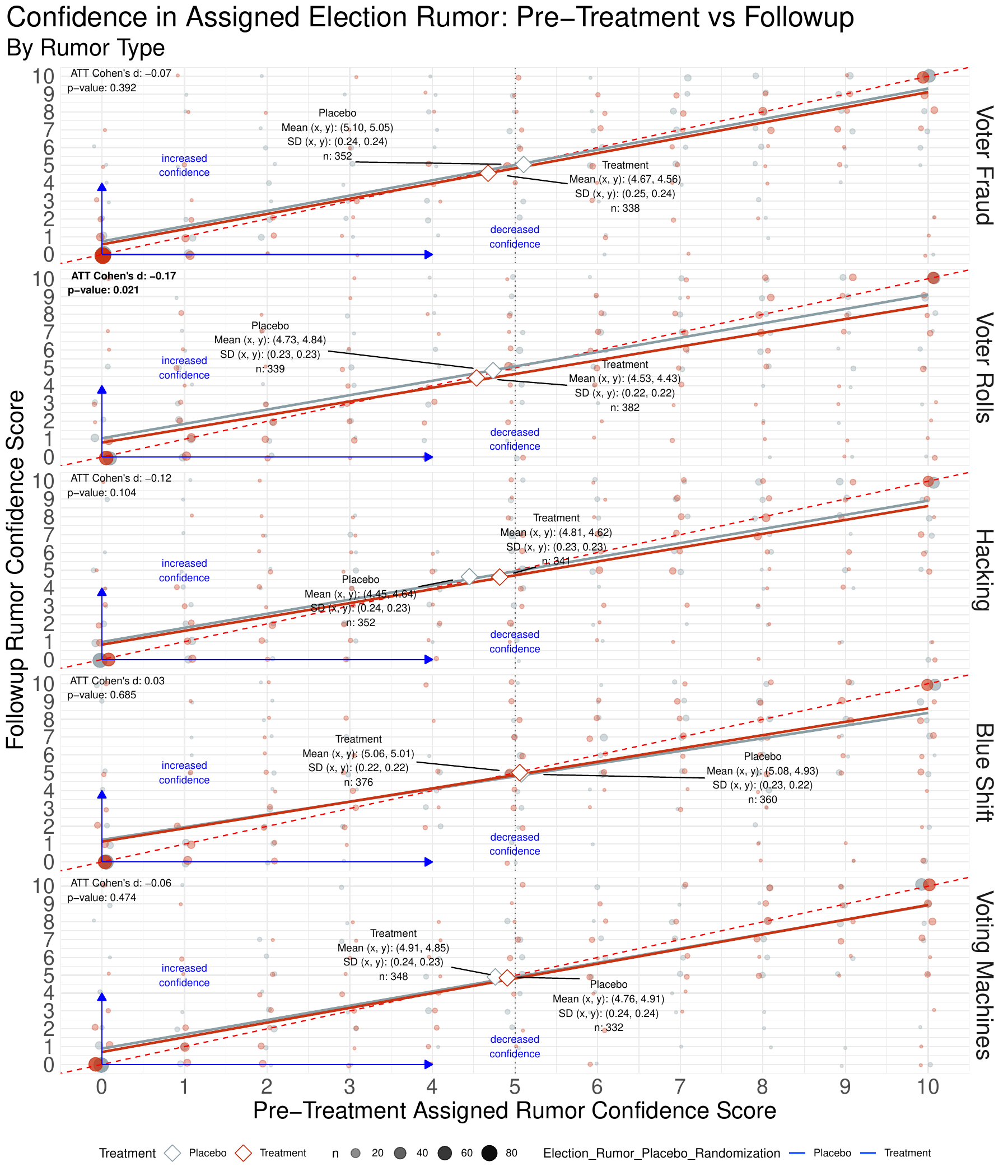}
    \caption{\textbf{Pre-treatment vs. recontact confidence in assigned election rumor. }}
    \label{fig:pre_post_cisa_rumor_recontact}
\end{figure}

\newpage
\clearpage

\begin{figure}[ht]
    \centering
    \includegraphics[width=0.9\linewidth]{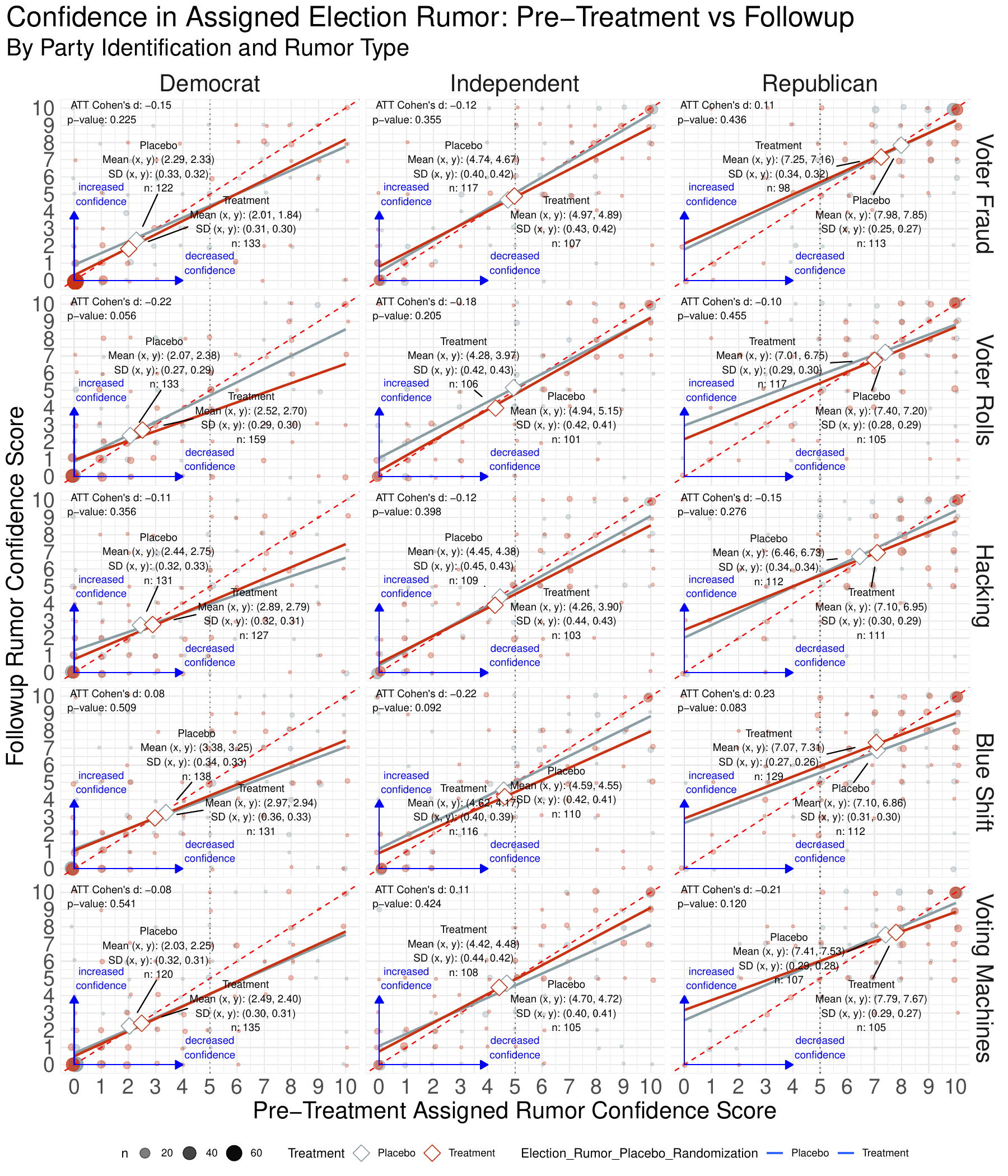}
    \caption{\textbf{Pre-treatment vs. recontact confidence in assigned election rumor, by party. }}
    \label{fig:pre_post_cisa_rumor_party_recontact}
\end{figure}

\subsection*{Data Files}\label{sec:data}
\paragraph{Caption for Data S1.}
\textbf{YouGov Data File}
File: ``prebunk\_full.csv". See variable definition below. Relevant data and analysis code will be made available on Harvard DataVerse upon publication \citep{linegar2024prebunkingdata}: \url{https://doi.org/10.7910/DVN/BLC8KO}.

See variable definitions below.

The following variables were included in Data S1:

\begin{itemize}
\item \textbf{Demographic Variables}
  \begin{itemize}
    \item \texttt{age4}: Age group categories (Under 30, 30-44, 45-64, 65+)
    \item \texttt{gender}: Binary gender (Male, Female)
    \item \texttt{race4}: Race/ethnicity categories (White, Black, Hispanic, Other)
    \item \texttt{educ4}: Education level (HS or less, Some college, College grad, Postgrad)
    \item \texttt{pid3}: Party identification (Democrat, Independent, Republican)
    \item \texttt{ideo3}: Political ideology (Liberal, Moderate, Conservative)
    \item \texttt{region}: Geographic region (Northeast, Midwest, South, West)
    \item \texttt{urbancity}: Residential area type (City, Suburb, Town, Rural area)
  \end{itemize}

\item \textbf{Political Engagement and Beliefs}
  \begin{itemize}
    \item \texttt{newsint}: Political interest (Most of the time, Some of the time, Only now and then, Hardly at all)
    \item \texttt{populism}: Aggregate score from populism belief items
    \item \texttt{conspiracy}: Aggregate score from conspiracy belief items
    \item \texttt{mist}: total number of Misinformation Susceptibility Test (MIST-8) scores correct, from 0-8.
  \end{itemize}

\item \textbf{Ballot Confidence Measures}
  \begin{itemize}
    \item \texttt{ballotcount\_scale}: Pre-treatment confidence in own ballot counting
    \item \texttt{ballotcounty\_scale}: Pre-treatment confidence in county ballot counting
    \item \texttt{ballotcountry\_scale}: Pre-treatment confidence in country ballot counting
    \item \texttt{ballotcount\_2\_scale}: Post-treatment confidence in own ballot counting
    \item \texttt{ballotcounty\_2\_scale}: Post-treatment confidence in county ballot counting
    \item \texttt{ballotcountry\_2\_scale}: Post-treatment confidence in country ballot counting
  \end{itemize}

\item \textbf{Recontact Measures}
  \begin{itemize}
    \item \texttt{ballotcount\_scale\_recontact}: Follow-up confidence in own ballot counting
    \item \texttt{ballotcounty\_scale\_recontact}: Follow-up confidence in county ballot counting
    \item \texttt{ballotcountry\_scale\_recontact}: Follow-up confidence in country ballot counting
  \end{itemize}

\item \textbf{CISA Related Measures}
  \begin{itemize}
    \item \texttt{cisa\_rel}: Initial rumor belief
    \item \texttt{cisa\_rel\_recontact}: Follow-up rumor belief
    \item \texttt{cisa\_fake}: Belief in all rumors
    \item \texttt{cisa\_fake\_recontact}: Follow-up belief in all rumors
    \item \texttt{cisa\_true}: Belief in all facts
    \item \texttt{cisa\_true\_recontact}: Follow-up belief in all facts
  \end{itemize}

\item \textbf{Treatment and Experimental Variables}
  \begin{itemize}
    \item \texttt{electionrumor\_rand}: Randomized rumor assignment (Voter Fraud, Voter Rolls, Hacking, Blue Shift, Voting Machines)
    \item \texttt{electionrumor\_placebo\_rand}: Treatment vs. placebo assignment
    \item \texttt{article\_recalled}: Whether participant recalled article at follow-up
  \end{itemize}

\item \textbf{Survey Weights}
  \begin{itemize}
    \item \texttt{weight}: Survey weight for main sample, as defined in Section \ref{sec:surveymethods_app}
    \item \texttt{weight\_recontact}: Survey weight for recontact sample, as defined in Section \ref{sec:surveymethods_app}
  \end{itemize}

\end{itemize}

Note: All scale variables were coded such that higher values indicate greater agreement or confidence unless otherwise noted.

\end{document}

%% file: data/main_models.tex
\begin{table}[h] \centering 
  \caption{Regression Results for Main Models} 
  \label{tab:main} 
\tiny 
\begin{tabular}{@{\extracolsep{-10pt}}lcccc} 
\\[-1.8ex]\hline 
\hline \\[-1.8ex] 
 & Rumor Post & Rumor Recontact & Post Confidence Country Ballots & Recontact Confidence Country Ballots \\ 
 Election\_Rumor\_Placebo\_RandomizationTreatment & $-$0.551$^{***}$ (0.066) & $-$0.167$^{*}$ (0.071) & 0.189$^{***}$ (0.046) & 0.015 (0.054) \\ 
  Rumor & 0.462$^{***}$ (0.018) & 0.536$^{***}$ (0.020) &  &  \\ 
  Pre\_Confidence\_Country\_Ballots &  &  & 0.729$^{***}$ (0.013) & 0.677$^{***}$ (0.015) \\ 
  Age\_Group30-44 & $-$0.326$^{**}$ (0.115) & $-$0.259$^{*}$ (0.127) & $-$0.015 (0.082) & 0.129 (0.084) \\ 
  Age\_Group45-64 & $-$0.481$^{***}$ (0.109) & $-$0.110 (0.123) & 0.042 (0.080) & $-$0.026 (0.081) \\ 
  Age\_Group65+ & $-$0.392$^{***}$ (0.119) & $-$0.189 (0.129) & 0.030 (0.084) & $-$0.047 (0.090) \\ 
  GenderFemale & 0.177$^{**}$ (0.069) & 0.108 (0.073) & $-$0.012 (0.047) & 0.096 (0.054) \\ 
  Race\_EthnicityBlack & 0.088 (0.126) & 0.157 (0.142) & 0.095 (0.083) & $-$0.019 (0.095) \\ 
  Race\_EthnicityHispanic & $-$0.265$^{*}$ (0.110) & $-$0.279$^{*}$ (0.117) & $-$0.049 (0.081) & $-$0.091 (0.084) \\ 
  Race\_EthnicityOther & 0.074 (0.157) & 0.109 (0.150) & 0.003 (0.094) & $-$0.168 (0.104) \\ 
  Education\_LevelSome college & $-$0.090 (0.088) & $-$0.076 (0.094) & $-$0.142$^{*}$ (0.065) & 0.008 (0.077) \\ 
  Education\_LevelCollege grad & 0.007 (0.093) & $-$0.135 (0.099) & $-$0.142$^{*}$ (0.066) & $-$0.052 (0.076) \\ 
  Education\_LevelPostgrad & $-$0.016 (0.109) & $-$0.082 (0.115) & $-$0.099 (0.071) & $-$0.070 (0.080) \\ 
  Party\_IdentificationIndependent & $-$0.057 (0.094) & 0.034 (0.102) & $-$0.310$^{***}$ (0.062) & $-$0.307$^{***}$ (0.066) \\ 
  Party\_IdentificationRepublican & 0.355$^{**}$ (0.125) & 0.430$^{**}$ (0.131) & $-$0.384$^{***}$ (0.083) & $-$0.409$^{***}$ (0.093) \\ 
  IdeologyModerate & 0.681$^{***}$ (0.098) & 0.398$^{***}$ (0.107) & $-$0.147$^{*}$ (0.061) & 0.0003 (0.068) \\ 
  IdeologyConservative & 1.121$^{***}$ (0.135) & 0.851$^{***}$ (0.145) & $-$0.285$^{**}$ (0.089) & $-$0.145 (0.106) \\ 
  RegionMidwest & $-$0.271$^{*}$ (0.107) & $-$0.252$^{*}$ (0.111) & 0.120 (0.070) & 0.047 (0.082) \\ 
  RegionSouth & $-$0.152 (0.095) & $-$0.179 (0.102) & 0.019 (0.067) & 0.050 (0.078) \\ 
  RegionWest & $-$0.232$^{*}$ (0.106) & $-$0.097 (0.112) & 0.010 (0.071) & 0.171$^{*}$ (0.083) \\ 
  Urban\_RuralSuburb & 0.034 (0.085) & 0.085 (0.092) & 0.029 (0.058) & $-$0.137$^{*}$ (0.067) \\ 
  Urban\_RuralTown & 0.001 (0.114) & $-$0.032 (0.118) & $-$0.061 (0.081) & $-$0.158 (0.095) \\ 
  Urban\_RuralRural area & 0.020 (0.102) & 0.147 (0.112) & $-$0.095 (0.074) & $-$0.211$^{*}$ (0.091) \\ 
  Political\_InterestPol Interest: Some of the time & $-$0.091 (0.079) & $-$0.050 (0.085) & 0.113$^{*}$ (0.053) & 0.097 (0.061) \\ 
  Political\_InterestPol Interest: Only now and then & 0.153 (0.123) & $-$0.016 (0.140) & 0.004 (0.088) & 0.111 (0.105) \\ 
  Political\_InterestPol Interest: Hardly at all & $-$0.015 (0.161) & 0.142 (0.197) & 0.024 (0.131) & $-$0.060 (0.148) \\ 
  Populism\_Score & $-$0.011 (0.010) & $-$0.035$^{**}$ (0.011) & 0.012 (0.007) & 0.012 (0.008) \\ 
  Conspiracy\_Score & $-$0.104$^{***}$ (0.009) & $-$0.088$^{***}$ (0.010) & 0.046$^{***}$ (0.006) & 0.037$^{***}$ (0.006) \\ 
  MIST\_Correct & $-$0.052$^{*}$ (0.026) & $-$0.119$^{***}$ (0.028) & 0.072$^{***}$ (0.019) & 0.099$^{***}$ (0.022) \\ 
  Constant & 5.290$^{***}$ (0.307) & 4.804$^{***}$ (0.332) & 0.370$^{*}$ (0.162) & 1.001$^{***}$ (0.201) \\ 
 \hline \\[-1.8ex] 
Observations & 4,084 & 3,376 & 4,084 & 3,378 \\ 
Log Likelihood & $-$8,827.042 & $-$7,205.407 & $-$7,360.194 & $-$6,278.493 \\ 
Akaike Inf. Crit. & 17,712.080 & 14,468.810 & 14,778.390 & 12,614.990 \\ 
\hline 
\hline \\[-1.8ex] 
\textit{Note:}  & \multicolumn{4}{r}{$^{*}$p$<$0.05; $^{**}$p$<$0.01; $^{***}$p$<$0.001} \\ 
\end{tabular} 
\end{table} 

%% file: data/Human_in_the_Loop_models.tex
\begin{table}[h] \centering 
  \caption{Regression Results for Human in the Loop Models} 
  \label{tab:human_in_the_loop} 
\tiny 
\begin{tabular}{@{\extracolsep{-10pt}}lcccc} 
\\[-1.8ex]\hline 
\hline \\[-1.8ex] 
 & Rumor Post & Rumor Recontact & Post Confidence Country Ballots & Recontact Confidence Country Ballots \\ 
 Election\_Rumor\_Placebo\_RandomizationTreatment & $-$0.477$^{***}$ (0.125) & $-$0.266 (0.145) & 0.311$^{**}$ (0.101) & 0.103 (0.120) \\ 
  Rumor & 0.462$^{***}$ (0.018) & 0.536$^{***}$ (0.020) &  &  \\ 
  Pre\_Confidence\_Country\_Ballots &  &  & 0.729$^{***}$ (0.013) & 0.677$^{***}$ (0.015) \\ 
  Human\_In\_The\_LoopOnly AI & 0.238$^{*}$ (0.101) & $-$0.081 (0.122) & 0.033 (0.079) & $-$0.006 (0.097) \\ 
  Age\_Group30-44 & $-$0.329$^{**}$ (0.115) & $-$0.254$^{*}$ (0.128) & $-$0.020 (0.082) & 0.126 (0.083) \\ 
  Age\_Group45-64 & $-$0.482$^{***}$ (0.109) & $-$0.107 (0.123) & 0.039 (0.080) & $-$0.028 (0.081) \\ 
  Age\_Group65+ & $-$0.390$^{**}$ (0.119) & $-$0.188 (0.129) & 0.027 (0.084) & $-$0.047 (0.090) \\ 
  GenderFemale & 0.180$^{**}$ (0.069) & 0.108 (0.072) & $-$0.013 (0.047) & 0.095 (0.054) \\ 
  Race\_EthnicityBlack & 0.090 (0.126) & 0.158 (0.142) & 0.093 (0.083) & $-$0.020 (0.095) \\ 
  Race\_EthnicityHispanic & $-$0.269$^{*}$ (0.110) & $-$0.280$^{*}$ (0.117) & $-$0.046 (0.081) & $-$0.088 (0.084) \\ 
  Race\_EthnicityOther & 0.073 (0.157) & 0.109 (0.150) & 0.001 (0.094) & $-$0.168 (0.104) \\ 
  Education\_LevelSome college & $-$0.095 (0.088) & $-$0.076 (0.094) & $-$0.141$^{*}$ (0.065) & 0.009 (0.077) \\ 
  Education\_LevelCollege grad & $-$0.001 (0.093) & $-$0.134 (0.099) & $-$0.140$^{*}$ (0.066) & $-$0.050 (0.076) \\ 
  Education\_LevelPostgrad & $-$0.017 (0.109) & $-$0.082 (0.115) & $-$0.099 (0.071) & $-$0.070 (0.080) \\ 
  Party\_IdentificationIndependent & $-$0.058 (0.093) & 0.033 (0.102) & $-$0.309$^{***}$ (0.062) & $-$0.306$^{***}$ (0.066) \\ 
  Party\_IdentificationRepublican & 0.351$^{**}$ (0.125) & 0.429$^{**}$ (0.131) & $-$0.382$^{***}$ (0.083) & $-$0.407$^{***}$ (0.093) \\ 
  IdeologyModerate & 0.685$^{***}$ (0.098) & 0.399$^{***}$ (0.107) & $-$0.148$^{*}$ (0.061) & $-$0.002 (0.068) \\ 
  IdeologyConservative & 1.130$^{***}$ (0.135) & 0.851$^{***}$ (0.145) & $-$0.287$^{**}$ (0.089) & $-$0.148 (0.106) \\ 
  RegionMidwest & $-$0.272$^{*}$ (0.107) & $-$0.252$^{*}$ (0.110) & 0.120 (0.070) & 0.045 (0.082) \\ 
  RegionSouth & $-$0.154 (0.094) & $-$0.180 (0.102) & 0.020 (0.067) & 0.050 (0.078) \\ 
  RegionWest & $-$0.230$^{*}$ (0.106) & $-$0.096 (0.112) & 0.008 (0.071) & 0.168$^{*}$ (0.083) \\ 
  Urban\_RuralSuburb & 0.033 (0.085) & 0.086 (0.092) & 0.028 (0.058) & $-$0.138$^{*}$ (0.067) \\ 
  Urban\_RuralTown & $-$0.010 (0.114) & $-$0.031 (0.118) & $-$0.059 (0.082) & $-$0.155 (0.096) \\ 
  Urban\_RuralRural area & 0.018 (0.102) & 0.149 (0.113) & $-$0.096 (0.074) & $-$0.212$^{*}$ (0.091) \\ 
  Political\_InterestPol Interest: Some of the time & $-$0.092 (0.078) & $-$0.051 (0.085) & 0.113$^{*}$ (0.053) & 0.099 (0.061) \\ 
  Political\_InterestPol Interest: Only now and then & 0.157 (0.123) & $-$0.018 (0.140) & 0.003 (0.088) & 0.112 (0.105) \\ 
  Political\_InterestPol Interest: Hardly at all & $-$0.018 (0.161) & 0.144 (0.197) & 0.022 (0.131) & $-$0.061 (0.148) \\ 
  Populism\_Score & $-$0.011 (0.010) & $-$0.035$^{**}$ (0.011) & 0.012 (0.007) & 0.012 (0.008) \\ 
  Conspiracy\_Score & $-$0.103$^{***}$ (0.009) & $-$0.088$^{***}$ (0.010) & 0.046$^{***}$ (0.006) & 0.037$^{***}$ (0.006) \\ 
  MIST\_Correct & $-$0.051 (0.026) & $-$0.119$^{***}$ (0.028) & 0.072$^{***}$ (0.019) & 0.099$^{***}$ (0.022) \\ 
  Election\_Rumor\_Placebo\_RandomizationTreatment:Human\_In\_The\_LoopOnly AI & $-$0.096 (0.146) & 0.123 (0.166) & $-$0.150 (0.114) & $-$0.108 (0.134) \\ 
  Constant & 5.098$^{***}$ (0.318) & 4.867$^{***}$ (0.350) & 0.347$^{*}$ (0.172) & 1.011$^{***}$ (0.215) \\ 
 \hline \\[-1.8ex] 
Observations & 4,084 & 3,376 & 4,084 & 3,378 \\ 
Log Likelihood & $-$8,824.241 & $-$7,205.142 & $-$7,359.127 & $-$6,277.791 \\ 
Akaike Inf. Crit. & 17,710.480 & 14,472.280 & 14,780.250 & 12,617.580 \\ 
\hline 
\hline \\[-1.8ex] 
\textit{Note:}  & \multicolumn{4}{r}{$^{*}$p$<$0.05; $^{**}$p$<$0.01; $^{***}$p$<$0.001} \\ 
\end{tabular} 
\end{table} 

%% file: data/party_models.tex
\begin{table}[h] \centering 
  \caption{Regression Results for Party Models} 
  \label{tab:party} 
\tiny 
\begin{tabular}{@{\extracolsep{-10pt}}lcc} 
\\[-1.8ex]\hline 
\hline \\[-1.8ex] 
 & Post Confidence Country Ballots & Recontact Confidence Country Ballots \\ 
 Election\_Rumor\_Placebo\_RandomizationTreatment & 0.171$^{*}$ (0.067) & 0.092 (0.071) \\ 
  Pre\_Confidence\_Country\_Ballots & 0.729$^{***}$ (0.013) & 0.676$^{***}$ (0.015) \\ 
  Party\_IdentificationIndependent & $-$0.316$^{***}$ (0.083) & $-$0.206$^{*}$ (0.089) \\ 
  Party\_IdentificationRepublican & $-$0.408$^{***}$ (0.099) & $-$0.384$^{***}$ (0.117) \\ 
  Age\_Group30-44 & $-$0.014 (0.082) & 0.127 (0.084) \\ 
  Age\_Group45-64 & 0.043 (0.080) & $-$0.025 (0.081) \\ 
  Age\_Group65+ & 0.030 (0.084) & $-$0.048 (0.090) \\ 
  GenderFemale & $-$0.012 (0.047) & 0.095 (0.054) \\ 
  Race\_EthnicityBlack & 0.095 (0.083) & $-$0.020 (0.095) \\ 
  Race\_EthnicityHispanic & $-$0.048 (0.081) & $-$0.094 (0.084) \\ 
  Race\_EthnicityOther & 0.003 (0.094) & $-$0.169 (0.104) \\ 
  Education\_LevelSome college & $-$0.143$^{*}$ (0.065) & 0.007 (0.077) \\ 
  Education\_LevelCollege grad & $-$0.142$^{*}$ (0.066) & $-$0.052 (0.076) \\ 
  Education\_LevelPostgrad & $-$0.100 (0.072) & $-$0.071 (0.080) \\ 
  IdeologyModerate & $-$0.147$^{*}$ (0.061) & 0.003 (0.068) \\ 
  IdeologyConservative & $-$0.287$^{**}$ (0.090) & $-$0.146 (0.106) \\ 
  RegionMidwest & 0.121 (0.071) & 0.040 (0.082) \\ 
  RegionSouth & 0.019 (0.067) & 0.045 (0.078) \\ 
  RegionWest & 0.010 (0.071) & 0.165$^{*}$ (0.082) \\ 
  Urban\_RuralSuburb & 0.029 (0.058) & $-$0.136$^{*}$ (0.067) \\ 
  Urban\_RuralTown & $-$0.060 (0.081) & $-$0.157 (0.095) \\ 
  Urban\_RuralRural area & $-$0.095 (0.074) & $-$0.212$^{*}$ (0.091) \\ 
  Political\_InterestPol Interest: Some of the time & 0.113$^{*}$ (0.053) & 0.098 (0.061) \\ 
  Political\_InterestPol Interest: Only now and then & 0.004 (0.088) & 0.113 (0.105) \\ 
  Political\_InterestPol Interest: Hardly at all & 0.024 (0.131) & $-$0.057 (0.148) \\ 
  Populism\_Score & 0.012 (0.007) & 0.012 (0.008) \\ 
  Conspiracy\_Score & 0.046$^{***}$ (0.006) & 0.037$^{***}$ (0.006) \\ 
  MIST\_Correct & 0.072$^{***}$ (0.019) & 0.099$^{***}$ (0.022) \\ 
  Election\_Rumor\_Placebo\_RandomizationTreatment:Party\_IdentificationIndependent & 0.010 (0.108) & $-$0.203 (0.118) \\ 
  Election\_Rumor\_Placebo\_RandomizationTreatment:Party\_IdentificationRepublican & 0.047 (0.112) & $-$0.048 (0.135) \\ 
  Constant & 0.379$^{*}$ (0.164) & 0.971$^{***}$ (0.199) \\ 
 \hline \\[-1.8ex] 
Observations & 4,084 & 3,378 \\ 
Log Likelihood & $-$7,360.096 & $-$6,277.225 \\ 
Akaike Inf. Crit. & 14,782.190 & 12,616.450 \\ 
\hline 
\hline \\[-1.8ex] 
\textit{Note:}  & \multicolumn{2}{r}{$^{*}$p$<$0.05; $^{**}$p$<$0.01; $^{***}$p$<$0.001} \\ 
\end{tabular} 
\end{table} 

%% file: data/rumor_models.tex
\begin{table}[h] \centering 
  \caption{Regression Results for Rumor Models} 
  \label{tab:rumor} 
\tiny 
\begin{tabular}{@{\extracolsep{-10pt}}lcccccccc} 
\\[-1.8ex]\hline 
\hline \\[-1.8ex] 
 & Rumor Post rumor & Rumor Recontact rumor & Post Confidence Own Ballot rumor & Recontact Confidence Own Ballot rumor & Post Confidence County Ballots rumor & Recontact Confidence County Ballots rumor & Post Confidence Country Ballots rumor & Recontact Confidence Country Ballots rumor \\ 
 Election\_Rumor\_Placebo\_RandomizationTreatment & $-$0.476$^{***}$ (0.124) & $-$0.266 (0.145) & $-$0.033 (0.107) & 0.003 (0.112) & 0.177 (0.104) & $-$0.073 (0.125) & 0.310$^{**}$ (0.101) & 0.103 (0.120) \\ 
  Rumor & 0.463$^{***}$ (0.018) & 0.536$^{***}$ (0.020) &  &  &  &  &  &  \\ 
  Pre\_Confidence\_Own\_Ballot &  &  & 0.737$^{***}$ (0.016) & 0.711$^{***}$ (0.018) &  &  &  &  \\ 
  Pre\_Confidence\_County\_Ballots &  &  &  &  & 0.723$^{***}$ (0.016) & 0.680$^{***}$ (0.018) &  &  \\ 
  Pre\_Confidence\_Country\_Ballots &  &  &  &  &  &  & 0.729$^{***}$ (0.013) & 0.677$^{***}$ (0.015) \\ 
  Election\_Rumor\_RandomizationVoter Rolls & 0.776$^{***}$ (0.139) & 0.115 (0.159) & $-$0.033 (0.106) & $-$0.022 (0.111) & $-$0.012 (0.103) & $-$0.067 (0.114) & $-$0.098 (0.105) & $-$0.048 (0.120) \\ 
  Election\_Rumor\_RandomizationHacking & 0.249 (0.130) & $-$0.153 (0.157) & $-$0.273$^{*}$ (0.113) & 0.063 (0.111) & $-$0.077 (0.107) & 0.058 (0.120) & $-$0.021 (0.099) & 0.055 (0.124) \\ 
  Election\_Rumor\_RandomizationBlue Shift & $-$0.104 (0.142) & $-$0.209 (0.159) & 0.133 (0.113) & 0.188 (0.108) & 0.033 (0.104) & 0.094 (0.111) & 0.176 (0.099) & 0.098 (0.118) \\ 
  Election\_Rumor\_RandomizationVoting Machines & 0.030 (0.137) & $-$0.069 (0.160) & $-$0.093 (0.104) & $-$0.070 (0.108) & 0.029 (0.095) & $-$0.134 (0.122) & 0.077 (0.101) & $-$0.139 (0.128) \\ 
  Age\_Group30-44 & $-$0.354$^{**}$ (0.115) & $-$0.256$^{*}$ (0.128) & 0.011 (0.084) & 0.027 (0.078) & $-$0.103 (0.077) & 0.077 (0.086) & $-$0.011 (0.082) & 0.126 (0.083) \\ 
  Age\_Group45-64 & $-$0.500$^{***}$ (0.108) & $-$0.113 (0.124) & 0.089 (0.079) & $-$0.034 (0.075) & 0.036 (0.074) & 0.031 (0.084) & 0.042 (0.080) & $-$0.025 (0.081) \\ 
  Age\_Group65+ & $-$0.413$^{***}$ (0.119) & $-$0.193 (0.129) & 0.247$^{**}$ (0.086) & 0.077 (0.081) & 0.110 (0.078) & 0.041 (0.090) & 0.033 (0.084) & $-$0.043 (0.090) \\ 
  GenderFemale & 0.170$^{*}$ (0.068) & 0.108 (0.073) & $-$0.094 (0.049) & $-$0.015 (0.051) & $-$0.074 (0.047) & 0.030 (0.052) & $-$0.012 (0.047) & 0.094 (0.055) \\ 
  Race\_EthnicityBlack & 0.096 (0.125) & 0.166 (0.142) & 0.063 (0.084) & $-$0.070 (0.083) & 0.134 (0.079) & $-$0.070 (0.087) & 0.095 (0.083) & $-$0.020 (0.095) \\ 
  Race\_EthnicityHispanic & $-$0.256$^{*}$ (0.109) & $-$0.273$^{*}$ (0.117) & 0.092 (0.084) & 0.053 (0.077) & 0.057 (0.078) & $-$0.019 (0.086) & $-$0.048 (0.080) & $-$0.089 (0.084) \\ 
  Race\_EthnicityOther & 0.082 (0.158) & 0.112 (0.150) & 0.100 (0.099) & 0.008 (0.097) & 0.196$^{*}$ (0.094) & 0.128 (0.106) & $-$0.003 (0.094) & $-$0.171 (0.104) \\ 
  Education\_LevelSome college & $-$0.092 (0.088) & $-$0.078 (0.094) & $-$0.032 (0.069) & 0.001 (0.072) & 0.008 (0.065) & 0.031 (0.072) & $-$0.144$^{*}$ (0.065) & 0.008 (0.077) \\ 
  Education\_LevelCollege grad & 0.001 (0.093) & $-$0.136 (0.099) & $-$0.085 (0.069) & $-$0.131 (0.070) & $-$0.037 (0.068) & $-$0.096 (0.075) & $-$0.140$^{*}$ (0.066) & $-$0.051 (0.076) \\ 
  Education\_LevelPostgrad & $-$0.022 (0.108) & $-$0.087 (0.116) & $-$0.012 (0.075) & $-$0.081 (0.077) & 0.016 (0.072) & $-$0.077 (0.079) & $-$0.097 (0.072) & $-$0.064 (0.080) \\ 
  Party\_IdentificationIndependent & $-$0.061 (0.093) & 0.033 (0.102) & $-$0.199$^{**}$ (0.062) & $-$0.240$^{***}$ (0.057) & $-$0.216$^{***}$ (0.060) & $-$0.213$^{***}$ (0.061) & $-$0.308$^{***}$ (0.061) & $-$0.305$^{***}$ (0.066) \\ 
  Party\_IdentificationRepublican & 0.359$^{**}$ (0.124) & 0.430$^{**}$ (0.131) & $-$0.346$^{***}$ (0.087) & $-$0.195$^{*}$ (0.089) & $-$0.284$^{***}$ (0.085) & $-$0.264$^{**}$ (0.089) & $-$0.385$^{***}$ (0.083) & $-$0.404$^{***}$ (0.093) \\ 
  IdeologyModerate & 0.681$^{***}$ (0.097) & 0.396$^{***}$ (0.106) & $-$0.128$^{*}$ (0.062) & $-$0.128$^{*}$ (0.060) & $-$0.106 (0.059) & $-$0.049 (0.060) & $-$0.146$^{*}$ (0.060) & $-$0.0002 (0.068) \\ 
  IdeologyConservative & 1.119$^{***}$ (0.135) & 0.847$^{***}$ (0.145) & $-$0.155 (0.094) & $-$0.172 (0.099) & $-$0.156 (0.092) & $-$0.139 (0.094) & $-$0.281$^{**}$ (0.089) & $-$0.146 (0.106) \\ 
  RegionMidwest & $-$0.276$^{**}$ (0.106) & $-$0.247$^{*}$ (0.110) & 0.207$^{**}$ (0.075) & 0.050 (0.078) & 0.169$^{*}$ (0.072) & 0.072 (0.081) & 0.120 (0.070) & 0.040 (0.082) \\ 
  RegionSouth & $-$0.143 (0.094) & $-$0.174 (0.102) & 0.217$^{**}$ (0.069) & 0.096 (0.074) & 0.191$^{**}$ (0.066) & 0.095 (0.075) & 0.014 (0.067) & 0.048 (0.078) \\ 
  RegionWest & $-$0.213$^{*}$ (0.105) & $-$0.088 (0.112) & 0.121 (0.074) & 0.031 (0.077) & 0.072 (0.071) & 0.093 (0.077) & $-$0.0002 (0.071) & 0.170$^{*}$ (0.083) \\ 
  Urban\_RuralSuburb & 0.050 (0.085) & 0.090 (0.093) & 0.040 (0.060) & $-$0.181$^{**}$ (0.061) & $-$0.010 (0.057) & $-$0.148$^{*}$ (0.063) & 0.024 (0.058) & $-$0.141$^{*}$ (0.067) \\ 
  Urban\_RuralTown & 0.006 (0.113) & $-$0.018 (0.118) & 0.133 (0.081) & $-$0.185$^{*}$ (0.089) & 0.127 (0.078) & $-$0.113 (0.091) & $-$0.060 (0.081) & $-$0.159 (0.095) \\ 
  Urban\_RuralRural area & 0.039 (0.102) & 0.158 (0.113) & $-$0.038 (0.079) & $-$0.073 (0.085) & $-$0.027 (0.079) & 0.095 (0.083) & $-$0.099 (0.074) & $-$0.219$^{*}$ (0.091) \\ 
  Political\_InterestPol Interest: Some of the time & $-$0.100 (0.078) & $-$0.055 (0.085) & $-$0.005 (0.056) & 0.038 (0.057) & $-$0.036 (0.053) & 0.030 (0.059) & 0.115$^{*}$ (0.053) & 0.100 (0.061) \\ 
  Political\_InterestPol Interest: Only now and then & 0.149 (0.123) & $-$0.020 (0.139) & $-$0.023 (0.090) & $-$0.037 (0.093) & $-$0.189$^{*}$ (0.093) & $-$0.031 (0.090) & 0.005 (0.088) & 0.116 (0.105) \\ 
  Political\_InterestPol Interest: Hardly at all & $-$0.004 (0.160) & 0.156 (0.197) & $-$0.108 (0.165) & 0.014 (0.155) & $-$0.001 (0.144) & $-$0.133 (0.164) & 0.017 (0.131) & $-$0.072 (0.148) \\ 
  Populism\_Score & $-$0.010 (0.010) & $-$0.035$^{**}$ (0.011) & 0.010 (0.007) & 0.011 (0.007) & 0.008 (0.007) & 0.006 (0.007) & 0.011 (0.007) & 0.013 (0.008) \\ 
  Conspiracy\_Score & $-$0.103$^{***}$ (0.009) & $-$0.089$^{***}$ (0.010) & 0.041$^{***}$ (0.006) & 0.034$^{***}$ (0.006) & 0.039$^{***}$ (0.006) & 0.036$^{***}$ (0.006) & 0.046$^{***}$ (0.006) & 0.037$^{***}$ (0.006) \\ 
  MIST\_Correct & $-$0.053$^{*}$ (0.026) & $-$0.120$^{***}$ (0.028) & 0.100$^{***}$ (0.021) & 0.039 (0.021) & 0.091$^{***}$ (0.020) & 0.068$^{**}$ (0.021) & 0.073$^{***}$ (0.019) & 0.100$^{***}$ (0.022) \\ 
  Election\_Rumor\_Placebo\_RandomizationTreatment:Election\_Rumor\_RandomizationVoter Rolls & $-$0.612$^{**}$ (0.197) & $-$0.098 (0.220) & 0.266 (0.148) & 0.176 (0.157) & 0.024 (0.146) & 0.190 (0.163) & 0.016 (0.148) & $-$0.132 (0.166) \\ 
  Election\_Rumor\_Placebo\_RandomizationTreatment:Election\_Rumor\_RandomizationHacking & $-$0.081 (0.191) & 0.117 (0.215) & 0.106 (0.157) & $-$0.178 (0.163) & $-$0.116 (0.152) & 0.017 (0.176) & $-$0.153 (0.145) & $-$0.144 (0.174) \\ 
  Election\_Rumor\_Placebo\_RandomizationTreatment:Election\_Rumor\_RandomizationBlue Shift & 0.317 (0.201) & 0.334 (0.219) & $-$0.072 (0.154) & $-$0.170 (0.152) & $-$0.218 (0.147) & $-$0.024 (0.161) & $-$0.360$^{*}$ (0.142) & $-$0.239 (0.165) \\ 
  Election\_Rumor\_Placebo\_RandomizationTreatment:Election\_Rumor\_RandomizationVoting Machines & $-$0.015 (0.193) & 0.127 (0.213) & 0.206 (0.151) & 0.060 (0.161) & $-$0.129 (0.145) & 0.174 (0.176) & $-$0.103 (0.143) & 0.103 (0.172) \\ 
  Constant & 5.107$^{***}$ (0.316) & 4.872$^{***}$ (0.350) & 0.123 (0.198) & 1.454$^{***}$ (0.215) & 0.470$^{**}$ (0.181) & 1.421$^{***}$ (0.218) & 0.343$^{*}$ (0.172) & 0.998$^{***}$ (0.214) \\ 
 \hline \\[-1.8ex] 
Observations & 4,084 & 3,376 & 4,085 & 3,378 & 4,085 & 3,377 & 4,084 & 3,378 \\ 
Log Likelihood & $-$8,802.801 & $-$7,202.192 & $-$7,588.478 & $-$6,031.355 & $-$7,431.664 & $-$6,165.258 & $-$7,353.488 & $-$6,274.646 \\ 
Akaike Inf. Crit. & 17,679.600 & 14,478.380 & 15,250.960 & 12,136.710 & 14,937.330 & 12,404.520 & 14,780.980 & 12,623.290 \\ 
\hline 
\hline \\[-1.8ex] 
\textit{Note:}  & \multicolumn{8}{r}{$^{*}$p$<$0.05; $^{**}$p$<$0.01; $^{***}$p$<$0.001} \\ 
\end{tabular} 
\end{table} 

%% file: data/Other_Election_Integrity_models.tex
\begin{table}[h] \centering 
  \caption{Regression Results for Other Election Integrity Models} 
  \label{tab:other_election_integrity} 
\tiny 
\begin{tabular}{@{\extracolsep{-10pt}}lcccccc} 
\\[-1.8ex]\hline 
\hline \\[-1.8ex] 
 & Rumor Post & Rumor Recontact & Post Confidence Own Ballot & Recontact Confidence Own Ballot & Post Confidence County Ballots & Recontact Confidence County Ballots \\ 
 Election\_Rumor\_Placebo\_RandomizationTreatment & $-$0.551$^{***}$ (0.066) & $-$0.167$^{*}$ (0.071) & 0.074 (0.049) & $-$0.019 (0.049) & 0.089 (0.047) & $-$0.002 (0.052) \\ 
  Rumor & 0.462$^{***}$ (0.018) & 0.536$^{***}$ (0.020) &  &  &  &  \\ 
  Pre\_Confidence\_Own\_Ballot &  &  & 0.736$^{***}$ (0.016) & 0.711$^{***}$ (0.018) &  &  \\ 
  Pre\_Confidence\_County\_Ballots &  &  &  &  & 0.723$^{***}$ (0.016) & 0.680$^{***}$ (0.018) \\ 
  Age\_Group30-44 & $-$0.326$^{**}$ (0.115) & $-$0.259$^{*}$ (0.127) & 0.001 (0.085) & 0.017 (0.077) & $-$0.105 (0.077) & 0.071 (0.086) \\ 
  Age\_Group45-64 & $-$0.481$^{***}$ (0.109) & $-$0.110 (0.123) & 0.091 (0.080) & $-$0.039 (0.074) & 0.040 (0.074) & 0.025 (0.084) \\ 
  Age\_Group65+ & $-$0.392$^{***}$ (0.119) & $-$0.189 (0.129) & 0.246$^{**}$ (0.087) & 0.067 (0.081) & 0.113 (0.079) & 0.035 (0.090) \\ 
  GenderFemale & 0.177$^{**}$ (0.069) & 0.108 (0.073) & $-$0.090 (0.050) & $-$0.011 (0.051) & $-$0.072 (0.047) & 0.031 (0.052) \\ 
  Race\_EthnicityBlack & 0.088 (0.126) & 0.157 (0.142) & 0.058 (0.084) & $-$0.067 (0.083) & 0.133 (0.079) & $-$0.066 (0.087) \\ 
  Race\_EthnicityHispanic & $-$0.265$^{*}$ (0.110) & $-$0.279$^{*}$ (0.117) & 0.094 (0.084) & 0.057 (0.077) & 0.052 (0.078) & $-$0.012 (0.086) \\ 
  Race\_EthnicityOther & 0.074 (0.157) & 0.109 (0.150) & 0.103 (0.101) & 0.010 (0.096) & 0.199$^{*}$ (0.095) & 0.130 (0.106) \\ 
  Education\_LevelSome college & $-$0.090 (0.088) & $-$0.076 (0.094) & $-$0.027 (0.069) & 0.0002 (0.072) & 0.008 (0.065) & 0.030 (0.072) \\ 
  Education\_LevelCollege grad & 0.007 (0.093) & $-$0.135 (0.099) & $-$0.087 (0.069) & $-$0.132 (0.070) & $-$0.042 (0.068) & $-$0.095 (0.075) \\ 
  Education\_LevelPostgrad & $-$0.016 (0.109) & $-$0.082 (0.115) & $-$0.011 (0.075) & $-$0.084 (0.077) & 0.019 (0.072) & $-$0.082 (0.079) \\ 
  Party\_IdentificationIndependent & $-$0.057 (0.094) & 0.034 (0.102) & $-$0.197$^{**}$ (0.062) & $-$0.241$^{***}$ (0.057) & $-$0.216$^{***}$ (0.060) & $-$0.214$^{***}$ (0.061) \\ 
  Party\_IdentificationRepublican & 0.355$^{**}$ (0.125) & 0.430$^{**}$ (0.131) & $-$0.343$^{***}$ (0.087) & $-$0.193$^{*}$ (0.089) & $-$0.284$^{***}$ (0.085) & $-$0.265$^{**}$ (0.089) \\ 
  IdeologyModerate & 0.681$^{***}$ (0.098) & 0.398$^{***}$ (0.107) & $-$0.129$^{*}$ (0.062) & $-$0.133$^{*}$ (0.060) & $-$0.107 (0.059) & $-$0.052 (0.060) \\ 
  IdeologyConservative & 1.121$^{***}$ (0.135) & 0.851$^{***}$ (0.145) & $-$0.160 (0.094) & $-$0.175 (0.100) & $-$0.158 (0.093) & $-$0.140 (0.094) \\ 
  RegionMidwest & $-$0.271$^{*}$ (0.107) & $-$0.252$^{*}$ (0.111) & 0.205$^{**}$ (0.076) & 0.049 (0.078) & 0.168$^{*}$ (0.072) & 0.073 (0.080) \\ 
  RegionSouth & $-$0.152 (0.095) & $-$0.179 (0.102) & 0.225$^{**}$ (0.069) & 0.102 (0.074) & 0.195$^{**}$ (0.066) & 0.098 (0.075) \\ 
  RegionWest & $-$0.232$^{*}$ (0.106) & $-$0.097 (0.112) & 0.130 (0.074) & 0.039 (0.076) & 0.081 (0.071) & 0.094 (0.077) \\ 
  Urban\_RuralSuburb & 0.034 (0.085) & 0.085 (0.092) & 0.041 (0.060) & $-$0.176$^{**}$ (0.061) & $-$0.011 (0.057) & $-$0.146$^{*}$ (0.063) \\ 
  Urban\_RuralTown & 0.001 (0.114) & $-$0.032 (0.118) & 0.125 (0.082) & $-$0.175$^{*}$ (0.089) & 0.122 (0.078) & $-$0.104 (0.091) \\ 
  Urban\_RuralRural area & 0.020 (0.102) & 0.147 (0.112) & $-$0.049 (0.079) & $-$0.070 (0.085) & $-$0.031 (0.079) & 0.098 (0.083) \\ 
  Political\_InterestPol Interest: Some of the time & $-$0.091 (0.079) & $-$0.050 (0.085) & $-$0.006 (0.057) & 0.032 (0.058) & $-$0.037 (0.053) & 0.027 (0.059) \\ 
  Political\_InterestPol Interest: Only now and then & 0.153 (0.123) & $-$0.016 (0.140) & $-$0.022 (0.090) & $-$0.036 (0.093) & $-$0.184$^{*}$ (0.093) & $-$0.031 (0.091) \\ 
  Political\_InterestPol Interest: Hardly at all & $-$0.015 (0.161) & 0.142 (0.197) & $-$0.120 (0.164) & 0.015 (0.155) & $-$0.0001 (0.144) & $-$0.130 (0.164) \\ 
  Populism\_Score & $-$0.011 (0.010) & $-$0.035$^{**}$ (0.011) & 0.010 (0.007) & 0.011 (0.007) & 0.008 (0.007) & 0.006 (0.007) \\ 
  Conspiracy\_Score & $-$0.104$^{***}$ (0.009) & $-$0.088$^{***}$ (0.010) & 0.041$^{***}$ (0.006) & 0.034$^{***}$ (0.006) & 0.039$^{***}$ (0.006) & 0.036$^{***}$ (0.006) \\ 
  MIST\_Correct & $-$0.052$^{*}$ (0.026) & $-$0.119$^{***}$ (0.028) & 0.101$^{***}$ (0.021) & 0.039 (0.021) & 0.092$^{***}$ (0.020) & 0.067$^{**}$ (0.021) \\ 
  Constant & 5.290$^{***}$ (0.307) & 4.804$^{***}$ (0.332) & 0.065 (0.188) & 1.488$^{***}$ (0.208) & 0.454$^{*}$ (0.177) & 1.424$^{***}$ (0.208) \\ 
 \hline \\[-1.8ex] 
Observations & 4,084 & 3,376 & 4,085 & 3,378 & 4,085 & 3,377 \\ 
Log Likelihood & $-$8,827.042 & $-$7,205.407 & $-$7,603.269 & $-$6,037.614 & $-$7,435.890 & $-$6,168.358 \\ 
Akaike Inf. Crit. & 17,712.080 & 14,468.810 & 15,264.540 & 12,133.230 & 14,929.780 & 12,394.720 \\ 
\hline 
\hline \\[-1.8ex] 
\textit{Note:}  & \multicolumn{6}{r}{$^{*}$p$<$0.05; $^{**}$p$<$0.01; $^{***}$p$<$0.001} \\ 
\end{tabular} 
\end{table} 